\documentclass[final,authoryear,3p,times]{elsarticle}
\usepackage{amssymb}
\usepackage{graphicx}
\usepackage{color}
\usepackage[polutonikogreek,english]{babel}
\newcommand*{\tg}[1]{\textgreek{#1}}

\journal{Studies in History and Philosophy of Modern Physics}

\begin{document}

\begin{frontmatter}

\title{Viscosity from Newton to Modern Non-equilibrium Statistical Mechanics}
\author{S\'ebastien Viscardy}\ead{sebastien.viscardy@aeronomie.be}
\address{Belgian Institute for Space Aeronomy, 3, Avenue Circulaire, B-1180 Brussels, Belgium}
\date{\today}

\begin{abstract}
In the second half of the 19th century, the kinetic theory of gases has probably raised one of the most impassioned debates in the history of science. The so-called reversibility paradox around which intense polemics occurred reveals the apparent incompatibility between the microscopic and macroscopic levels. While classical mechanics describes the motion of bodies such as atoms and molecules by means of time reversible equations, thermodynamics emphasizes the irreversible character of macroscopic phenomena such as viscosity. Aiming at reconciling both levels of description, Boltzmann proposed a probabilistic explanation. Nevertheless, such an interpretation has not totally convinced generations of physicists, so that this question has constantly animated the scientific community since his seminal work. In this context, an important breakthrough in dynamical systems theory has shown that the hypothesis of microscopic chaos played a key role and provided a dynamical interpretation of the emergence of irreversibility. Using viscosity as a leading concept, we sketch the historical development of the concepts related to this fundamental issue up to recent advances. Following the analysis of the Liouville equation introducing the concept of Pollicott-Ruelle resonances, two successful approaches --- the \textit{escape-rate formalism} and the \textit{hydrodynamic-mode method} --- establish remarkable relationships between transport processes and chaotic properties of the underlying Hamiltonian dynamics.
\end{abstract}

\begin{keyword}
statistical mechanics \sep viscosity \sep reversibility paradox \sep chaos \sep dynamical systems theory
\end{keyword}

\end{frontmatter}

\tableofcontents


\section{Introduction}

The remarkable advances made in the engineering of steam engines over the 18th century induced a deep transformation of the economic activities, which is usually called industrial revolution. The critical importance of optimizing the engine efficiency resulted in the development of a new science: \textit{thermodynamics}, which emphasized the \textit{irreversible} character of macroscopic phenomena (such as heat dissipation). On the other hand, the development of chemistry and, later, of the kinetic theory of gases, transformed the concept of matter from a continuous to a discrete description. In other words, matter should be considered as made of very small particles: the atoms and the molecules. However, the motion of these entities is governed by the \textit{reversible} equations of Newton. As a result appeared the paradox opposing macroscopic and microscopic descriptions, which has given rise after the seminal works of Maxwell and Boltzmann to a large number of contributions addressing this paradox.

In the sixties a new revolutionary field of the science of the 20th century was born: {\em Chaos}. Sometimes compared to the scientific revolutions implied by quantum mechanics and special relativity, the phenomenon of chaos has been discovered in most of the natural sciences such as physics, chemistry, meteorology, and geophysics. Dynamical systems theory describes systems with few degrees of freedom which, although ruled by deterministic equations, exhibit a stochastic behavior. This remarkable feature led to the idea that the tools of statistical mechanics could be applied to such systems, and as result to the hypothesis of microscopic chaos as the key element of the emergence of irreversibility from the reversible equations of mechanics.

The purpose of this paper is to sketch the historical development of non-equilibrium statistical mechanics by using viscosity as the leading concept, from Newton -- the founder of classical mechanics in its modern form on the one hand, and of the viscosity law on the other hand -- to the recent advances providing a dynamical interpretation of the irreversibility of macroscopic processes without any stochastic assumptions -- such as the {\em Sto\ss zahlansatz} largely used in the frame of the kinetic theory of gases.
 
The paper is organized as follows: we start in section \ref{irreversibility} with the presentation of the discovery of the concept of irreversibility as a fundamental property in nature. Section \ref{hydro-visc} is dedicated to the development of hydrodynamics as the domain of physics being particularly concerned by the process of viscosity. Further, in section \ref{kinetic-theory}, we focus on the kinetic theory that has a long history and has played an important role in the establishment of statistical mechanics, that is to say, in the understanding of the relationship between the microscopic dynamics and macroscopic behaviors. In section \ref{micro-chaos}, we cover the brief history of the development of the theory of chaos and of dynamical systems. Finally, section \ref{mnesm} is dedicated to the recent theories of non-equilibrium statistical mechanics based on the hypothesis of microscopic chaos. Conclusions are drawn in section \ref{conclusions}.


\section{Irreversibility}\label{irreversibility}

\subsection{Mechanics. Energy conservation and reversibility}

The history of modern science holds its richness in the variety of conceptions developed successively throughout the years.\footnote{For general information on this section, see, for example, \citet{brush-SP} and \citet{holton-brush}.} In the 17th century, it was the doctrine of the \textit{clockmaker God} or \textit{mechanical philosophy} which prevailed above the others.\footnote{For a history of the establishment of this doctrine, see, for example, \citet{BOAS1952}.} The development of these ideas began with the work of Boyle (1627-1691) and others. According to his doctrine, the world would work like a perfect clock: once created and energized, it can run forever in a deterministic way and without any need for a divine intervention. To ensure that this ``clockwork universe" never runs down, Descartes (1596-1650) more or less intuitively introduced the statement that the total amount of  quantity as the \textit{scalar} momentum $mv$ (quantity of matter\footnote{ More precisely, Descartes considered the volume of the bodies instead of their mass, the volume being for him the real measure of quantity of matter to be considered. See \citet{descartes-laws}, note 14.} multiplied by its scalar speed).\footnote{In 1633, Descartes had already worked out his main concepts on motion in a treatise entitled {\em Le Monde}. However, this was rejected for publication because of its positiµon in favor of the heliocentrism. His views on motion were later published in one of his masterpieces {\em Principia Philosophiae} \citep{descartes_1644}, whereas {\em Le Monde} finally appeared  after his death \citep{descartes_1664}.} However, as Descartes himself later observed, experiments did not confirm his enunciation of the conservation law of motion. Huygens (1629-1695) corrected it in 1668 by modifying it into the \textit{vector} form. Huygens also worked on the problem of collisions.\footnote{This work appeared in his {\em De motu corporum ex percussione} published for the first time in the {\em Opuscula posthuma} eight years after his death \citep{huygens_motion}. According to \citet{blackwell_huygens}, the exact date of the manuscript version of {\em De motu} is unknown, although it is later to 1673.} This revised law allowed him to claim that the vector sum of the product of the mass by the vector velocity $m\mathbf{v}$ remained unchanged after a collision even if it was inelastic and with dissipation of energy after the collision itself. Introducing this modification, it then appeared that the world could stop after a certain amount of time. This was totally contradictory with the clockworld concept.

The only way to avoid this possibility was to postulate that the matter be composed of elastic particles. If macroscopic objects lose some motion after a collision, it is only in appearance because the motion is transferred to the objects' invisible particles. Therefore, the idea emerged that heat was related to a rapid motion of the invisible parts inside the macroscopic bodies.

In the 17th century, the study of the motion of bodies and its change after collisions was of great importance. In addition to this study, another quantity $mv^2$\label{original_vis_viva} was introduced for the first time. Remarkably, its conservation law in elastic collisions was proved by the same who helped to formulate the conservation of momentum: namely, Huygens.
Later, in his {\em Specimen dynamicum}, Leibniz (1646-1716) called this quantity \textit{vis viva} (living force), to which he added the {\em vis mortua} (dead force) which was associated with the tendency of bodies to begin motion \citep{leibniz_1695}. He precisely asserted that:
\begin{quote}
Force is twin. The elementary force, wich I call {\em dead} because motion does not yet exist in it, but only a solicitation to motion, is like that of a sphere in a rotating tube or a stone in a sling.

The other is the ordinary force associated with actual motion, an I call it {\em living}.

Examples of dead force are provided by centrifugal force, by gravity or centripetal force, and by the force with which a stretched spring starts to contract.

But in percussion that is produced by a body which has been falling for some time, or by an arc which has been unbending for some time, or by any other means, {\em the force is living and born of an infinity of continued impressions of the dead force}.\footnote{Quoted in \citet[p. 221]{dugas_book} from \citet{leibniz_1695}'s {\em Specimen dynamicum}.}
\end{quote}
In the 18th century, Emilie du Ch\^{a}telet (1706-1749) strongly defended Leibniz's concepts in asserting that the {\em vis mortua} could be transformed into {\em vis viva}, and vice versa \citep{chatelet_1740}. These major contributions induced the first step to the central quantity acknowledged in mechanics: the energy. The concept of \textit{vis viva} was used up to the 19th century, when the factor $\frac{1}{2}$ was added and it then became {\em kinetic energy}\footnote{See footnote \ref{vis_viva_kinetic_energy}.} while the name of {\em vis mortua} was replaced by {\em potential energy}.

The world-machine concept is often attributed to Newton (1642-1727) because of the importance played by his \textit{Principia} \citep{newton-principia-1ed}.\footnote{The celebrated {\em Philosophi\ae \ Naturalis Principia Mathematica} originally published in latin in 1687 has been translated on several occasions, e.g. by I. B. Cohen and A. Whitman in \citet{newton-principia}.} In his masterpiece, the Newtonian dynamics is depicted for the first time although it clearly appeared that the author was opposed to it as he pointed out in his \textit{Opticks} \citep{newton-opticks-1ed} 
\begin{quote}
By reason of the [...] weakness of elasticity in solids, motion is much more apt to be lost than got, and is always upon the decay.\footnote{Quoted from  \citet[Query 31, p. 397]{newton-opticks}.}
\end{quote}
Therefore, in a certain sense, he suggested the existence of the dissipation of motion in the world. However, it should be specified that this viewpoint has to be considered in the light of the polemic between Newton\footnote{More precisely, his student and faithful friend Clarke (1675-1729).} and Leibniz concerning the role of God.\footnote{For further information about this debate, see, for example, \citet{koyre-book} and \citet{vailati_1997}.}

During the 18th century, the concept of the ``Newtonian world-machine" continued to dominate the thought of scientists. However, because the Principia were difficult to be applied, the Newtonian paradigm was many times reformulated in order to be easier in practice for calculation. In particular, avoiding the use of the notion of {\em force} introduced in Newton's formalism, Lagrange (1736-1813) proposed the concept of minimization of {\em action} in his {\em M\'ecanique analytique} \citep{lagrange_1788}. Later, in the 1830s, using his variational principle, Hamilton (1805-1865) derived from Lagrangian mechanics the so-called {\em Hamiltonian formalism} based on a function, namely the {\em Hamiltonian} $H$, representing the total energy (nowadays known as the kinetic and potential energies) of the $N$-particle system \citep{hamilton_1834,hamilton_1835}:\footnote{Note that modern notations are used in this paper.}
\begin{equation}
H = \sum_{a=1}^N \frac{p_a^2}{2m} + V({\bf r}_1,{\bf r}_2, \dots , {\bf r}_N),
\label{hamiltonian}
\end{equation}
where ${\bf r}_1,{\bf r}_2, \dots , {\bf r}_N$ and ${\bf p}_1,{\bf p}_2, \dots , {\bf p}_N$ are respectively the positions and momenta of the $N$ particles. Hence, Newton's equations can be substituted by more efficient Hamilton's equations of motion\footnote{For a history of mechanics, see, for example, \citet{dugas_book}.	}:
\begin{eqnarray}
\frac{d {\bf r}_a}{dt}  & = & \frac{\partial H}{\partial {\bf p}_a} \label{hamilton_equations1} \\
\frac{d {\bf p}_a}{dt}  & = & - \frac{\partial H}{\partial {\bf r}_a} \label{hamilton_equations2}
\end{eqnarray}

These equations have the important property of time reversibility. Indeed, the transformation $t \rightarrow -t$, which also implies the transformation ${\bf p}_a \rightarrow - {\bf p}_a$, leaves Eqs. (\ref{hamilton_equations1}-\ref{hamilton_equations2}) unchanged. Consequently, the dynamics described by Hamiltonian equations with energy conservation ($H = {\rm const}$) are reversible in time: past and future play the same role.


\subsection{Thermodynamics. Energy conservation and irreversibility}\label{thermodynamics}

Although the mechanical philosophy claiming that the world works like a machine was triumphant in the 18th century, at the same time, the question of Earth interior's temperature gave rise to another viewpoint and to a new science: Geology.\footnote{For further information, see, for example, \citet{brush-geophysics} and \citet{gohau_1990}.} Yet in 1693 Leibniz thought that the Earth was originally hotter and cooled down, at least on the outside. Later Buffon (1707-1788) studied this question by leading some experiments on the cooling of heated spheres of iron. By considering a molten Earth at the origin he found that our planet was about 75,000 years \citep{buffon}.

At the opposite, James Hutton (1726-1797), one of the founders of Geology, disagreed with this theory of cooling. Defending his {\em Uniformitarianism}\footnote{This principle tells that the geological processes in the past have to be explained by using only the laws and physical processes that can now be observed.} and accepting the hypothesis that the interior of the Earth is much hotter than its surface, he thought this situation had been like that forever \citep{hutton_1795}. To him the geological processes are cyclic: alternance of periods of erosion and denudation implying the destruction of the mountains, and periods of uplifts of new continents as a result of the subterranean fires. Actually one of his disciples Playfair (1748-1819) promoted the uniformitarianist viewpoint by referring the mathematical works by Lagrange and Laplace showing the cyclic movement of the planets around the sun \citep{playfair_1802}. During the 19th century, the doctrine of uniformitarianism was perpetuated by Lyell (1797-1875) who publicized it for society at large \citep{lyell_1830}. In particular, Darwin (1809-1882) relied heavily on this viewpoint of Geology in establishing his theory developed in his famous book {\em On the Origin of Species by Means of Natural Selection} \citep{darwin}. Uniformitarianism, which thus rejected the theory of Earth's cooling, was attacked by Thomson (1824-1907).\footnote{Recall that William Thomson is also known under Lord Kelvin, title received in 1892.} His arguments were mainly based on the dissipation principle mathematically expressed by Fourier \citep{thomson_1862,thomson_1866}. In the early 19th century, Fourier (1768-1830) was also interested in the problem of Earth's cooling and developed the first mathematical theory describing the propagation of heat. In 1819, in his \textit{M\'emoire sur le refroidissement s\'eculaire du globe terrestre} \citep{fourier-memoire}, he came up with an equation:
\begin{equation}
\frac{\partial T}{\partial t} = \alpha \; \nabla^2 T
\label{fourier_equation}
\end{equation}
where $T$ is the temperature and $\alpha$ the thermal diffusivity. Eq. (\ref{fourier_equation}) presents a significant feature: unlike Hamilton's equations (\ref{hamilton_equations1}-\ref{hamilton_equations2}), Fourier's equation is \textit{irreversible} since the transformation $t \rightarrow -t$ introduces a minus sign. Consequently, any system in which a temperature difference exists presents a heat flow from the high to low temperature. This was the discovery of an equation describing processes in which the past did not play the same role as the future. His theory presents an important turning point in the history of physics due to the powerful analysis he developed --- what we now call Fourier's analysis ---, and because it is explicitly based on the postulate of irreversibility. In the first half of the 19th century the eventual contradiction between the reversible Newtonian dynamics and the irreversible heat conduction did not appear because the most studied problem at this time was at the phenomenological level and the Newtonian mechanics had already been successfully applied to problems with dissipative forces (friction, etc.). The contradiction appeared during the first attempt to explain macroscopic processes in terms of the assumed reversible Newtonian dynamics of the particles composing the system. 

A few years after Fourier's work on the propagation of heat, Carnot (1796-1832) published in 1824 his essay \textit{R\'{e}flexions sur la force motrice du feu} \citep{carnot}. While the original issue that stimulated him was the efficiency of steam engine, he rather imagined all ``possible" engines which would be able to produce work. He then considered an abstract ``fire machine" which consisted of a hot and a cold reservoir separated by any gas of which the volume varied with temperature. Working in the framework of the \textit{caloric theory}\footnote{\label{caloric_theory} The caloric theory appeared in the 18th century and was developed by Lavoisier (1743-1794). This theory supposed that heat was a fluid composed of particles independent of the rest of matter. These particles repel each other but are attracted to the particles of ordinary matter. According to this theory, it has to be an eventual material substance and thus should be limited. Hence the increase of the temperature of water by rotating a drill should be due to the transfer of the so-called \textit{caloric} from the drill to the water. But, in 1798, Thompson, Count Rumford (1753-1814) realized that one can produce heat without any limit \citep{thompson_1798}. He then concluded that heat is not a chemical substance or a material substance but is the expression of a movement. However, since he did not propose any alternative theory of heat, the caloric theory had an important influence until the 1830s and the works carried out especially by Joule.} and assuming the conservation of caloric\footnote{This assumption was justified by the fact that the caloric was viewed as a material substance (see note \ref{caloric_theory}), which could thus neither be created nor destroyed.}, Carnot drew an analogy with hydraulics in which a waterfall can be used to do work. He then concluded that this is the temperature difference existing between two bodies  (and not the transport of steam) that gave the possibility of doing work by allowing heat to expand a gas as the heat flows from the hot body to the cold one.

Aiming to determine the maximum efficiency of an engine in terms of the temperature of the two heat reservoirs, Carnot imagined an ideal engine in which the subsequent transformations making the famous {\em Carnot cycle} are ``perfect" (i.e., we would speak today of transformations without energy dissipation, or simply reversible transformations). He hence showed that only such an engine could reach the maximum efficiency which was proportional to the temperature difference between the two reservoirs and lower when the temperature of the hot body was higher.

Nevertheless, neither engineers nor scientists payed attention to Carnot's memoir. This lack of interest might be explained by the fact that, on the one hand, the concept of ideal engines was considered too abstract by the former and, on the other hand, he avoided to provide any mathematical analysis of his engines. As a result, his masterpiece was completely forgotten for the next ten years, until Clapeyron (1799-1864) thought further investigation based on Carnot's ideas would be valuable \citep{clapeyron_1834}. Assuming the heat conservation as Carnot did, he then made easier their understanding by providing a graphical representation and trying to give a mathematical description of cycles. Although he did not reach his purpose, it is thanks to him that Thomson and Clausius later became aware of Carnot's essay.

The heat conservation and thus the caloric theory assumed by Carnot and Clapeyron were more and more discarded in the next years, especially when Mayer (1814-1878) \citep{mayer} and later by Joule (1818-1889) \citep{joule} refuted it and brought up a fundamental statement: work and heat are actually two different expressions of the same quantity: namely, the energy.\footnote{The discovery of the principle of energy conservation is a typical case of simultaneous works. This question has been investigated in \citet{kuhn_1959}.} Indeed, although heat and mechanical work seem to be two independent quantities, we now understand that heat can be transformed into work and vice versa. As a result of Joule's results, in 1850, Clausius (1822-1888) introduced another quantity which remained conserved over a cyclical process, that is the {\em internal energy} or simply the {\em energy} $E$ \citep{clausius_1850}. By means of this new quantity, the equivalence between heat and work discovered by Joule can be reformulated as:
\begin{equation}
dE = dQ+dW \; .
\end{equation}
Hence, this new quantity is a state function, depending on the state variables of the system, so that the energy change $dE$ over a cycle or in an isolated system is zero: 
\begin{equation}
dE = 0 \qquad {\rm (isolated \ systems)}.
\label{first-principle}
\end{equation}
A science of transformations was born: \textit{thermodynamics}. 

A few years later, \citet{clausius-1854} considered that the efficiency of real engines was between the maximum efficiency of Carnot's engine and that of systems in which an irreversible heat flow occurred from a hot body to a colder one without producing work (vanishing efficiency). He concluded that the loss of efficiency was due to the irreversible character of heat conductivity. It then became necessary to define another quantity $dQ/T$ as a measure of the quantity of work lost during the heat transfer \citep{clausius-1854}. Later, he gave it the famous name \textit{entropy}\footnote{Let us quote \citet{clausius-1865} himself:
\begin{quote}If we wish to designate $S$ by a proper name we can say of it that it is the transformation content of the body, in the same way that we say of the quantity [$E$] that it is the heat and work content of the body. However, since I think it is better to take the names of such quantities as these, which are important for science, from the ancient languages, so that they can be introduced without change into all the modern languages, I proposed to name the magnitude $S$ the {\em entropy} of the body, from the Greek word \tg{<h trop`h}, a transformation. I have intentionally formed the word {\em entropy} so as to be as similar as possible to the word {\em energy}, since both these quantities, which are to be known by these names, are so nearly related to each other in their physical significance that a certain similarity in their names seemed to me advantageous. (\citet[p.~390]{clausius-1865}, translation from \citet[p.~234]{magie_1935}.) 
\end{quote}
} \citep{clausius-1865}
\begin{equation}
dS = \frac{dQ}{T} \qquad {\rm (reversible \ transformations)}.
\label{clausius-inequality}
\end{equation}
Extending his results to the entire universe, he ended his seminal paper with the enunciation of the first and second laws of thermodynamics:
\begin{quote}
For the present I will confine myself to announcing as a result of my argument that if we think of that quantity which with reference to a single body I have called its entropy, as formed in a consistent way, with consideration of all the circumstances, for the whole universe, and if we use in connection with it the other simpler concept of energy, we can express the fundamental laws of the universe which correspond to the two fundamental laws of the mechanical theory of heat in the following simple form.

1) The energy of the universe is constant. \\
2) The entropy of the universe tends toward a maximum.\footnote{Quoted from \citet[p.~400]{clausius-1865}, translation from \citet[p.~236]{magie_1935}.}
\end{quote}
In mathematical terms, this may thus be expressed as follows:
\begin{equation}
\left \lbrace
\begin{array}{l}
dE = 0  \\
dS \ge  0 \\
\end{array}
\right.
\qquad {\rm (isolated \ systems)}.
\label{thermo_laws}
\end{equation}

As a result, the original problem giving birth to thermodynamics led to the separation between the concepts of conservation and reversibility. In mechanical transformations, the conservation of energy --- at the beginning the \textit{vis viva}\footnote{Whereas the {\em vis viva} (living force) was initially $m v^2$ (see p.~\pageref{original_vis_viva}), Coriolis (1792-1843) redefined it as $\frac{1}{2} m v^2$ in order to be equal to the work it can produce \citep[p.~III]{coriolis_1829}. In 1853 and 1855, Rankine (1820-1872) proposed before the Philosophical Society of Glasgow to call this quantity {\em actual energy}. The latter together with {\em potential energy} constitute the two forms that the energy may take \citep{rankine_1855_1,rankine_1855_2}. In a discourse given in 1856, Thomson --- now known as Lord Kelvin --- upheld this proposition:
\begin{quote}
The energy of motion may be called either ``dynamical energy" or ``actual energy". The energy of a material system at rest, in virtue of which it can get into motion, is called ``potential energy", or, generally, motive power possessed among different pieces of matter, in virtue of their relative positions, has been called potential energy by Rankine (from the conference entitled {\em On the Origin and Transformation of Motive Power} (1856), reprinted in \citet[vol. II, p.~425]{thomson_1894}).
\end{quote}
However shortly later, he and Tait (1831-1901) replaced the term {\em actual energy} by {\em kinetic energy} which is now in general use \citep{tait_thomson_1862}.\label{vis_viva_kinetic_energy}} --- coincides with the idea of reversibility. In contrast, the physico-chemical transformations can preserve the energy while they cannot be reversed. In other words, the two laws of thermodynamics combine the two concepts of energy conservation and nonreversibility of macroscopic phenomena.\footnote{For further information on the history of thermodynamics, see, for example, \citet{locqueneux_2009}.}

As we shall see in Section \ref{kinetic-theory}, the atomic theory increased its influence during the 19th century. The development of the kinetic theory of gases played an important role proving the discrete character of matter. One of the most important leaders in this area was Boltzmann (1844-1906). He particularly proposed a more general definition of entropy, in terms of the probabilities of molecular arrangements, that can change even when there is no heat flow (e.g. the system becomes more disordered). His theory based on a statistical description was a first understanding of the link between the atomic level and the macroscopic phenomena, that is between the reversible microscopic dynamics and the macroscopic irreversible processes.

In the nineteenth-century society, in which social and economical activities, science, and technology induced a progress never observed before in the history of Humanity, arose the idea of the evolution of species up to Darwin (1809-1882). According to him, the species that we know today are the result of a long process of natural selection \citep{darwin}. This mechanism has induced the production of increasingly complex and organized beings, from unicellular bacteries to mammals such as human beings. Whereas thermodynamics introduced a quantity measuring the continuous growth of disorder, biology put in evidence the continuous growth of order and organization in the biological world. These two points of view influenced some philosophers and writers. On the one hand, the Darwinism strongly influenced philosophers like Spencer (1819-1903) who thought that all in universe goes gradually from a state of confused homogeneity to a differentiated heterogeneous state. On the other hand, Adams (1838-1918), influenced by the discovery of constant dissipation of energy in the universe, expressed a pessimistic vision concerning the future of the society, as argued in \textit{The Degradation of the Democratic Dogma} \citep{adams-book}, and made the second law of thermodynamics an explicit basis for the tendency of history.\footnote{For further information on Adams' concepts on history inspired by the second law of thermodynamics, see, for example, \citet{burich_1987}.} In the literature, the dissipation principle was at the origin of novels such as \citet{flammarion}'s {\em La fin du monde}.\footnote{For further information on the impact of the second law of thermodynamics on the French literature, see, \citet{thiher_2001}. On the metaphoric use of the concept of entropy in art, literature, and linguistics, see \citet{shaw_1983}. \citet{hiebert_1966} discusses the influence of thermodynamics on religion.} Furthermore, \citet{zola}'s {\em Le Docteur Pascal} explicitly addressed the problem of reconciling the increasing organization of the biological systems with the second law of thermodynamics.\footnote{See \citet[p.~156-157]{thiher_2001}.} These two points seemed to be contradictory. One had yet to wait for the development of non-equilibrium thermodynamics, especially attributed to Prigogine (1917-2003)  and his coworkers, to resolve this apparent contradiction. They showed that, out of thermodynamic equilibrium, i.e., in {\em open} systems, matter is able to exhibit self-organization, i.e., the so-called {\em dissipative structures}, although the internal production of entropy is absorbed by the environment.\footnote{A non-technical introduction can be found in \citet{prigogine_1972a,prigogine_1972b}. For further information, see \citet{Glansdorff-prigogine} and \citet{nicolis-prigogine,nicolis-prigogine2}. In addition, note that the out-of-equilibrium condition is not necessary for observing a spontaneous organization. For example, phospholipids which consist of hydrophobic and hydrophilic ends form a stable structure in an aqueous solution whereas the system is at the thermodynamic equilibrium. Such an instance is associated with the so-called {\em self-assembly} instead of {\em self-organization}. The main criterion used to distinguish these two kinds of spontaneous organization is their thermodynamic features. While the self-organization is observed in far-from-equilibrium systems with energy dissipation, one speaks of self-assembly in closed systems which minimize their free energy. See, for example, \citet{jones_2004}.} These recent  works have pointed out the importance of non-equilibrium phenomena and have led to important advances on the properties of non-equilibrium systems, as will be seen in next sections.


\section{Hydrodynamics and viscosity}\label{hydro-visc}

\subsection{Euler's equations}

The vital importance of water throughout all civilizations induced  a special interest for the study of the properties and behaviors of fluids (in particular of water). The earliest quantitative application of ``real fluid" or viscosity effects was by the ancient Egyptian Amenemhet ($\sim$1600 BC). He made a $7$ degree correction to the drainage angle of a water clock to account for the viscosity change of water with temperature (which can be significant between day and night in this region). Moreover, in general the early centres of civilization in Egypt, Mesopotamia, India and China systematically used various machines for irrigation and water supply.\footnote{For further information, see, for example, \citet{scott-blair}.} In particular, Archimedes (287-212 BC) is not only considered as the father of hydrostatics for his famous law, but has also developed the so-called {\em Archimedes screw}, a water elevating machine, which has been used for different purposes (e.g. to extract water from rivers). It is worth also noting that Hero of Alexandria (first century A.D.) treated in \textit{Mechanica} the problem of friction between a body and a plane and, considering a horizontal, frictionless plane, he said 
\begin{quote}
We demonstrate that a weight in this situation can be moved by a force less than any given force.\footnote{Quoted by \citet[p.~289]{russo}.}
\end{quote}

The earliest extensive treatise devoted to practical hydraulics is a manual on urban water supply written by Roman soldier and engineer Sextus Julius Frontinus (first century BC), inspector of the aqueducts and the public fontains in Rome. In his book {\em De Aquaeductibus urbis Romae}\footnote{His book was translated by R. H. Rodgers in \citet{frontinus}.} he noted that the amount of water discharged through an orifice in a given interval of time depends, not only on the size of the orifice, but also on its depth $h$ below the surface of the water in the reservoir from which the water is supplied. Starting upon the results of Galileo's experiments with falling bodies, Torricelli (1608-1647) came to the conclusion that the velocity $v$ of the water exiting out of the reservoir is proportional to the square root of the depth $h$ \citep{torricelli_1641}. This is \textit{Torricelli's theorem} which has been mathematically expressed as
\begin{equation}
v=\sqrt{2gh}
\end{equation}
later by Daniel Bernoulli (1700-1782). On the other hand, the centuries since the \textit{Renaissance} have been characterized by the major influence of the engineerings on the development of physics.\footnote{For further information on the relationships between engineers and scientists in the Renaissance, see, for example, \citet{gille_1964} and \citet{thuillier_1976}.} More specifically, the construction of bridges and canals induced a lot of theoretical studies on the flow of fluids. By observing the behavior of the flow of water in rivers, da Vinci (1425-1519) already came to the conclusion that, when the river becomes shallower or narrower, water flows faster. Later Castelli (1577-1644) confirmed this result and came up with the so-called \textit{velocity-area law}:
\begin{equation}
vA=\mathrm{const}
\end{equation}
where $v$ is the velocity of water and $A$ the cross-sectional area of the flow. This is the first idea of continuity of flow (for an incompressible fluid), which was later developed by Euler (1707-1783).

The term \textit{hydrodynamics} was first used by Daniel \citet{bernoulli} as the title of his celebrated book \textit{Hydrodynamica}.\footnote{For a  presentation of the treatise, see \citet{mikhailov_2002}.} His theory was original because he was the first scientist to combine the sciences of hydrostatics (considering the pressure) and hydraulics (considering the motion of fluids). From the conservation of \textit{vis viva} he reached the famous \textit{Bernoulli's principle} relating the velocity of flow at a point in a pipe to the pressure $P$ there:
\begin{equation}
\frac{\rho v^{2}}{2} + P = \mathrm{const}
\end{equation}
where $\rho$ is the mass density. However, in the modern sense, hydrodynamics began with the work of d'Alembert (1717-1783) and especially Euler.

\begin{figure}[h!]
\centerline{\mbox{\scalebox{.55}{\rotatebox{0}{\includegraphics{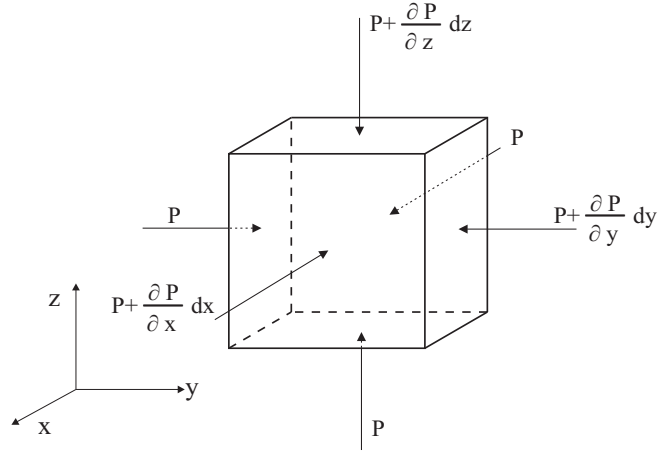}}}}}
\caption{Pressures acting ont Euler's fluid infinitesimal parallelopiped.}
\label{euler-pression}
\end{figure}

During his Berlin period as an engineer, Euler was in charge of the construction of canals, the supply of water for King's Sans-Soucis Palace, and the improvement of the water turbine. However, his strong interest for mathematics, mechanics and physics was much greater. Hydrodynamics was among  his numerous topics of interest. In 1755, he derived the fundamental equations of hydrodynamics \citep{euler} by introducing the new and main concept of \textit{fluid particle}.\footnote{ A fluid particle is imagined as an infinitesimal body, small enough to be treated mathematically as a point, but large enough to possess such physical properties as volume, mass, density, and so on.} He first proposed the modern and general form of the equation of continuity (\ref{continuity-equation}), which expresses the conservation of matter as\footnote{We use Einstein's convention of summation over repeated indices. For instance, $v_{j} \frac{\partial v_{i}}{\partial r_{j}}$ has to be understood as $\sum_{j} v_{j} \frac{\partial v_{i}}{\partial r_{j}}$.}
\begin{equation}
\frac{d\rho}{dt} + \rho \frac{\partial v_{j}}{\partial r_{j}} = 0 \; ,
\label{continuity-equation}
\end{equation}
where $v_{i}$ the $i$-component of the velocity of the considered fluid particle. Furthermore, considering a small parallelepiped and pressure $P$ acting on its different faces depicted in Fig. \ref{euler-pression},
starting from Newton's second law, he derived the general equations for the motion of an ideal fluid, the so-called  \textit{Euler equations of motion}
\begin{equation}
\rho \frac{dv_{i}}{dt} =  \rho \left ( \frac{\partial v_{i}}{\partial t} + v_{j}
\frac{\partial v_{i}}{\partial r_{j}}\right )
=  - \frac{\partial P}{\partial r_{i}} + f_{i} \; ,
\label{euler_equations}
\end{equation}
where $f_{i}$ is the $i$-component of the external force (e.g. gravity, etc.).


\subsection{Navier-Stokes equations}

In the 19th century, the internal friction ---{\em viscosity} in modern terminology--- became a subject of increasing interest.  However, this was not yet included within these equations.
	
As a matter of fact, the first studies on this property go back already to Newton's Book II of his \textit{Principia} \citep{newton-principia-1ed}. In concrete terms, Newton made precisely the following hypothesis:
\begin{quote}
The resistance which arises from the lack of lubricity or slipperiness of the parts of a fluid is, other things being equal, proportional to the velocity with which the parts of the fluid are separated from one another.\footnote{Quoted from \citet[Book II, Section 9]{newton-principia}.}
\end{quote}
His theory thus stated that, if a portion of fluid is kept moving, the motion gradually communicates itself to the rest of the fluid. He ascribed to this effect the term \textit{defectus lubricitatis}, that is, the \textit{lack of slipperiness}. This is nothing but viscosity. Consider two particles of a viscous flow sliding one over another. Then friction ---or viscous resistance--- is observed along the surface of contact. The force of this resistance per unit area is nowadays known as the \textit{shear stress} $\tau$. Hence, the shear stress should depend on the speed with which the two layers slide one over another. By extension, we obtain that the shear stress is proportional to the velocity gradient in the perpendicular direction of the movement of the fluid layers:
\begin{equation}
\tau \sim \frac{\partial v_{x}}{\partial y}.
\label{newton_law}
\end{equation}

At the end of the 18th century and the beginning of the 19th century, numerous and extensive investigations on the flow of pipes and open channels started to be carried out by scientists and engineers such as Du Buat, Girard, Prony, Coulomb, and especially Poiseuille (1797-1869). Interested in the circulation of blood in capillary vessels, he then used glass capillaries of very much smaller bore that any of his predecessors. Long series of measurements of the quantity $Q$ of liquid discharged in unit time were carried out in function of the different factors. He then obtained in the 1840s the relation 
\begin{equation}
Q = K \frac{P D^4}{L}
\end{equation}
where $P$ is the pressure, $D$ and $L$ respectively the diameter and the length of the capillary \citep{poiseuille-40a,poiseuille-40b,poiseuille-46}. And in 1856, Wiedemann (1826-1899) as well as Hagenbach in 1860, deduced mathematically, by using Newton's hypothesis (\ref{newton_law}), that the constant $K$ is such that\footnote{Their works were published in \citet{wiedemann} and \citet{hagenbach}.}
\begin{equation}
Q = \frac{\pi P R^4}{8 \eta L}
\end{equation}
where $R$ is the radius of the capillary. A new factor $\eta$ appears which \citet{wiedemann} proposed to call \textit{Z\"ahigkeitconstante der Fl\"ussigkeiten}, that is, the \textit{viscosity coefficient of the liquids}.\footnote{The word viscosity derives from the latin word \textit{viscum} for mistletoe. The mistletoe berries produce a viscous glue which was used to lime birds. Viscosity is expressed in $Pa . s$ or in \textit{Poise} ($10^{-1} Pa . s$) in the honor of Poiseuille (1797-1869).} For the first time, the term viscosity was used in a technical sense.

In parallel, the first scientist having theoretically used the property of viscosity in the fundamental equations of hydrodynamics was the civil engineer Navier (1785-1836) in a {\em m\'emoire} read in 1822 to the {\em Acad\'emie des Sciences} \citep{navier}. His idea was that any pressure tends to reduce the distances between the fluid's molecules. Taking into account the intermolecular repulsive forces produced by this action, which gave him the explanation of Newton's \textit{lack of slipperiness}, he added to the pressure acting on a small volume the contribution (\ref{newton_law}) of this effect
\begin{equation}
P \delta_{ij} \rightarrow P \delta_{ij} - \eta \frac{\partial v_{i}}{\partial r_{j}},
\end{equation}
that is a dissipative term implying the presence of the viscosity coefficient $\eta$. Rediscovered by Stokes (1819-1903) in 1845 without introducing any molecular behavior \citep{stokes}, the history has given the name of these two persons to the famous \textit{Navier-Stokes equations}
\begin{equation}
\rho \frac{dv_{i}}{dt} =  -  \frac{\partial P}{\partial r_{i}} + \eta \left (
\frac{\partial^2 v_{i}}{\partial r_{j} \; \partial r_{j}}
 \right ) + f_{i} \; .
\label{Navier-Stokes-equations}
\end{equation}

On the other hand, in 1829, Poisson (1781-1836) derived an equation in a {\em m\'emoire} devoted to the motion of solid bodies and fluids in which appeared a second constant \citep{poisson}, in addition to the viscosity introduced by Navier. These two constants measure the response of the substance to two different types of forces: the first tending to shear it, and the second tending to expand or contract it. We saw above that the first type is characterized by the viscosity coefficient $\eta$ which is more precisely called \textit{shear viscosity}. However, until seventy years ago, most of the standard works did not consider the other constant $\eta'$ called the \textit{second viscosity coefficient} by making it simply proportional to the shear viscosity. Indeed, since \citet{stokes}, it was usually admitted that
\begin{equation}
\eta' + \frac{2}{3} \eta = 0 \; ,
\label{stokes-relation}
\end{equation}
which is generally called \textit{Stokes' relation}. Stokes used this relation by arguing that, in most cases of practical interest, the rate of expansion would be so small that the value assigned to this sum would be unimportant. Before seventy years ago, no direct determination of $\eta'$ had yet been made. Indeed it is only in 1942 that Tisza (1907-2009) made this determination by evaluating the ratio $\eta' / \eta$ when ultrasonic vibrations are passed through a liquid \citep{tisza}. He then showed that this vanishing relation is not justified in general and that the re-introduction of the second viscosity coefficient is necessary to get an agreement between the theory and experiments.\footnote{For further information, see, for example, \citet{markham}, \citet{karim}, and \citet{brush-viscosity}.} Consequently, a new coefficient $\zeta$ can be defined as
\begin{equation}
\zeta = \eta' + \frac{2}{3} \eta
\label{def-bulk-viscosity}
\end{equation}
which is known as the \textit{bulk viscosity coefficient}.

The bulk viscosity is more difficult to be measured experimentally than the shear viscosity. Moreover, its measure is generally less known. The different ways to measure it imply the phenomenon of absorption of sound in liquids and gases. Before \citet{tisza}, it was well-known that the absorption of sound in polyatomic gases and in liquids obtained experimentally exhibited great disagreements with the absorption predicted by the classical theory based on viscosity. Previously, it was usually admitted that the absorption was given by
\begin{equation}
\alpha_{\mathrm{cl}} = \frac{2 \pi ^{2} \nu ^{2}}{\rho v_s ^3} \left \lbrack
\left ( \frac{1}{c_V} - \frac{1}{c_P} \right ) \kappa
+ \frac{4}{3} \eta \right \rbrack ,
\label{absorption-classic}
\end{equation}
where $\nu$ is the acoustic frequency, $\rho$ the mass density, $v_s$ the sound velocity
\begin{equation}
v_s = \sqrt{\left ( \frac{\partial P}{\partial \rho} \right )_s} ,
\label{acoustic_velocity}
\end{equation}
$c_V$ and $c_P$ the specific heats respectively at constant volume and constant pressure, and $\kappa$ the thermal conductivity. Hence, only the shear viscosity was considered. In his paper, \citet{tisza} pointed out that Stokes' assumption (\ref{stokes-relation}) is not justified, except in dilute monoatomic gases. However, for polyatomic gases and liquids, the bulk viscosity should be quantitatively quite important, like for hydrogen ${\rm H}_2$ for which $\zeta$ is 32 times greater than $\eta$ (see Table \ref{table-bulk-viscosity}). Consequently, Tisza obtained a modified expression for the absorption coefficient in which the bulk viscosity is added in the expression (\ref{absorption-classic}) as follows
\begin{equation}
\alpha_{\mathrm{obs}} = \frac{2 \pi ^{2} \nu ^{2}}{\rho v_s ^3} \left \lbrack \left ( \frac{1}{c_V} - \frac{1}{c_P} \right ) \kappa 
+ \frac{4}{3} \eta + \zeta \right \rbrack
\label{absorption-observed}
\end{equation}
or by considering the ratio of the viscosity coefficients $\frac{\zeta}{\eta}$
\begin{equation}
\frac{\zeta}{\eta} = \frac{4}{3} \frac{\alpha_{\mathrm{obs}} -
\alpha_{\mathrm{cl}}}{\alpha_{\mathrm{cl}}}
\left ( 1 + \frac{3}{4} \left ( \frac{1}{c_V} - \frac{1}{c_P} \right )
\frac{\kappa}{\eta} \right ) .
\end{equation}

\begin{table} [!t]
\begin{center}
\begin{tabular}{|c|c|c|}
\hline
Fluids  &  $\eta \ \lbrack 10^5$ Pa s\rbrack  &  $\zeta $/$ \eta$  \\
\hline
\hline
He & 1.98 & 0 \\
Ar & 2.27  & 0 \\
$\mathrm{H_2}$ & 0.887 & 32 \\
$\mathrm{N_2}$ & 1.66 & 0.8 \\
$\mathrm{O_2}$ & 2.07 & 0.4 \\
$\mathrm{CO_2}$ & 1.50 & 1.000 \\
Air & 1.85 & 0.6 \\
$\mathrm{H_2O}$ (liquid) & 85.7 & 3.1 \\
Ethyl alcohol & 110 & 4.5 \\
Glycerine & 134.000 & 0.4 \\
\hline
\end{tabular}
\end{center}
\caption{Shear and bulk viscosities obtained experimentally at 300 K and 1 Atm in monoatomic, diatomic and polyatomic gases, and in liquids;  values given by \citet{thompson}.\label{table-bulk-viscosity}} 
\end{table}

In addition, the tangential force acting for example on the plane $xy$ should be related to the sliding in the two perpendicular directions composing this plane, that is $\frac{\partial v_x}{\partial z} + \frac{\partial v_z}{\partial x}$ in the direction of $x$, and $\frac{\partial v_y}{\partial z} + \frac{\partial v_z}{\partial y}$ in the direction of $y$. This is why Saint-Venant (1797-1886) combined both of them in 1843 and obtained more precise expressions for the pressure tensor, e.g. \citep{saint-venant}
\begin{equation}
P_{xx} = P
+ 2 \eta \frac{\partial v_x}{\partial x}
\end{equation}
and
\begin{equation}
P_{xy} = \eta \; \left ( \frac{\partial v_x}{\partial y} + \frac{\partial
v_y}{\partial x} \right ).
\end{equation}
It allows us to write the Navier-Stokes equations in a more complete form as\footnote{ For further information on the history of hydrodynamics, see, for example, \citet{stokes-report}, \citet{rouse}, \citet{tokaty}, \citet{mikhailov}, and \citet{darrigol-book}. A more rigorous derivations of Navier-Stokes equations can be found in \citet{L-L}.}
\begin{equation}
\rho \frac{d v_{i}}{dt}  =  - \frac{\partial}{\partial r_{j}} \left \lbrack P \; \delta_{ij}- \eta \left (\frac{\partial v_{i}}{\partial r_{j}}
+ \frac{\partial v_{j}}{\partial r_{i}} \right ) + \left(\frac{2}{3} \eta -	 \zeta \; \right) \delta_{ij} \frac{\partial v_{l}}{\partial r_{l}} \right \rbrack + f_{j} .
\label{NS-equations-intro}
\end{equation}

When the fluid is considered incompressible ($\frac{\partial v_l}{\partial r_l} = 0$), one recovers the original Navier-Stokes equations (\ref{Navier-Stokes-equations}). In \ref{phen_appr_visc} are provided the Navier-Stokes equations in their modern form and their thermodynamic analysis leading to the definition of the positivity of the two viscosity coefficients $\eta$ and $\zeta$. Let us also mention that these equations are valid in so far as Newton's assumption is reasonable. A wide range of fluids (water, air, etc.) follows the linear relation between the shear stress $\tau$ and the velocity gradient. But liquids composed by lengthened molecules (polymers, etc.) obey to different viscosity laws, which are called {\em non-Newtonian}.

As we shall see in the next section, the 19th century and the beginning of the 20th century before the First World War have seen the decisive development of the scientific atomistic conception of matter and the will of deriving macroscopic phenomena like viscosity in terms of the Newtonian microscopic dynamics. These works gave rise to the powerful kinetic theory of gases which led to the elaboration of statistical mechanics.


\section{Kinetic theory of gases and statistical mechanics}\label{kinetic-theory}

\subsection{Atoms and the early kinetic theories of gases}

The idea of atomism goes back to  the fifth century BC with the ancient Greek philosophers Leucippus and Democritus. However, the history of the kinetic theory of gases does not really begin until the seventeenth century when Torricelli, Pascal (1623-1662) and Boyle first established the physical nature of the \textquotedblleft sea\textquotedblright \ of air that exerts mechanical pressure on surfaces in contact with it.\footnote{As a matter of fact, Hero of Alexandria's {\em Pneumatica} (first century A.D.) translated into latin in the 16th century would have influenced the work of Torricelli and Boyle. For Hero, air is composed of very small particles between which there is only vacuum. See \citet{boas_1949} for further information.} This development of the concept of air pressure can be considered as part of the change in scientific attitude which led to explanations based simply on matter and motion instead of postulating \textquotedblleft occult forces\textquotedblright \ or teleological principles. In this context, \citet{boyle} described several discoveries on pneumatics in his {\em New Experiments}, especially, the fact that air was permanently elastic. In an appendix of the second edition of his 1662 book, he developed this discovery into the well-known quantitative relationship
\begin{equation}
PV=\mathrm{const}
\end{equation}
now called \textit{Boyle law}.\footnote{This law is sometimes associated with Mariotte (1620-1684) together with Boyle. In 1679, in his work entitled {\em De la nature de l'air} reprinted in \citet{mariotte_1717}, he enunciated the same law without giving any credit to Boyle, neither making any claims of originality. As a matter of fact, he rather treated the law as one of a series of well-known properties of air.}

Daniel Bernoulli is usually considered as the first scientist having proposed a kinetic theory of gases. In his famous treatise \textit{Hydrodynamica} \citep{bernoulli}, he gave a derivation of Boyle's gas law from {\em billiard ball} model, assuming that gas consists of a very large number of small particles in rapid motion. Using the principle of \textit{vis viva} conservation, he concluded that pressure was proportional to the square of the velocities of gas particles, and thus proportional to the temperature.\footnote{For further information, see \citet[p.~62-67]{mikhailov_2002}.} In other terms, heat is nothing but atomic motion. However, by proposing these remarkable results, Bernoulli was about a century ahead of his time, and his theory model was forgotten. The latter success of Lavoisier's caloric theory buried it by proposing another conception of matter based on equilibrium between caloric repulsion (due to the atmosphere of caloric which density increases with temperature) and gravitational attraction between matter particles.\footnote{See note \ref{caloric_theory}.}

In chemistry, the atomic theory emerged at the beginning of the 19th century with Dalton (1766-1844). Thanks to the laws of chemical discontinuities -- that is (i) the law of \textit{constant proportions}\footnote{ The proportion between two combining elements cannot vary continuously  \citep{Proust1806}.} discovered by Proust (1754-1822) and (ii) the law of \textit{multiple proportions}\footnote{When two elements can combine according to different ratios, the ratios of their weight in the different cases are simple, that is integer numbers \citep{Dalton1805}.} found in by Dalton himself --, he proposed in 1808 a discrete view of matter composed by indivisible entities, namely the atoms, and characterized by different \textit{atomic weights} \citep{Dalton1808}. In the meantime these laws and experiences did not reject the concept of \textit{equivalents}\footnote{The concept of {\em equivalents} was presented in the first volume of his {\em Anfangsgrunde der St\"ochiometrie}  \citep{richter_book}. This notion played a key role in chemistry since this allowed determining the amount of a chemical substance (e.g. an acid) needed to neutralize a given quantity of another substance (e.g. a given base). In this context, Richter defined the {\em stoichiometry}. For further detail on Richter and his theory, see, for example, \citet{partington_richter}.} first proposed in 1792 by Richter (1762-1807). Because of various ambiguities amongst the partisans of the atomic theory, the {\em equivalentism} had a great success during the 19th century. However, many new phenomena and laws were discovered that only the atomic hypothesis could explain. Among these, the law of Gay-Lussac (1778-1850) discovered in 1809,\footnote{The ratios of volumes of reacting gases are integer numbers \citep{Gay-Lussac1809}.} which allowed Avogadro (1776-1850) to emit his famous hypothesis saying that all gases considered in the same conditions of temperature and pressure contain, for equal volumes, the same number of molecules \citep{Avogadro1811}.\footnote{In addition, let us notice that Avogadro was the first scientist having used the term \textit{molecule} in its modern sense. The word is derived from the french word \textit{mol\'ecule} meaning \textquotedblleft extremely minute particle\textquotedblright \ which comes from the Latin word \textit{molecula}. This term is the diminutive of \textit{mole} meaning \textquotedblleft mass, cluster, great number, multitude\textquotedblright. On the other hand, the word \textit{atom} is derived from the Latin term \textit{atomus} used by Lucretius, and from the Greek word \tg{'atomos}, which means uncut.} We can also cite the law of Dulong (1785-1838) and Petit (1791-1820)\footnote{The atomic weight multiplied by the specific heat of an element is independent of the nature of the element \citep{Dulong_Petit1819}.}; the emergence of structural chemistry, especially the development of the theory of valence by Kekul\'{e} (1829-1896), organic chemistry and the concept of isomerism, and stereochemistry. Above all, one of the most important works having played a fundamental role in the development of the atomic theory and its acceptance was the establishment of the periodic table of the elements in 1869 by Mendeleev (1834-1907). By assuming his periodic law in the properties such as the valence of elements when these are listed in order of atomic weight, \citet{mendeleev1,mendeleev2} predicted the existence of unknown elements corresponding to the gaps in his famous table and with which a given atomic weight is associated.\footnote{For further information on the rise of Mendeleev's table and its reception, see, for example, \citet{brush-mendeleev}.} The discovery of some of them in the following years confirmed the periodic property of elements. In the polemic between atomists and equivalentists, Mendeleev's brilliant idea heavily contributed to the acceptance of the atomic conception of matter amongst chemists.\footnote{ For further information on the history of atomic theory in chemistry, see, for example, \citet{leicester} and \citet{pullman}.}

While the caloric theory was being brought to its final stage of perfection by Laplace, Poisson, Carnot, and others, Herapath (1790-1868) in the late 1810s,\footnote{See \citet{brush-herapath}, \citet[p.~107-133]{brush-kind-motion2}, \citet{talbot_pacey_1966}, and \citet{mendoza_1975} for further information.} as well as Waterston (1811-1883) in 1845 introduced their kinetic theory.\footnote{For further information, see, for example, \citet{brush-waterston} and \citet[p.~134-159]{brush-kind-motion2}.} However their papers were rejected by the Royal Society. What needed to be established was not simply a connection between heat and molecular motion, for that was already admitted by many scientists and was not considered incompatible with the caloric theory. It was rather the notion that heat is \textit{nothing but} molecular motion, and the idea that molecules \textit{move freely through space} in gases rather than simply vibrating around fixed positions. This  statement could not yet be accepted.

As seen in Section \ref{thermodynamics}, in the period between 1842 and 1847, the general scientific and intellectual climate implied simultaneous works by different scientists on one of the most important concepts inducing a revival of kinetic theory: namely, the \textit{conservation of energy}.\footnote{For further detail on these simultaneous works, see \citet{kuhn_1959}.} The German romanticism and  \textit{Naturphilosophie} strongly influenced scientists, especially by claiming that there must be a single unifying principle underlying all natural phenomena.\footnote{For an exciting overview of the relationships between culture and thermodynamics in the 19th century, see, for example, \citet{brush-thermo}. On the influence of philosophical ideas of Kant (1724-1804), Schelling (1775-1854), Novalis (1772-1801), and others on science of the early 19th century, see, for example, \citet{thuillier} and \citet[p.~104-114]{locqueneux_2009}.} Furthermore, this period inherited different discoveries of various conversion processes\footnote{Oersted (1777-1851)'s discovery of electromagnetism (1820), Seebeck (1770-1831)'s discovery of thermo-electricity (1822), Faraday (1791-1867)'s many discoveries in electricity and magnetism, and many others.} and gave birth to the principle usually attributed to Mayer and Joule. \citet{mayer} emphasized the philosophical generality of the principle while \citet{joule} provided the experimental verification in specific cases. From that point on, heat, mechanical work, electricity, and other apparently different entities were considered as different forms of the same thing, now called \textit{energy}. The same year, Helmholtz (1821-1894) indicated quite clearly that \textit{mechanical} energy was to be regarded as the basic entity \citep{helmholtz_1847}. As a result, it became more and more natural that heat was nothing but a mechanical energy, which made the kinetic theory appear to be an obvious consequence of the principle of energy conservation.


\subsection{Kinetic theory of gases and the mean-free path}

The real breakthrough for the kinetic theory took place when Kr\"{o}nig (1822-1879) assumed in 1856 that the molecules of gas move with constant velocity in a straight line until they hit other molecules, or the surface of the container \citep{kronig}.\footnote{Actually his publication did not represent a real advance compared to previous works by Bernoulli and Herapath. However, his influence in the physics community was important and induced a special interest among the physicists for the kinetic theory. See \citet{brush-clausius} and \citet[p.~160-182]{brush-kind-motion2}. Moreover, according to \citet{daub_1971}, Kr\"onig would have read \citet{waterston_1852}'s abstract of a paper read at the British Association for the Advancement of Science and would be clearly influenced by this in writing his 1856 paper.} Nevertheless, the kinetic theory was still confronted to objections like those of Buijs-Ballot (1817-1890). Since the kinetic theory claimed that the velocities of molecules were of the order of several hundred meters per second, he pointed out that one would expect two gases to mix with each other very rapidly. However, the experience shows that the process takes a certain time, of the order of several minutes \citep{buijs-ballot_1858}. To answer this objection, he showed that, in real gases, the molecules could not travel without colliding with other molecules. Consequently, \citet{clausius-1858} introduced a new concept: the so-called \textit{mean free path} $\lambda$ of a molecule between two successive collisions
\begin{equation}
\lambda = \frac{3}{4} \frac{1}{n \pi \sigma^2}
\label{mfp_clausius}
\end{equation}
where $n$ is the number of molecules per unit volume. This concept was major not only for its further developments, but also because it establishes in concrete terms one of the most fundamental ideas of the kinetic theory of gases rejected in the past: molecules can move freely through space and yet collide with each other.

The early kinetic theorists assumed that molecules tending to equilibrium move all at the same velocity. Maxwell (1831-1879) was the first scientist who introduced the idea of random motion of the molecules ---hence {\em statistical} considerations.\footnote{See \citet{brush-maxwell} and \citet[p.~283-230]{brush-kind-motion2} for further information on Maxwell's conceptual contributions to the kinetic theory of gases.} Actually Maxwell has probably been influenced by works on statistics, especially those realized by Qu\'etelet (1796-1874) on the height distribution of a population of soldiers.\footnote{In 1857-1858, Maxwell would have read \citet{herschel_1850}'s review of Qu\'etelet, which probably helped him to elaborate a statistical approach of molecules. See \citet[p.~151-152]{brush_1967_foundations}, \citet[p.~183-188]{brush-kind-motion2}, and \citet[Section~4]{uffink_2010}.} \citet{quetelet} stated the idea of the {\em average man} with {\em average height}, etc., as the result of an experimentally established normal distribution function. Maxwell worked out the idea of an {\em average molecule} in analogy with that of the average man. In 1860, in his first paper on kinetic theory entitled \textit{Illustrations of the dynamical theory of gases},\footnote{The name of {\em kinetic} theory was not adopted before \citet{thomson_tait_1867}'s {\em Treatise on Natural Philosophy}:
\begin{quote}
We adopt the suggestion of Amp\`ere, and use the term {\em Kinematics} for the purely geometrical science of motion in the abstract. Keeping in view the proprieties of language, and following the example of the most logical writers, we employ the term {\em Dynamics} in its true sense as the science which treats of the action of force, whether it maintains relative rest, or produces acceleration of relative motion. The two corresponding divisions of Dynamics are thus conveniently entitled {\em Statics} and {\em Kinetics} (Thomson, W. and Tait, P. G., 1879. Treatise.... New Edition, Cambridge University Press, Cambridge, p.~VI).
\end{quote}
Maxwell himself would have rapidly adopted the terminology of kinetic theory of gases, at least in a 1871 letter to Thomson. See \citet{bernstein_1963}. Nevetheless, some scientists continued to use the old phrase, such as Jeans (1877-1946) in his {\em The Dynamical Theory of Gases} \citep{jeans2}. It is only in another book {\em Introduction to the Kinetic Theory of Gases} published in 1840 that he adopted the general usage \citep{jeans_1940}.} \citet{maxwell-1860} suggested that, instead of tending to equalize the velocities of all molecules, the successive collisions would produce a {\em statistical distribution of velocities} in which all of them might occur, with a known probability. Concretely he considered a system of $N$ colliding particles. Denoting by $v_x$, $v_y$, and $v_z$ the components of the velocity of each particle, he expressed as $N \psi(v_x) dv_x$ the number of particles for which $v_x$ was in the range $\lbrack v_x, v_x + dv_x \rbrack$, where $\psi(v_x)$ was a ``function of $v_x$ to be determined". Assuming that (i) the existence of this velocity did not affect those in the two other directions (which let him write that $f_1({\bf v}) = \psi(v_x)\psi(v_y)\psi(v_z)$), and (ii) $f_1({\bf v})$ would be spherically symmetric ($f_1({\bf v}) = \phi(v_x^2 + v_y^2 + v_z^2)$), he could then derive his famous distribution function for thermal equilibrium, which, in modern notation, took the form:\footnote{The subscript $1$ is used to distinguish this one-particle distribution function and the probability distribution $f({\bf \Gamma})$ defined in the phase space. See below.}
\begin{equation}
f_1({\bf v})= \left ( \frac{m}{2 \pi k_B T} \right )^{3/2} \exp \left ( -\frac{mv^2}{2k_B T} \right ) \; ,
\label{maxwell_distribution}
\end{equation}
where $m$ and $v$ are the mass and velocity of the particles, and $k_B$ the Boltzmann constant. Therefore the statistical character appeared to be a fundamental element of kinetic theory.

In 1859, Maxwell came to the kinetic theory as an exercise in mechanics involving the motions of systems of particles acting on each other only throught impact. Being interested in viscosity, in the same paper \citep{maxwell-1860}, he proposed a mechanism which allowed him to establish a relation between the (shear) viscosity and the mean free path. Considering gas as divided into parallel layers and supposing that the motion is uniform in each layer but varying from one to another, he demonstrated that the viscosity $\eta$ should be proportional to the mean free path $\lambda$, the mass density $\rho$ and the mean molecular velocity $\left \langle v \right \rangle$, relation written as
\begin{equation}
\eta_M = \frac{1}{3} \; \left \langle v \right \rangle \; \rho \; \lambda 
\label{maxwell-viscosity}
\end{equation}
where $\rho = n m$ and the mean free path
\begin{equation}
\lambda = \frac{1}{\sqrt{2} n \pi \sigma^2}
\label{mfp_maxwell}
\end{equation}
derived by \citet{maxwell-1860} differs with respect to Clausius' formula by a factor as a result of the assumption on the velocity distribution. However, Eq. (\ref{mfp_clausius}) as well as Eq. (\ref{mfp_maxwell}) are inversely proportional to the density, which implied the surprising result that
\begin{quote}
[...] if this explanation of gaseous friction be true, the coefficient of friction is independent of density. Such a consequence of a mathematical theory is very startling, and the only experiment I have met with on the subject does not seem to confirm it.\footnote{Quoted from {\em The Scienfitic Papers of James Clerk Maxwell} \citep[p.~391]{maxwell_scientific_papers}.}
\end{quote}
Influenced by Stokes' opinion who found that $\sqrt{\eta/\rho}=0.116$,\footnote{This value was reported by Maxwell himself. See \citet{maxwell_scientific_papers}, p.~391.} Maxwell believed his predictions were absurd and therefore that the kinetic theory was wrong, or at least inadequate. At that time a few experiments on the gas viscosity had already been achieved. The first studies probably began in the 1840s with, for example, \citet{graham1,graham2}. However, accurate experiments had not yet been carried out on the viscosity of \textit{gases}. Therefore, a few years later, \citet{maxwell-1866} himself carried out his own experiment and found out that viscosity remained constant over a large range of pressure.\footnote{This experimental result was confirmed later, in particular by \citet{Meyer1873} and \citet{KW1875}.} This work played an important role in the development of the kinetic theory and its acceptance by most of scientists who so far had been doubting it. For instance, in 1865, Loschmidt (1821-1895) gave the first convincing estimate of the diameter of an air molecule -- about $10 \; \mathrm{\r{A}}$ which is about four times too large -- as well as the Avogadro number -- about $6.025 \times 10^{23}$ \citep{loschmidt}. On the other hand, Maxwell's formula, combined with the equation of state for real gases that van der Waals (1837-1923) derived later,\footnote{In order to obtain this celebrated equation of state, \citet{van-der-waals} used the \textit{virial theorem} introduced by \citet{clausius1870} who demonstrated that the mean kinetic energy of the system was equal to the mean of its virial multiplied by $-\frac{1}{2}$: $$\overline{E}= -\frac{1}{2} \left \langle \sum_{a=1}^{N}\mathbf{r}_a \cdot \mathbf{F}_a\right \rangle,$$ where $\mathbf{r}_a$ and $\mathbf{F}_a$ are respectively the position of the $a$th particle and the force acting on it.  \label{virial_footnote}} belongs to the list of thirteen different phenomena mentioned in 1913 by Perrin (1870-1942) in {\em Les atomes} and allowing one to evaluate the atomic magnitudes  as well as the Avogadro number \citep{perrin-atomes}.

On the other hand, \citet{maxwell-1860} derivation of viscosity suggested that the latter would depend on the square root of the absolute temperature. However, his experiments showed instead the viscosity was simply proportional to the temperature. This disagreement motivated him to develop a theory of transport based on the assumption that the molecules repelled each other with a force inversely proportional to the $n$th power of the distance between their centres \citep{maxwell_1867}. In the case of $n=5$, the viscosity thereby satisfied a linear dependence on the temperature. Accordingly, molecules with this kind of interaction are usually called {\em Maxwellian}.


\subsection{Boltzmann's statistical interpretation of irreversibility}\label{Boltzmann_stat_irrev}

As the kinetic theory, which establishes a microscopic mechanical basis to the macroscopic processes such as viscosity, was attracting more and more attention from physicists, the idea of reconciling the second law of thermodynamics with the principles of mechanics emerged. It is in this context that Boltzmann (1844-1906), who played a tremendous role in the elaboration of statistical mechanics, started to be concerned by this fundamental issue.\footnote{For Boltzmann's biography and an analysis of his work, see, for example, \citet{cercigniani}.} One year after \citet{clausius-1865}'s paper in which he defined the concept of entropy, \citet{boltzmann_1866} observed that, while the first principle (\ref{first-principle}) corresponded to the conversation of energy\footnote{At this time, this was still called the principle of {\em vis viva} ({\em lebendige Kraft} in German).} in mechanical system, no analogy was found for the second principle (\ref{clausius-inequality}). He thus proposed a theorem of pure mechanics which corresponded with the second law, namely, the principle of least action in a somewhat more general form.\footnote{For further infomation, see, for example, \citet{daub_1969}, \citet{Klein1973}, and \citet{uffink_2005}.}

His first major achievement came two years later when he extended Maxwell's statistical concepts. He especially payed attention to Maxwell's treatment of colliding-particle systems. \citet{maxwell_1867} considered two kinds of particles with $n_1$ and $n_2$ denoting their respective number in unit of volume, and selected those with initial velocities ${\bf v}_1$ and ${\bf v}_2$ (within the ranges $d{\bf v}_1$ and $d{\bf v}_2$, respectively). After he introduced (i) a relative impact parameter comprised between $b$ and $b + db$, and (ii) a relative azimuthal angle comprised between $\phi$ and $\phi + d\phi$, he obtained that the number $n({\bf v}_1, {\bf v}_2)$ of encounters between the two kinds of molecules occuring within the time interval $\delta t$ is given by:
\begin{equation}	
n({\bf v}_1, {\bf v}_2) = n^2 \vert {\bf v}_1 - {\bf v}_2 \vert \; \delta t \; b \; db \; d\phi \; f_1({\bf v}_1) \; d{\bf v}_1 \; f_1({\bf v}_2) \; d{\bf v}_2 \; ,
\label{maxwell_colliding_particles}
\end{equation}
where $n f_1({\bf v}) d{\bf v}$ is the number of particles with a velocity lying in the range $d {\bf v}$ around $\bf v$.\footnote{Note that the subscript $1$ in $f_1$ indicates that one deals with a function associated with {\em one} particle.} To obtain Eq. (\ref{maxwell_colliding_particles})
Maxwell made an important assumption: the number of collisions $n({\bf v}_1, {\bf v}_2)$ observed between the two kinds of particles in a time $\delta t$ is proportional to the  product of the the number of molecules $n f_1({\bf v}_1)d {\bf v}_1$ and $n f_1({\bf v}_2)d {\bf v}_2$. This assumption will be largely used by Boltzmann in his work, and was later coined \label{SZA}{\em Sto\ss zahlansatz}\footnote{The {\em Sto\ss zahlansatz} means {\em Assumption about the number of collisions}.} by Paul (1880-1933) and Tatiana (1876-1964) Ehrenfest who carried out a deep analysis of Boltzmann's work in a paper published in 1911 in which they emphasized the hypothetical character of this assertion \citep{ehrenfest}.

Starting with the approach developed in Maxwell's paper, \citet{boltzmann-1868} considered $f_1({\bf v}) d{\bf v}$ as Maxwell did. He however also interpreted it, at the same time, as the fraction of time spent by a given particle in $d{\bf v}$ around $\bf v$ after a very long time had elapsed. Along his paper, he assumed the equivalence between these two meanings of $f_1({\bf v}) d{\bf v}$.

However, a fundamental change of approach appeared at the end of the same paper. Instead of using the space of velocities of individual particles, he rather considered the mechanical system as whole, which is described by its state
\begin{equation}
{\bf \Gamma} = {\bf r}_1, {\bf r}_2, \cdots , {\bf r}_N, {\bf p}_1, {\bf p}_2, \cdots , {\bf p}_N
\end{equation}
which is usually called {\em phase}.\footnote{See \citet{nolte_2010} for further information on the history of this terminology.}
Accordingly, in the case of Hamiltonian systems characterized by the Hamiltonian $H({\bf \Gamma})$, the time evolution of the system is described by a trajectory confined to the energy shell $H({\bf \Gamma}) = E$.

Furthermore, he introduced the probability distribution $f({\bf \Gamma}, t)$, which was defined in the so-called {\em phase space} $\cal M$,\footnote{Actually, Boltzmann never attributed a specific phrase to this concept. See Section \ref{gibbs_stat_mech} for some information on the introduction of the term {\em phase space} for further information.} and thus gave the probability of finding the system in $d\bf \Gamma$ around $\bf \Gamma$ at time $t$. In considering such a function, this was the first time in the history of physics that a probabilistic approach was applied to describe a mechanical system as whole, instead of individual particles. As a result, \citet{boltzmann-1868}'s paper is sometimes regarded as the one in which the transition was made between kinetic theory of gases and statistical mechanics \citep{ehrenfest, Klein1973}.

Finally, Boltzmann proved in the same paper a theorem playing a key role in statistical mechanics, that is the conservation of the probability density under time evolution. The origin of this theorem is based on a fundamental theorem which goes back to the late 1830s when Liouville (1809-1882) found a general property in pure mathematics \citep{liouville_1838}. Considering a system of $n$ first-order differential equations:
\begin{equation}
\frac{d {\bf X}}{dt} = {\bf F} ({\bf X}, t) \; ,
\end{equation}
he showed that, if ${\bf X} = X_1, X_{2} \dots, X_n$ forms a complete set of solutions
\begin{equation}
{\bf X} = {\bf X}({\bf a}, t)
\end{equation}
where ${\bf a} =  a_1, a_2 \dots, a_n $ are arbitrary constants, then the function:
\begin{equation}
u(t) = \det \left ( \frac{\partial {\bf X}}{\partial {\bf a}} \right )
\label{jacobian_det}
\end{equation}
satisfies
\begin{equation}
\frac{du}{dt} = \left ( \nabla_{\bf X} \cdot {\bf F} \right ) u \; .
\label{liouville_th_1}
\end{equation}
As a result, if $\nabla_{\bf X} \cdot {\bf F}$ vanishes, then $u$ remains constant with respect to $t$. In addition, the values of the variables ${\bf X}(0)$ at $t=0$ can be chosen as the arbitrary constants ${\bf a}$, so that Eq. (\ref{jacobian_det}) becomes
\begin{equation}
u(t) = \det \left ( \frac{\partial {\bf X}(t)}{\partial {\bf X}(0)} \right ) = 1 \; .
\end{equation}

Although \citet{hamilton_1834,hamilton_1835} had already elaborated his formalism in which the fundamental equations (\ref{hamilton_equations1}-\ref{hamilton_equations2}) belonged to this class of differential equations, Liouville did not make any reference to his work. Actually, it is to Jacobi (1804-1851) that we owe the first application of this result to the Hamiltonian mechanics.\footnote{The connection of Liouville's work with mechanics was done in lectures that Jacobi gave in 1842-1843, although there were published only several years later \citep{jacobi_1866}.} Let us consider a volume element $d {\bf \Gamma} (0)$ around ${\bf \Gamma} (0)$. This volume contains different copies of the same mechanical system that are distinct only by their initial conditions. After these systems have moved according to the Hamiltonian equations of motion for the same time $t$, the coordinates and momenta lie into the volume element $d {\bf \Gamma} (t)$. Jacobi obtained the result that $\nabla_{\bf X} \cdot {\bf F}$ in Eq. (\ref{liouville_th_1}) was zero for Hamiltonian systems. Consequently, he proved the famous {\em Liouville's theorem}\footnote{This is Boltzmann who attributed this theorem to Liouville. In his early papers, \citet{boltzmann-1868,boltzmann_1871c} did not mention the works of Liouville and Jacobi and proposed his own proof of the theorem. In his very next paper, he refered to Jacobi \citep{boltzmann_1871}. However, he made no mention of Liouville's paper. It was only in the second volume of his {\em Lectures} that he first placed Liouville's name to the theorem of volume conservation \citep{boltzmann-LGT2}.}
\begin{equation}
d {\bf \Gamma} (t) = d {\bf \Gamma} (0) \; 
\label{volume_preserve}
\end{equation}
which expressed mathematically a fundamental property of Hamiltonian systems: phase-space volume is preserved under time evolution. Consequently Jacobi can be considered as the first scientist who derived a fundamental theorem of mechanics.

\citet{boltzmann-1868} showed that this property had crucial implications on a statistical approach of mechanical systems and let him prove that the probability distribution $f({\bf \Gamma}, t)$ remained constant along the trajectory in phase space:
\begin{equation}
\frac{d f({\bf \Gamma},t)}{dt} = 0 \; .
\label{liouville_theorem}
\end{equation}

Finally, making a crucial assumption, he showed that $f$ was uniform over the entire energy shell, and can be written in a modern notation as
\begin{equation}
f_{\rm mc} ({\bf \Gamma}) = \frac{1}{\omega (E)}\; \delta ( H({\bf \Gamma}) - E)
\label{mc_dens_prob}
\end{equation}
where $\delta$ is Dirac's delta function and $\omega (E)$ the volume of the energy shell. He later called it {\em Ergode} \citep{boltzmann-1884}, what we nowadays call {\em microcanonical ensemble} according to Gibbs (see Section \ref{gibbs_stat_mech}). 

The assumption needed to obtain such a result can be expressed as follows: the mechanical system in question must pass through every point of the phase space of the energy hypersurface $H({\bf \Gamma})$. This assumption was later emphasized by the Ehrenfests who coined it {\em ergodic hypothesis}. Section \ref{ergodicity_mixing} will be devoted to this hypothesis which is central in modern statistical mechanics and at the basis of the emergence of a new field in pure mathematics; namely, the ergodic theory.

Three years later, \citet{boltzmann_1871b} showed that the probability density of small subsystems interacting with its environment is given by
\begin{equation} 
f_{\rm c} = \frac{1}{Z} \exp \left ( -\frac{H}{k_{\rm B}T}  \right )
\label{can_dens_prob}
\end{equation}
where $Z$ is a normalizing factor. He later called it {\em Holode} \citep{boltzmann-1884}, which corresponds to the {\em canonical} ensemble derived later by \citet{gibbs}.

These advances realized by Boltzmann between 1868 and 1871 played a key role in the development of his statistical approach that led him to his 1872 paper, which is commonly considered as one of his main achievements.\footnote{\citet{uffink_2007} and \citet{badino_2010} have recently emphasized the importance of Boltzmann's early studies as a preliminary to the 1872 memoir. \citet{badino_2010} especially argues against the common idea that Boltzmann developed his statistical approach in response to Loschmidt's critics (See below).} \citet{boltzmann-1872} started this seminal memoir by discussing the velocity distribution obtained by Maxwell a few years before. He started it recognizing that
\begin{quote}
[\citet{maxwell_1867}] gave a very elegant proof that, if the [Eq. (\ref{maxwell_distribution})] has once been established, it will not be changed by collisions.\footnote{Quoted from \citet{boltzmann-1872}, translation in \citet[p.~92]{brush-kinetic-theory-1}.}
\end{quote}
However, he pointed out that
\begin{quote}
It has still not yet been proved that, wathever the initial state of the gas may be, it must always approach the limit found by Maxwell.\footnote{Quoted from \citet{boltzmann-1872}, translation in \citet[p.~92]{brush-kinetic-theory-1}.}
\end{quote}
Aiming to tackle this problem, he considered a one-particle distribution function $f_1(\mathbf{r}, \mathbf{v}, t)$ so that $f_1(\mathbf{r}, \mathbf{v}, t) \; \delta \mathbf{r} \delta \mathbf{v}$ gives the average number of molecules in the infinitesimal volume $\delta \mathbf{r} \delta \mathbf{v}$ around the position $\mathbf{r}$ and the velocity $\mathbf{v}$.\footnote{We here present the part of his paper that is more familiar today. In fact, he rather started his paper by introducing a distribution function depending on the kinetic energy of the particles. Since particles are expected to be homogeneously distributed in the vessel when the equilibrium is reached, he considered that this spatial configuration could be chosen initially. It was only later in this memoir that he dealt with the distribution function $f_1(\mathbf{r}, \mathbf{v}, t)$.} He attempted to establish an equation describing the changes in $f_1$ resulting from collisions between molecules. As a result he found that the time evolution of this distribution function is governed by the integro-differential equation which is now called \textit{Boltzmann equation} \citep{boltzmann-1872}
\begin {equation}
\frac{\partial f_1}{\partial t} = -\mathbf{v} \cdot \frac{\partial f_1}{\partial \mathbf{r}} + J_{B}(f_1, f_1) \; .
\label{boltzmann-equation}
\end{equation}
Here $J_{B}$ is the {\em binary collision term}\footnote{
$J_B(f_1, f_1^*) = \int d \mathbf{v} \int b \; db \int d\phi \; |\mathbf{v}^* - \mathbf{v}| \; \left ( f_1^{\prime}f_1^{* \prime} - f_1f_1^{*} \right )$ where $f_1$ and $f_1^*$ ($f_1^{\prime}$ and $f_1^{* \prime}$) are the functions of the two particles before (after) the collision. The other terms are the same as those used by \citet{maxwell_1867} in Eq.~(\ref{maxwell_colliding_particles}). For a modern, detailed derivation of the Boltzmann equation, see, for example, \citet{huang} and \citet{liboff_1969,liboff_2003}.} taking account of only two particle collisions, which is a good approximation for dilute gases. Here again, the \textit{Sto\ss zahlansatz} is assumed by Boltzmann, which implies that the velocities ${\bf v}$ and ${\bf v}^*$ of the two colliding particles must be uncorrelated: $f_2({\bf v},{\bf v}^*) = f_1({\bf v}) f_1({\bf v}^*)$.\footnote{Note that Jeans (1877-1946) later developed this statement and introduced the assumption of \textit{molecular chaos} \citep{jeans1,jeans2}. For further information on the distinctions between both assumptions, see, for example, \citet{ehrenfest}.} He then introduced the quantity:\footnote{Actually, Boltzmann denoted it by the letter $E$ in his 1872 paper. It is Burbury (1831-1911) who replaced Boltzmann's $E$ by the letter $H$ \citep{burbury_1890}, which was later adopted by \citet{boltzmann_1895} himself. It has been suggested afterwards that Burbury's $H$ was actually the capital Greek letter $\eta$. This has probably risen because $\eta$ has often been used at this time for the entropy (e.g. \citet{gibbs} defined the index of probability denoted by $\eta$, of which the ensemble average gave the entropy. See Section \ref{gibbs_stat_mech}). \citet{chapman_1937} and, later, \citet{brush_1967} pointed out that Burbury did not provide any argument justifying this change. Although the explanation in favor of the capital Greek letter is plausible, this remains unclear until now \citep[p.~182]{brush_kin_th_2003}.}
\begin{equation}
H=\int f_1(\mathbf{r}, \mathbf{v}, t) \ln f_1 (\mathbf{r}, \mathbf{v}, t) \; d\mathbf{r} d\mathbf{v}
\end{equation}
always decreased with time unless $f_1$ was the Maxwell distribution, in which case $H$ remained constant:
\begin{equation}
\frac{dH}{dt}\leq 0 \ .
\end{equation}
In other words, with the help of this fundamental result, which is well known as \textit{Boltzmann's H-theorem}, he showed successfully that, whatever the initial distribution function, the collisions always pushed $f_1$ toward the equilibrium Maxwell distribution (\ref{maxwell_distribution}).

\citet{boltzmann-1872} suggested that $H$ could be considered as a generalized entropy having a value for any state, contrary to the thermodynamic entropy defined only for equilibrium states.\footnote{For further information on the $H$-theorem and objections against it, see, for example, \citet{dugas_1959} and \citet{brown_myrvold}.} As we shall see, he later explained the reason why the Maxwell distribution law is the one corresponding to the thermal equilibrium by showing that this distribution was the most likely to be found, because it corresponded to the largest number of microstates.\footnote{A microstate describes a specific microscopic configuration of a thermodynamic system. By contrast, a macrostate is defined by the macroscopic properties of the same system.}

The importance of the philosophical impact among physicists necessarily implied some serious criticisms against this theorem. One of them was made by \citet{LoschA,LoschB}. His objection stated that, in the case of any system approaching the equilibrium, if the molecular velocities are reversed, the $H$ function should increase instead of decreasing, which would invalidate Boltzmann's theorem. This is the famous {\em reversibility paradox}
or {\em Loschmidt paradox}.

Boltzmann replied to this by comparing the kinetic approach and the one based on probabilities. He could then conclude that the process of irreversible approach to equilibrium, which is a typical example of entropy-increasing process, corresponded to a transition from less probable to more probable microstates. Entropy itself can therefore be interpreted as a measure of probability. Concretely, \citet{boltzmann_1877} introduced the notion of {\em Complexion},
which is equivalent to what we nowadays call a microstate. By defining $W$ as the number of microstates or Complexions associated with a specific macrostate, he proposed the generalized entropy of this macrostate\footnote{For a historical outline of the concept of entropy up to Boltzmann, see, for example, \citet{darrigol_2003}.}
\begin{equation}
S_B= k_B \ln W ,
\label{boltzmann_entropy}
\end{equation}
now called the \textit{Boltzmann entropy}.\footnote{Although Boltzmann himself established this relation between entropy and probability, he never introduced a constant of proportionality. It is to Planck (1858-1947) that we owe this famous relationship in which the so-called {\em Boltzmann constant} $k_B$ appeared for the first time \citep{planck_1901}.} \citet{boltzmann_1877} interpreted the physical meaning of the second law of thermodynamics as follows: 
\begin{quote}
In most cases the initial state will be very improbable; the system passes from this through ever more
probable states, reaching finally the most probable state, that is the state of thermal equilibrium.\footnote{Quoted from \citet[p.~374]{boltzmann_1877}.}
\end{quote}

Another well-known objection against his theory was due to \citet{Zermelo1896}. Constructing his criticism of the mechanistic world view by means of the {\em recurrence theorem} that Poincar\'e (1854-1912) formulated a few years before \citep{Poincare1890}, he claimed that any system of finite volume returned as closely as wanted to any given initial positions and velocities of the particles of the system. \citet{boltzmann_1896_zermelo} answered Zermelo by using arguments that he presented in his {\em Lectures} as follows:
\begin{quote}
One should not however imagine that two gases in a $\frac{1}{10}$ liter container, initially unmixed, will mix, then again after a few days separate, then mix again, an so forth. On the contrary, one finds by the same principles which I used\footnote{See \citet{boltzmann-1896}.} for a similar calculation that not until after a time enormously long compared to $10^{10^{10}}$ years will there be any noticeable unmixing of the gases. One may recognize that this is practically equivalent to {\em never}, if one recalls that in this length of time, according to the laws of probability, there will have been many years in which every inhabitant of a large country committed suicide, purely by accident, on the same day, or every building burned down at the same time ---yet the insurance companies get along quite well by ignoring the possibility of such events. If a much smaller probability than this is not practically equivalent to impossibility, then no one can be sure that today will be followed by a night and then a day.\footnote{Quoted from \citet[p.~444]{boltzmann-LGT}.}
\end{quote}
As a result, Boltzmann interpreted Loschmidt's and Zermelo's paradoxes in a probabilistic manner by Boltzmann. Later, objections raised, in particular, the one due to Tatiana Ehrenfest-Afanassjewa who formulated it as follows:
\begin{quote}
The very important irreversibility of all observable processes can be fitted into the picture in the following way. The period of time in which we live happens to be a period in which the $H$-function of the part of the world accessible to observation decreases. This coincidence is really not an accident, since the existence and the functioning of our organisms, as they are now, would not be possible in any other period. To try to explain the coincidence by any kind of probability considerations will, in my opinion, necessarily fail. The expectation that the irreversible behavior will not stop suddenly is in harmony with the mechanical foundations of the kinetic theory.\footnote{Quoted from Tatania Ehrenfest-Afanassjewa's preface to the English translation \citep[p.~X]{ehrenfest_1959} of the German paper \citep{ehrenfest}.}
\end{quote}

Those criticisms outlined above and Boltzman's responses to them were the source of some ideas that gave rise later to a revolution in physics, e.g. initiated by Planck in his derivation of the law of black-body radiation \citep{planck_1901} and Einstein (1879-1955) in his treatment of the photoelectric effect \citep{einstein_1905_photoelectric_effect}.\footnote{See \citet{dugas_1959} for further information on the influence of Boltzmann's ideas on the modern physics, and \citet{klein_1966} and \citet{kuhn_1978} especially for their role in the rise of quantum theory.} However, such a probabilistic viewpoint, although convincing enough to some contemporary physicists,\footnote{See, for example, \citet{Lebowitz1993,lebowitz1999}.} is not satisfactory with respect to the fundamental character of the irreversibility in nature and the underlying Newtonian dynamics. Einstein himself emphasized the necessity of deriving entropy in terms of the microscopic dynamics:
\begin{quote}
$W$ is commonly equated with the number of different possible ways (complexions) in which the state considered --- which is incompletely defined in the sense of a molecular theory by observable parameters of a system --- can conceivably be realized. In order to be able to calculate $W$, one needs a {\em complete} theory (perhaps a complete molecular-mechanical theory) of the system under consideration. Given this kind of approach, it therefore seems questionable whether Boltzmann's principle {\em by itself} has any meaning whatsoever, i.e., without a {\em complete} molecular-mechanical or other theory that completely represents the elementary processes (elementary theory). If not supplemented by an elementary theory or ---  to put it differently --- considered from a phenomenological point of view, Eq. (\ref{boltzmann_entropy}) appears devoid of content.\footnote{Quoted from \citet[p.~1276]{einstein_1910}, translated in \citet[p.~232]{einstein_collected_papers_vol3}. }
\end{quote}
The aim of understanding in a dynamical sense how irreversibility emerges from the reversible microscopic dynamics is the source of recent advances in non-equilibrium statistical mechanics. These will be discussed in Section \ref{mnesm}.


\subsection{Gibbs' statistical mechanics}\label{gibbs_stat_mech}

It is to thermodynamics that Gibbs devoted his first activities as a scientist.\footnote{See \citet{klein_1983} for an overview of his early career and contributions to equilibrium thermodynamics. A biography of Gibbs can be found in \citet{bumstead_1903} and \citet{hastings_1909}.}
At this time (the early 1870s), Clausius, Thomson, and a few others were elaborating thermodynamics as a mechanical theory of heat in which the main interest concerned the exchange of heat and work between a system and its surroundings. By contrast, Gibbs interpreted thermodynamics as a theory of the properties of matter at equilibrium.\footnote{For a discussion about Gibbs' interpretation of thermodynamics with respect to Clausius', see \citet[p.~149-150]{klein_1983} and \citet[Section 2]{uffink_2007}.} He naturally emphasized the relevance of the state functions (energy and entropy) instead of work $W$ and heat $Q$. In his first contributions, \citet{gibbs_1873a} generalized geometrical representations of thermodynamic concepts which was already used successfully at that time. In order to do that, he considered what the relevant quantities are for describing a thermodynamic system in any state, i.e., the energy $E$, the pressure $p$, the volume $V$, the temperature $T$, and the entropy $S$. Knowing that the mechanical work and heat take respectively the form: $dW = -pdV$ and $dQ = TdS$ in terms of these relevant quantities, Eq. (\ref{first-principle}) can be rewritten as
\begin{equation}
dE = TdS - pdV \; .
\label{thermo_fund_eq}
\end{equation}
After he obtained this formula, he considered that
\begin{quote}
An equation giving $E$ in terms of $S$ and $V$, or more generally any finite equation between $E$, $S$ and $V$ for a definite quantity of any fluid, may be considered as the fundamental thermodynamic equation of that fluid [...].\footnote{Quoted from the last note in \citet[vol.~I, p.~2]{gibbs_1906}.}
\end{quote}
In his second paper, which appeared the same year, he developed this geometrical method with the purpose of analysing the conditions of thermodynamic equilibrium between phases \citep{gibbs_1873b}.

A few years later, \citet{gibbs_1876_1878} wrote an extensive memoir that appeared in two parts, in which he established the foundations of modern equilibrium thermodynamics of heterogeneous systems, so that, in the words of M. J. Klein, ``this book-length work [...] surely ranks as one of the true masterworks in the history of physical science".\footnote{Quoted from \citet[p.~9]{klein_1990}. Given that his memoir was quite long (about 300 pages), \citet{gibbs_1878} wrote a much shorter paper describing the basic concepts of his formalism.} Gibbs extended his theory in order to cover a large range of phenomena (such as chemical, elastic, surface, and electrochemical phenomena) by a unifying method. He proposed a general criterion for equilibrium in the following terms:
\begin{quote}
For the equilibrium of any isolated system it is necessary and sufficient that in all
possible variations of the state of the system which do not alter its energy, the
variation of its entropy shall either vanish or be negative.\footnote{Quoted from \citet[p.~56]{gibbs_1906}.}
\end{quote}
In order to characterize the chemical properties of the thermodynamic system, he introduced the concept of {\em chemical potential} (associated with chemical species in the different phases) as a new variable playing the same role as temperature and pressure. As a result, the thermodynamic equilibrium is characterized by a constant value of these variables through the heterogeneous system.\footnote{For a modern treatise of equilibrium thermodynamics, see, for example, \citet{denbigh_1997}.} Perhaps, he viewed his formalism as complete enough, so that he paid only little attention to thermodynamics afterwards, and decided to tackle various issues in other fields of physics and mathematics.

One of the questions he then asked himself concerned the rational foundations of thermodynamics. In 1884, Gibbs wrote an interesting paper entitled {\em On the Foundamental Formula of Statistical Mechanics, with Applications to Astronomy and Thermodynamics} \citep{gibbs_1884}. Unfortunately, only its abtract pronounced at a scientific meeting has survived and been reprinted in Gibbs' {\em Scientific Papers} \citep{gibbs_1906}. This can nevertheless provide some important information. Gibbs probably esteemed that the elaboration of recent works of Boltzmann and Maxwell deserved a new name, {\em statistical mechanics}, in order to make the distinction with the kinetic theory of gases. In addition, the {\em fundamental formula} which he refered to is another form of Liouville' theorem. From Eq. (\ref{liouville_theorem}), one can easily derive what we nowadays call {\em Liouville equation}:
\begin{eqnarray}
\frac{\partial f}{\partial t} & = &
- \sum_{a=1}^{N} \left ( \frac{\partial H}{\partial {\bf r}_a} \cdot \frac{\partial f}{\partial {\bf p}_a}
- \frac{\partial H}{\partial {\bf p}_a} \cdot \frac{\partial f}{\partial {\bf r}_a} \right ) \; .
\label{liouville_equation1}
\end{eqnarray}
He concluded his abstract in saying that:
\begin{quote}
The object of the paper is to establish this proposition (which is not claimed as new\footnote{Gibbs here refered likely to the use of Liouville's theorem by \citet{boltzmann-1868,boltzmann_1871c} and, then \citet{maxwell_1879}.}, but which has hardly received the recognition which it deserves) and to show its applications to astronomy and thermodynamics.\footnote{Quoted from \citet[vol.~II, p.~16]{gibbs_1906}.}
\end{quote}
As will be seen below, this formula has been at the basis of Gibbs' later developments of his statistical treatment of mechanics.

Little is known about how he proceeded afterwards.\footnote{See \citet{klein_1990} for a discussion about the reasons of this lack of information.} According to his former student Bumstead (1870-1920), Gibbs ``seldom, if ever, spoke of what he was doing until it was practically in its final and complete form".\footnote{Quoted from \citet[vol.~I, p.~xxvi]{gibbs_1906}.} He worked quite alone without the need of criticisms or encouragements. Thanks to notes of students who attended his lectures in the 1890s, it appears clearly that his Statistical Mechanics was almost already elaborated several years before his famous 1902 treatise. Moreover, Gibbs wrote papers only when he was completely satisfied by his results. It is only at the very begining of the 20th century, two decades after his 1884 paper, that the scientific community became eventually aware of his statistical approach of mechanics in a nowadays celebrating book entitled {\em Elementary Principles of Statistical Mechanics} \citep{gibbs}. In fact, in regard to his 1902 treatise, as M. J. Klein has suggested,
\begin{quote}
it is not impossible that the pressure to contribute to the Yale Bicentennial Series propelled Gibbs into completing a book that he might otherwise have considered not quite ready to publish.\footnote{See \citet[p.~15]{klein_1990}.}
\end{quote}

Roughly speaking, his Statistical Mechanics is based on the description in phase space\footnote{Note that Gibbs called it the {\em extension-in-phase}, probably because physicists before the advent of relativity still had some aversion to use the term {\em space} to a space with more than three dimensions. Ten years later, at the time of the Ehrenfest's article, these conceptual difficulties vanished. Although they explicitly introduced the phrase {\em Phasenraum}, that is {\em phase space} (\citet{ehrenfest_1959}, p.~18), they prefered the use of {\em $\Gamma$-space} along their paper. It is only after \citet{rosenthal_1913} and \citet{plancherel_1913} that {\em phase space} became common in the scientific literature. See \citet{nolte_2010} for further information.} of an ensemble of identical mechanical systems which differed in their initial conditions. In his Preface, he then started paying tribute to Boltzmann 
\begin{quote}
The explicit consideration of a great number of systems and their distribution in phase, and of the permanence or alteration of this distribution in the course of time is perhaps first found in Boltzmann's paper on the ``{{Zusammenhang zwischen den S\"atzen \"uber das Verhalten mehratomiger Gasmolek\"ule mit Jacobi's Princip des letzten Multiplicators}}" (1871).\footnote{Quoted from \citet[p.~viii]{gibbs}. He here refered to the first section of \citet{boltzmann_1871}'s paper.}
\end{quote}
As seen in Section \ref{Boltzmann_stat_irrev}, Boltzmann had already introduced the approach of ensemble of mechanical systems three years before \citep{boltzmann-1868}. However, Gibbs' purpose was ``to find in rational mechanics an {\em a priori} foundation for the principles of thermodynamics".\footnote{Quoted from \citet[p.~165]{gibbs}.} He thus constructed the well-known ``ensembles of systems" which led him to make some ``analogies" with thermodynamics.

On the other hand, important conceptual difficulties in kinetic theory, such as the problem concerning the anomalous values of specific heat of gases, were not yet overcome when Gibbs wrote his book. Furthermore, the intense debate on the atomic theory which animated the scientific community at the end of the 19th century (see Section \ref{vict_at_th}) would also justify the fact that Gibbs, unlike Maxwell and Boltzmann, avoided any hypothesis on the constitution of the matter\footnote{See \citet{deltete_1995} for a discussion about the relationships between Gibbs and anti-atomists such as the energetists.} and rather considered mechanical systems with an arbitrarily large number of degrees of freedom. It is only in the very last chapter that he discussed the application of statistical mechanics to systems composed of molecules.

As mentioned above, Gibbs indeed aimed at showing the key role that the Liouville equation could play. He then showed that Eq. (\ref{liouville_equation1}) provided directly the condition for {\em statistical equilibrium}:
\begin{equation}
\sum_{a=1}^{N} \left ( \frac{\partial H}{\partial {\bf r}_a} \cdot \frac{\partial f}{\partial {\bf p}_a}
- \frac{\partial H}{\partial {\bf p}_a} \cdot \frac{\partial f}{\partial {\bf r}_a} \right ) = 0 \; .
\label{gibbs_stat_cond}
\end{equation}
For that reason, he considered the Liouville equation as the {\em fundamental equation of statistical mechanics}.

We here note that Gibbs' condition of statistical equilibrium is quite different than that of Maxwell and Boltzmann. Whereas the last two considered that the equilibrium was reached when the velocity distribution was given by Eq. (\ref{maxwell_distribution}), Gibbs defined it when the probability $f$ remained stationary: $\frac{\partial f}{\partial t} = 0$. On the basis of Eq. (\ref{gibbs_stat_cond}), he distinguished different cases: the {\em microcanonical}, {\em canonical}, and {\em grand-canonical ensembles}, of which the first two had already been introduced by Boltzmann who called them respectively {\em Ergode} (\ref{mc_dens_prob}) and {\em Holode} (\ref{can_dens_prob}).

In Chapter XIV, \citet{gibbs} discussed the analogies between his formalism and thermodynamics. After Gibbs suggested to view statistical equilibrium as analogous of thermal equilibrium, he considered a closed system, i.e., system which exchanges only energy with its environment. He hence reformulated Eq. (\ref{thermo_fund_eq}) as
\begin{equation}
dE = TdS - \sum_i A_i da_i
\label{thermo_fund_eq2}
\end{equation}
where $a_i$ are what Gibbs called the {\em external co\"ordinates} (e.g., the volume), and $A_i$ their associated external forces (e.g., the pressure).

In Gibbs' statistical mechanical formalism, the canonical distribution is the most appropriate for describing ensemble of closed systems. He derived for this ensemble a relation with the same structure as Eq. (\ref{thermo_fund_eq2})
\begin{equation}
d \langle H \rangle = - \theta d \langle \eta \rangle - \sum_i \langle A_i \rangle da_i
\end{equation}
where $\langle A_i \rangle = \left \langle \frac{\partial H}{\partial a_i} \right \rangle$ are the external forces, $\theta$ the modulus, and $\eta$ the {\em index of probability} defined as the logarithm of the density probability: $\eta = \ln f$.
The analogy between the last two equations allows to associate the modulus $\theta$ with the temperature as well as the canonical ensemble averages with the corresponding thermodynamic quantities, so that the average of the index of probability $\langle \eta \rangle$ defines the famous {\em Gibbs entropy}:
\begin{equation}
S_G(t) = - \langle \eta \rangle =  - \int f({\bf \Gamma},t) \ln f({\bf \Gamma},t) \; d{\bf \Gamma}
\label{gibbs_entropy}
\end{equation}
The Gibbs entropy has the useful property of taking the highest value for any equilibrium ensemble (microcanonical, canonical or grand-canonical distribution). However, this definition involves quickly some limitation since Liouville's theorem confines $S_G(t)$ to constant values in the course of time: $S_G(t) = S_G(0)$. In other words, whatever the initial phase-space distribution, the entropy $S_G$ remains constant in time and, as a result, is not appropriate to describe an approach to equilibrium. To solve this problem, Gibbs used the analogy of the time evolution of an initial distribution with the stirring of a coloring matter in an incompressible fluid. The average density of the coloring matter also remains constant. Nevertheless, it is obvious that a uniform mixture is obtained after a certain time. Hence, this analogy tackles directly the notion of density. Indeed, the density is commonly obtained by the quantity of coloring matter in a volume element of which the size tends to zero. If such a limit is taken before stirring the fluid, $S_G$ will actually remain constant. On the contrary, if the fluid is stirred in considering a small but non-zero volume element, the density will become uniform, even if the limit is taken afterwards. Consequently, the result differs depending on whether the limit is taken before of after the fluid is stirred. This argument leads to reconsider the so-called {\em fine-grained entropy} (\ref{gibbs_entropy}) in replacing the {\em fine density} $f$ by a {\em coarse-grained density}. This operation consists in dividing the phase space into cells $\lbrace {\cal C}_i \rbrace$ (let us say of size $\epsilon$) over which $f$ is averaged: $f({\cal C}_i,t) = \int_{{\cal C}_i} f({\bf \Gamma},t) d{\bf \Gamma}$. One hence obtained the {\em coarse-grained entropy} which can be written as:
\begin{equation}
S_G(\epsilon, t) = - \sum_i f({\cal C}_i,t) \ln f({\cal C}_i,t)
\label{cg_gibbs_entropy}
\end{equation}
In such a way, the coarse-grained entropy can be used to describe the approach towards the statistical equilibrium. However, such a process requires that the system satisfy an important assumption which is usually called {\em mixing property}. This question will be addressed in the next section.

If Gibbs viewed the Liouville equation as the fundamental equation of statistical mechanics, it was mostly because it provided him the basic condition of statistical equilibrium. However, its validity is of course not restricted to equilibrium. Despite that, the study of non-equilibrium processes by means of the Liouville equation appeared much later. After an original idea of \citet{yvon}, Bogoliubov (1909-1992), Born (1882-1970), Green (1920-1999), and Kirkwood (1907-1952) investigated its relation with transport equations such as the Boltzmann equation, which led them to establish the so-called {\em BBGKY hierarchy}.\footnote{Their respective contributions were published in \citet{bogoliubov}, \citet{born-green}, \citet{kirkwood-1946}, and \citet{yvon}.} At the same time, Gibbs' formalism has been widely popularized in the 1930s by Tolman (1881-1948) who wrote an extensive treatise on its application to both classical and quantum systems \citep{tolman_1938}. This book then became a standard work for many years and had a significative impact, such that Eq. (\ref{liouville_equation1}) became more and more attractive. \citet{prigogine_1962} later attributed it a tremendous role in its investigations of non-equilibrium processes. This gave rise in the 1980s to the elaboration of a spectral approach of the Liouville equation and, then, to recent successful advances in the understanding of irreversible processes at the statistical level (see Section \ref{mnesm}).


\subsection{Ergodic hypothesis and mixing}\label{ergodicity_mixing}

As seen in Section \ref{Boltzmann_stat_irrev}, an assumption was underlying in Boltzmann's work from his 1868 paper on; namely, the ergodic hypothesis. This is commonly stated as follows: the phase-space trajectory associated with the evolution of an isolated mechanical system covers all the energy hypersurface. If so, the average of a physical quantity over a long trajectory (time average) might be replaced by an average over all points of the phase space satisfying the equation of energy (phase space average). This assumption discussed later by Boltzmann and Maxwell was in fact strongly emphasized by the Ehrenfests in their fundamental paper in which they coined it {\em ergodic hypothesis}.\footnote{As a matter of fact, \citet{boltzmann-1884} introduced the words {\em Ergode} and {\em ergodisch} without providing any information on its etymology. The Ehrenfests then proposed the combination of the Greek words \tg{{'<e}rgon} (= energy) and \tg{<od'os} (= path) (\citet[p.~89, note 93]{ehrenfest_1959}). The etymology of this word seems to remain unclear and has recently given rise to other speculations, e.g. due to \citet{gallavotti_1982,gallavotti_1995} and \citet{mathieu_1988}. On the other hand, as S. G. Brush has pointed out, the Ehrenfests made a mistake on the meaning of the phrase {ergodic} attributed to Boltzmann. The latter did not define the word {\em Ergode} as a single system which had the ergodic property but rather as an ensemble of systems described by a probability distribution that he called {\em ergodisch} (see \citet[p.~297]{boltzmann-LGT} and \citet[p.~169]{brush_1967_foundations}). However, although the Ehrenfests made some mistakes at the historical level, it was nevertheless justified to associate the phrase {\em ergodic} with this assumption since this is required to obtain an Ergode, that is the microcanonical ensemble. Note also that they claimed that the ergodic hypothesis was called by Maxwell the assumption of the {\em continuity of path} whereas this phrase should be attributed to \citet{jeans1} in a paper published more than two decades after Maxwell's death.}

In 1868, Boltzmann did not pay attention to this assumption that he viewed as ``not improbable". It is only three years later that he investigated more carefuly its validity. He considered a point moving in a plane when submitted to a potential function $a x^2 + b y^2$. The figures obtained by drawing such a trajectory in the ($x$,$y$)-configuration space are similar to the so-called {\em Lissajous figures}.\footnote{In 1857, Lissajous (1822-1880) proposed a method useful to visualize vibrating phenomena \citep{lissajous_1857}. In particular, he considered two acoustic waves in perpendicular directions: $x = A \sin(at + \phi)$ and $y = B \sin(bt)$. Thanks to this method, he obtained a series of closed curves for different rational values of $a/b$ and different phases $\phi$ which are known as the Lissajous figures. These are sometimes called Bowditch figures in reference to Bowditch (1773-1838) who, several decades before, studied the movement of a pendulum suspended from two points \citep{bowditch_1815}.} When the ratio $a/b$ is rational, the point starts over at the initial position after a finite time and then repeat the same motion. The motion is thus periodic. In contrast, when $a/b$ is irrational (i.e., the periods $a$ and $b$ are incommensurable), periodic curves could no longer be obtained, so that Boltzmann expected that the trajectory would ``gradually traverse the entire area of a rectangle".\footnote{\citet[vol.~I, p.~271]{boltzmann_WA}. For further detail, see \citet[p.~169-170]{brush_1967_foundations} and \citet[Section~7.6]{cercigniani}.}

However, as pointed out by Brush,\footnote{See \citet[p.~170]{brush_1967_foundations}.} what Boltzmann exactly meant by a trajectory that traverses the entire space is actually ambiguous. Indeed, it could be understood in two ways, i.e., the trajectory goes (i) through all points in the plane or (ii) arbitrarily close to every point. The distinction between these two interpretations remained actually unclear in the following papers in which he tackled this issue \citep{boltzmann_1887,boltzmann_1892}. Furthermore, he often considered the role of random external forces to support his hypothesis, and expressed some doubts on the validity for mechanical systems when left undisturbed. In a more explicit statement on the ergodic hypothesis, \citet{maxwell_1879} did address only systems interacting with their environment. According to Brush\footnote{See \citet[p.~174]{brush_1967_foundations}.}, Boltzmann did not likely distinguish both interpretations insofar as this distinction was meaningless for him.\footnote{See Part 1, Chapter 1 of \citet{dugas_1959}'s book in which he provided a deep analysis of Boltzmann's physical conception of nature, in particular his reject of the idea of infinity in real physical systems.} This was rather made later by the Ehrenfests who coined the second meaning {\em quasi-ergodic hypothesis} \citep{ehrenfest}. Although they did not discuss the validity of the latter hypothesis, they clearly exposed the issue to mathematicians and, as a result, provided the basis of a new mathematical field; namely, the {\em ergodic theory}.

The ergodic hypothesis, which claims that every point of the phase space is visited by the trajectory describing the system evolution, gives rise naturally to some conceptual difficulties: how can a curve (a one-dimensional object) pass through all the points of a region with more than one dimension (the phase space)? In the mathematical language, the question can be reformulated as follows: can we have a mapping from a one-dimensional manifold to another which has a higher dimensionality? This question animated several mathematicians from the 1870s. Cantor (1845-1918) developed the set theory and proved that a mapping existed between any manifolds with distinct dimensionality \citep{cantor_1878}, though this mapping did not satisfy the property of continuity\footnote{The property of continuity insures that two close points in a $d>1$ dimensional manifold are mapped into points that are still close together on the one dimensional manifold.} when a one-dimensional manifold is considered. In the continuity of these investigations, Brouwer (1881-1966) later proved that a bicontinuous one-to-one mapping existed only when the manifolds had the same dimensionality \citep{brouwer_1910}. Consequently this result gave a new mathematical tool to determine whether two manifolds have (or not) the same number of dimensions.

During the same period, Borel (1871-1956), and later, Lebesgue (1875-1941) completed Cantor's method of counting infinite sets of points by developing another approach to distinguish sets of different dimensionality; namely, the {\em measure theory} \citep{borel_1898,lebesgue_1902}. Let us focus on some basic concepts of {\em measure}: a point or a denumerable set of points has measure zero with respect to a one-dimensional line segment. Similarly, a phase-space trajectory has also measure zero with respect to a two-dimensional set, a fortiori with respect to the phase space. The measure theory is essential not only to modern probability theory but also to modern statistical mechanical theories, as we shall see below (see \ref{measure} for further detail on the concept of probability measure).

These mathematical advances played a key role for closing the debate on the existence of ergodic systems. Soon after the Ehrenfests' article, Plancherel (1885-1967) announced at the Bern meeting of the Swiss Physical Society that ``(except in the trivial case $d = 1$) the ergodic systems {\em does not exist}". \footnote{Quoted from \citet[p.~255]{plancherel_1912}.} However, in the report of the meeting, no indication of the method of proof can be found \citep{plancherel_1912}. The following year, Rosenthal (1887-1959) and Plancherel presented different ways of showing the {\em impossibility of ergodic systems} in the same volume of the {\em Annalen der Physik} \citep{rosenthal_1913,plancherel_1913}. Rosenthal based his own proof on Brouwer's results (see above), whereas Pancherel used the measure theory elaborated by Borel and Lebesgue.\footnote{For further information on their arguments, see \citet[p.~182]{brush_1967_foundations}.}
Consequently, these two papers confirmed the Ehrenfests' doubts on the validity of the ergodic hypothesis: the phase-space trajectory describing a mechanical system cannot pass through every point of the energy hypersurface.\footnote{For further detail on this discussion of the ergodic hypothesis and the proofs of its impossibility, see \citet{brush_1967_foundations,brush_1971} and \citet{plato_1992}.}

Despite this result, a revival of this discussion occurred in the 1930s. A new approach of the issues in question was made possible by crucial mathematical advances carried out by Koopman (1900-1981) in applying the Hilbert space formalism to the Hamiltonian dynamics \citep{koopman_1931}. Evolution operators $\hat{U}^t$ are then introduced and represented as a unitary group. By means of this new method, von Neumann (1903-1957) investigated the possibility of stating an equivalence between time average and phase space average for dynamical systems satisfying the {\em quasi-ergodic hypothesis}. In the early 1930s, he proposed the {\em Proof of the Quasi-Ergodic Hypothesis} \citep{von_neumann_1932}\footnote{Although the title of \citet{von_neumann_1932}'s paper explicitly addressed the quasi-ergodic hypothesis, he decided at the very begining of the paper to say ergodic hypothesis ``for brevity". This decision can be misleading in the light of the historical overview presented above. Nevertheless, following him and Birkhoff, the word ``ergodic" instead of ``quasi-ergodic" is now very common in the technical language of mathematicians and physicists. Despite the fact that this choice could involve some confusion, we here use the phrase ergodic hypothesis (or systems).} which was strengthened soon after by Birkhoff (1884-1944).\footnote{At the end of 1931, Birkhoff published a paper in which he refered to the ``not yet published" paper of von Neumann \citep{birkhoff_1931}. As a matter of fact, von Neumann's paper was communicated to PNAS nine days before Birkhoff's, but was finally published a few later.} This result which equates the time average and phase space average of an observable $A$ is now well known as the {\em Birkhoff ergodic theorem}\footnote{For further information on the advances made in the early 1930s, see \citet{birkhoff_koopman_1932}. See \citet{arnold_avez_1968} for further technical detail on the ergodic theory.}
\begin{equation}
\lim_{T \to \pm \infty} \frac{1}{T} \int_{0}^{T} A_t \; dt = \int_{{\cal M}} A({\bf \Gamma}) \; \mu_{\rm i}({d\bf \Gamma}) \quad {\rm for \ almost \ all \ }{\bf \Gamma} \ {\rm iff \ the \ system \ is} \ metrically \ transitive
\end{equation}
where $\mu_{\rm i}$ is {\em an invariant measure} (see \ref{measure}), $A_t = \hat{U}^t A_0$ is the observable at time $t$, of which the evolution is ruled by the transformation $\hat{U}^t$. According to \citet{birkhoff_smith_1928},
\begin{quote}a transformation will be called {\em metrically transitive} if there exists no measurable invariant set ${\cal A} \subset {\cal M}$  such that $0 < \mu({\cal A}) < \mu({\cal M})$.\footnote{Quoted from \citet[p.~365]{birkhoff_smith_1928}.}
\end{quote}
In other words, a system has the property of metrical transitivity iff, for any partition of the phase space $\cal M$ into disjoint sets $\cal A$, $\cal B$ such that $\hat{U}^t {\cal A} = {\cal A}$ and $\hat{U}^t {\cal B} = {\cal B}$, it holds that $\mu ({\cal A}) = 0$ or $\mu ({\cal B}) = 0$.\footnote{See \citet{seidel_1933} for further information on the metrical transitivity.}

Despite the importance of this mathematical result, it is legitimate for physicists to know whether this property can be found in models as realistic as possible. In this context, \citet{sinai} proved that a two-dimensional system composed of two hard disks satisfied the condition of metric transitivity. More recently, \citet{szasz_1996} extended the proof to the case of systems composed of many hard balls, so that the Birkhoff ergodic theorem appeared to be relevant to a class of models largely used in statistical mechanics. Although the hard-sphere potential in such models is very simple, it is expected by physicists that the metrical transitivity would remain plausible for systems of gas particles submitted to more sophisticated interaction potentials.\footnote{For a discussion on the use of metrically transitive systems in statistical mechanics, see, for example,\citet[Section~6.1]{uffink_2007}.}

In the previous section, we have seen that \citet{gibbs} used the argument of coarse graining to define an entropy which described the evolution of an ensemble of systems toward the equilibrium. This actually requires a property called {\em mixing}. To visualize how a point set in phase space evolved in time,
\begin{quote}
We may find a closer analogy with the case under consideration in the effect of stirring an incompressible liquid. In space of $2n$ dimensions the case might be made analytically identical with that of an ensemble of systems of $n$ degrees of freedom, but the analogy is perfect in ordinary space. Let us suppose the liquid to contain a certain amount of coloring matter which does not affect its hydrodynamic properties. Now the state in which the density of the coloring matter is uniform, i.e., the state of perfect mixture, which is a sort of state of equilibrium in this respect that the distribution of the coloring matter in space is not affected by the internal motions of the liquid, is characterized by a minimum value of the average square of the density of the coloring matter. Let us suppose, however, that the coloring matter is distributed with a variable density. If we give the liquid any motion whatever, subject only to the hydrodynamic law of incompressibility, --- it may be a steady flux, or it may vary with the time,--- the density of the coloring matter at any same point of the liquid will be unchanged, and the average square of this density will therefore be unchanged. Yet no fact is more familiar to us than that stirring tends to bring a liquid to a state of uniform mixture, or uniform densities of its components, which is characterized by minimum values of the average squares of these densities.\footnote{\citet[p.~144-145]{gibbs}.} 
\end{quote}
This property, now called {\em mixing}, was later formulated in a mathematically precise way by \citet{von_neumann_1932b}, and developed by Hopf (1902-1983) a few years later \citep{hopf_1934,hopf_1937}. In the language of measure theory, a system is said {\em mixing} if:
\begin{equation}
\lim_{t \to \infty} \mu \left ( \hat{U}^t {\cal A} \cap {\cal B} \right ) = \mu({\cal A}) \mu({\cal B}) \; .
\label{mixing_def}
\end{equation}
It is easy to show that mixing is a stronger property than ergodicity (i.e., metric transitivity). Let us consider $\cal A$ as a invariant set ($\hat{U}^t{\cal A} = {\cal A}$) and ${\cal B} = {\cal A}$. As a result, Eq. (\ref{mixing_def}) becomes
\begin{equation}
\mu ({\cal A}) = \mu({\cal A}) \mu({\cal A})  \; .
\end{equation}
This equation has two solutions:
\begin{eqnarray}
\mu ({\cal A}) & = & 0 \ : \ {\cal A} \ {\rm is \ a \ trivial \ invariant \ set \ of \ measure \ zero} ; \nonumber \\ 
\mu ({\cal A}) & = & 1 \ : \ {\cal A} \ {\rm is \ the \ whole \ energy \ hypersurface} . \nonumber
\end{eqnarray}
Consequently, when a system is mixing, the only invariant set with positive measure is the energy hypersurface. In other words, mixing implies ergodicity. As a matter of fact, contrary to ergodicity, mixing ensures that a non-equilibrium ensemble will relax toward equilibrium.\footnote{See for example \citet{lebowitz_penrose_1973}} As will be seen in Sections \ref{micro-chaos} and \ref{mnesm}, this concept will play an fundamental role in the modern investigations of irreversible processes.


\subsection{Victory of the atomic hypothesis}\label{vict_at_th}

Despite all the brillant discoveries due to the atomic hypothesis of matter accumulated at the end of the 19th century, a lot of scientists (amongst the most important) continued to reject the atomic theory. In general, they belonged to movements called \textit{equivalentism} represented especially by Berthelot (1827-1907), \textit{energetics} with Helm (1851-1923), Ostwald (1853-1932)\footnote{See \citet{holt_1970} for further information.}, and Duhem (1861-1916), or \textit{empiriocriticism}, the doctrine of Avenarius (1843-1896) and Mach (1838-1916).\footnote{For further information, see, for example, \citet{brush-thermo,brush_mach,brush-kind-motion2}, \citet{kubbinga}, and \citet{pullman}.} The debate then were still vigorous between anti-atomists and the partisans of the kinetic theory, especially Boltzmann. Fortunately, during the first decade of the 20th century, the triumph of the kinetic theory and the atomic hypothesis was finally recognized with the work done in 1905 by Einstein on the \textit{Brownian motion} \citep{einstein}. Aware of the attacks on kinetic theory by Mach, Ostwald, and others, Einstein started his article by contrasting the predictions made by thermodynamics and by the kinetic theory. In particular, he pointed out that thermodynamics distinguished the Brownian particle and the {\em hypothetic} molecules composing the liquid, whereas kinetic theory did not. Moreover he asserted that the colloidal particle should follow Clausius' \textit{equipartition theorem}.\footnote{The equipartition theorem is an extension of the \citet{clausius1870}'s virial theorem. See footnote \ref{virial_footnote}.} By combining Stokes' formula for the force on a sphere moving through a viscous fluid, and the formula for the osmotic pressure of dissolved molecules, he derived an expression for the mean-squared displacement of the Brownian particle submitted to the molecular agitation, and hence provided a way to evaluate the Avogadro number $N_{\rm Av}$. A few years later, \citet{perrin-1908} made different experiments applying to Einstein's Brownian-motion theory\footnote{\citet{newburgh_2006} have repeated the \citet{perrin-1908} experiment and set it back in the context of the first decade of the 20th century.} and found $N_{\rm Av} = 6.7 \times 10^{23}$. In 1913, he published his now celebrated book \textit{Les atomes} \citep{perrin-atomes} in which he proposed thirteen experiments based on different phenomena for evaluating Avogadro's number. The quite narrow range $6.0 - 7.5 \times10^{23}$ thereby obtained is such that we now attribute the demonstration of the discrete feature of the matter to Perrin, especially due to his efforts to convince the opponents of atoms' existence.\footnote{Let us mention that the recent measurements gives us $6.02214179(30) \times 10^{23}$ for the current value of Avogadro's number \citep{CODATA}.} Whereas Ostwald and others recognized it, some of them like Mach\footnote{For further information on the development of Mach's ideas about the atomic theory, see, for example, \citet{brush_mach}.} and Duhem continued to reject the reality of the atom. It is ironic to remark that those who were against the theory of atoms ---i.e., the most fundamental unit of matter--- became later convinced when J. J. Thomson (1856-1940) discovered the electron \citep{thomson_1897}, which disproved the indivible character of the atoms.


\subsection{Development of theories from Boltzmann's equation}

If the performances of the kinetic theory were already numerous at the time of Maxwell and Boltzmann, new works and predictions were added which emphasized its depth.

Although \citet{dufour_1873} already showed experimentally the so-called {\em diffusion thermo-effect} in the gaseous phase,\footnote{One deals with the diffusion thermo-effect when the diffusion of one gas into another at the same temperature results in the establishment of a temperature gradient.} the opposite cross-phenonenon had not yet been observed at the beginning of the 20th century.\footnote{By constrast, the partial separation of a {\em liquid} mixture in the presence of a temperature gradient was already highlighted by \citet{ludwig_1856} and \citet{soret_1879}.} In 1911, Enskog (1884-1947) predicted that diffusion would appear in a mixture when a temperature gradient is imposed, namely the \textit{thermal diffusion} \citep{enskog_1911}. A few years later, Chapman (1888-1970) derived independently an expression for the thermal diffusion coefficient \citep{chapman_1916} that Dootson and himself experimentally verified the next year \citep{chapman_dootson_1917}.

At the turning of the 20th century, different works based on the mean-free-path method were achieved by a lot of physicists (Maxwell, Boltzmann, Tait, Rayleigh, etc.) to improve the expression for the viscosity in low-density gases.\footnote{These studies especially investigated the dependence on temperature and the factor $\frac{1}{3}$ in Eq. (\ref{maxwell-viscosity}).} But it is only in 1917 that Chapman and Enskog derived independently an expression for the viscosityfrom the kinetic equation \citep{chapman,enskog-1917}.\footnote{Notice that Chapman started from Maxwell's transport theory whereas the Enskog's derivation was based on Boltzmann's. For a discussion on these two methods, see \citet{brush-kinetic-theory-VI}.} The derivation known today is that of Enskog, which is based on a series expansion of Eq. (\ref{boltzmann-equation}) for the distribution function by introducing a parameter $\lambda$. He therefore derived analytical expressions for the  transport coefficients for a general interaction potential. In the case of the hard-sphere potential, the viscosity coefficient takes the form~:
\begin{equation} 
\eta_B = 1.0162 \; \frac{5}{16 \; \sigma^2}\;  \sqrt{\frac{m k_BT}{\pi}}
\label{chapman-enskog-viscosity}
\end{equation}
where $\sigma$ and $m$ are respectively the diameter and the mass of the particles. This expression for the viscosity  is now called the \textit{Boltzmann viscosity} and it confirms the temperature-dependence and the absence of density-dependence predicted by earlier theories, especially that of Maxwell in Eq. (\ref{maxwell-viscosity}).\footnote{By substituting the mean velocity $\overline{v}$ and the mean free path $\overline{l}$ in Eq. (\ref{maxwell-viscosity}) by their expression in terms of temperature and atomic properties, one obtains $\eta_M = \frac{2}{3 \; \sigma^2} \sqrt{\frac{m k_B T}{\pi^3}}$.} Furthermore, available for any interaction potential, the Chapman-Enskog development turned out to be a method capable of determining the actual force law of real molecules by comparing the experimental and theoretical results of transport properties such as viscosity and thermal conductivity. This was the path followed by Lennard-Jones (1894-1954) until the early 1930s when he proposed the now called \textit{Lennard-Jones 6-12 potential} which is still considered today as the most realistic potential existing between particles \citep{jones}.\footnote{For further historical detail on the development of expressions for interatomic forces, see, for example, \citet{brush-LJ}.}

Maxwell and others showed both theoretically and experimentally that the viscosity coefficient is independent of the density in dilute gases. However, even in moderately dense gases, it appears that this property can no longer be observed. In 1922, Enskog proposed an extension of his previous method predicting this dependence \citep{enskog-dense}. Considering a hard-sphere system, he assumed that the collision rate in a dense gas would be changed by a factor $\chi$ which could be related to the equation of state. To do so, he modified the term $f_1 f_1^*$ in Eq. (\ref{boltzmann-equation}). Although $f_1$ and $f_1^*$ were initially evaluated in the same point in space, Enskog replaced them by $f_1(x,y,z,\dots) f_{1}^{*}(x^{*},y^{*},z^{*},\dots)$, where the points $(x,y,z)$ and $(x^{*},y^{*},z^{*})$ must be separated by the distance $\sigma$. In such a way, the Enskog shear viscosity is written as\footnote{For the development of Enskog's theory, see, for example, \citet{HCB} and \citet{CC}.}
\begin{equation}
\eta_E =\eta_{B} \left ( \frac{1}{\chi} + \frac{4}{5} \; b_0 n + 0.7614 \; b_0^2
n^2 \chi \right )
\label{enskog-viscosity}
\end{equation}
and the bulk viscosity as
\begin{equation}
\zeta_E = 1.002 \chi \eta_{B} (b \; n)^2
\label{enskog-bulk-viscosity}
\end{equation}
in which the factor $\chi = 1 + \frac{5}{8} b_0  n + 0.2869 \; (b_0 n)^2$, $b_0=\frac{2 \pi \sigma^3}{3}$ and $n$ is the number density. We thus obtain a density-dependent expression for the viscosity.\footnote{Let us mention that, in 1899-1900, by considering the analogy with the modification of the equation of state for ideal gas when the effect of finite molecular size is taken into account, and the correction of the mean free path due to the effect of excluded volumes on the collision rate, J\"ager (1865-1938) modified the dilute gas viscosity formula (\ref{maxwell-viscosity}). He thereby obtained $$\eta_J = \eta_{M} \left ( \frac{1}{A} + 8 \; b_0 n + 16 \; b_0 ^2 n^2 A \right ) $$ where $A = 1 + \frac{5b_0 n}{2} + \cdots$. The similarity with the Enskog viscosity (\ref{enskog-viscosity}) is remarkable. See \citet{brush-kinetic-theory-VI,brush-kind-motion2} for more detail.} The first experimental verification of Enskog's theoretical predictions came only a decade later. Indeed, \citet{michels-gibson} measured the viscosity of nitrogen gas at pressure up to 1000 atm (1013250 hPa) and obtained a good agreement with the semi-empirical Enskog theory. They also evaluated the dimensions of the gas molecule at different temperatures. Further, the only work allowing Enskog to test his theory was the one carried out by \citet{warburg-babo} for carbon dioxide. The agreement is quite good as detailed in Table \ref{comparison-enskog-viscosity}.

\begin{table}[!t]
\begin{center}
\begin{tabular}{|c|c|c|}
\hline
Mass density & Kinematic viscosity & Kinematic viscosity\\
 & (experimental) & (calculated) \\
\hline
\hline
0.100 & 1.80 & 1.91 \\
0.170 & 1.12  & 1.17 \\
0.240 & 0.908 & 0.899 \\
0.310 & 0.784 & 0.774 \\
0.380 & 0.724 & 0.719 \\
0.450 & 0.702 & 0.701 \\
0.520 & 0.704 & 0.705 \\
0.590 & 0.722 & 0.724 \\
0.660 & 0.756 & 0.754 \\
0.730 & 0.795 & 0.792 \\
\hline
\end{tabular}
\end{center}
\caption{Kinematic viscosity $\frac{\eta}{\rho}$ $\lbrack 10^3$ poises/(g/cm$^3$)$\rbrack$ vs. the mass density $\lbrack$ g/cm$^3 \rbrack$ at temperature $T=313.45$ K, near the critical point. Comparison between the experimental results obtained by \citet{warburg-babo} for carbon dioxide and Enskog's predictions. As the density increases, the kinematic viscosity goes through a minimum after which it increases. Enskog's predictions confirm this property (data provided in Enskog (1922)'s paper).
\label{comparison-enskog-viscosity}}
\end{table}

A few decades later, new experimental data for noble gases (for viscosity as well as for thermal conductivity) were added to allow a comparison and a test of theoretical predictions \citep{sengers1,sengers2,HMC72}. The particular advantage of Enskog's theory compared to other theories for dense gases is that it requires the adjustment of only a few parameters like the diameter of the spheres in order to observe an agreement between Enskog's shear viscosity and the experimental data for a larger range of density.\footnote{Indeed effective values may be attributed to the parameters $b$ and $\chi$. We owe to Michels and Gibson (1931) the introduction of such a procedure by identifying the pressure $P$ in the equation of state for rigid spheres with the thermal pressure $T (\partial p / \partial T)_V$ of the real gas $$bn \chi = \frac{1}{R} \left ( \frac{\partial PV}{\partial T} \right )_V - 1 .$$ For low densities, one should require that the Enskog viscosity (\ref{enskog-viscosity}) reduces to the Chapman-Enskog viscosity (\ref{chapman-enskog-viscosity}) by imposing $\lim_{n \to 0} \chi = 1$, so that we have \citep{HCB} $$b = B + T\frac{dB}{dT}$$ where $B$ is the second virial coefficient.}

In Table \ref{table-bulk-viscosity}, as predicted by the Enskog theory, it is shown that the bulk viscosity is vanishing in the dilute monoatomic gases (argon and helium). However, a non-vanishing ratio $\zeta / \eta$ has been measured in liquid argon \citep{naugle} as well as in dense gaseous argon \citep{madigosky}.\footnote{A comparison with the modified Enskog theory briefly presented above can be found in \citet{hanley-cohen}.} Consequently Stokes' relation is shown not to be justified against many cases, not only through theoretical predictions, but also experimentally.

In the 1960s, it was found that transport coefficients, in particular the viscosity coefficient $\eta$, cannot be expressed in a power series in terms of the density $n$ ($\eta = \eta_B + \eta_1 n + \eta_2 n^2 + \eta_3 n^3 + \dots $). Indeed it was shown that correlations between molecules was observed over large distances, larger than the range of the intermolecular interaction. As a result, \citet{dorfman-cohen-67} obtained theoretically that the coefficient of the quadratic term contained a contribution proportional to the logarithm of the density so that the previous density expansion should be rewritten as
\begin{equation} \eta_{DC} = \eta_B + \eta_1 \; n + \eta_2^{'} \; n^2 \ln n + \eta_3 \; n^3 + \dots  \label{viscosity-expansion-log-density} \end{equation} 
Nevertheless, although the contribution of the discovery of long-distance correlations is great in the modern kinetic theories, their contributions do not seem to be significant, as it was shown by \citet{sengers-1966} for the viscosity in a hard-disk gas, and later by \citet{KPS-1983} in a hard-sphere gas. Moreover, since the early 1970s, several attempts were made to detect this logarithmic-density dependence of experimental data, in particular for the viscosity coefficient \citep{kestin-et-al1,kestin-et-al2,kestin-et-al3}. Even if the viscosity is generally the transport coefficient measured with the highest precision, it was never shown that the addition of the logarithmic term was necessary.  Although we now know that Enskog viscosity is not exactly correct, the latter may still be used in order to compare numerical results.\footnote{A historical survey of the discovery of the logarithmic-density dependence is given by \citet[p.~72-80]{brush-kinetic-theory-3}.}

After Enskog proposed his theory in 1922, a revival of interest occurred only in the 1940s for the extension of kinetic theory to dense  gases and liquids. As already mentioned in Section \ref{gibbs_stat_mech}, after an original idea of \citet{yvon}, an exact hierarchy of coupled equations ruling the time evolution of the reduced $N$-particle  distribution functions was derived by \citet{bogoliubov}, \citet{born-green}, and \citet{kirkwood-1946} such that, by truncating this BBGKY hierarchy, kinetic equations were derived leading to expansions of the transport properties in terms of the particle density.

In 1928, Pauli (1900-1958) derived a \textit{master equation} to apply quantum mechanics to irreversible processes \citep{pauli}. This derivation also contained an assumption analogue to the \textit{Sto\ss zahlansatz} --- the repeated random phase assumption. Since then, different works have been completed by means of master equations, such as the generalization of the classical Boltzmann equation to quantum systems by \citet{UU33}. In the fifties, master equations were derived for weakly coupled systems by \citet{van-hove1,van-hove2,van-hove3} as well as \citet{brout-prigogine}. Kinetic equations were also obtained for plasmas thanks to \citet{landau_1936} and \citet{vlassov_1938}. In this context, \citet{balescu} and \citet{lenard} derived the so-called \textit{Balescu-Lenard} transport equation for plasmas. Kinetic theory has continued to generate many advances, especially in physics of plasmas \footnote{See, for example, \citet{balescu_1988,balescu_1997}.} and rarefied gases.\footnote{See, for example, \citet{beylich_1991}.}


\subsection{Irreversibility and microscopic fluctuations at equilibrium}\label{fluct_eq}

During the 17th century, the development in microscopy was more and more important, in particular due to the progress brought by van Leeuwenhoek (1632-1723). Thanks to the magnification provided by the microscopes, the microscopic world of cells called \textit{animalcules} was discovered. In this context, the observation of irregular motion of small grains immerged in a fluid became possible \citep{gray}. The interpretation of such a phenomenon until the 1820s was that these organic grains were endowed with living force. However, in 1828, Brown (1773-1858) observed that inorganic particles also exhibited the same kind of dancing motion \citep{brown}. He put in evidence that such a behavior had a physical rather than a biological cause, and thus opened the way to a new area in physics. This behavior of particles suspended in fluids is nowadays well known as \textit{Brownian motion}.

During the following decades, different interpretations were given to this phenomenon --- e.g. the motion would have been due to a local difference of temperature created by the light used to observe the particle. It is only in 1863 that these explanations were refuted by  Wiener (1826-1896)\footnote{Let us point out that we here refer to Ludwig Christian Wiener and not to Norbert Wiener (1894-1964) who contributed to the theory of stochastic processes and gave his name to the so-called {\em Wiener processes} \citep{wiener_1923}.} who proposed to look for the origin of the phenomenon in the liquid itself \citep{wiener}. For that reason, \citet{perrin-1909} considered him as the discoverer of the origin of the Brownian motion. It is to Gouy (1854-1926) however that goes the credit for having really prepared the way for our present point of view \citep{gouy}, since his experiments established conclusively
\begin{quote}
[...] i) that the Brownian movement appears for any particle, and the more viscous the liquid is, and the bigger the particles are, the lower the magnitude of the movement is; ii) that the phenomenon is perfectly regular, appears at constant temperature and in absence of any cause of external movement.\footnote{Quoted from \citet[p.~104-105]{gouy}.}
\end{quote}

However it was not before \citet{einstein} and \citet{smoluchowski} that a successful theory was proposed for the Brownian motion.\footnote{Note that, in his thesis \textit{Th\'eorie de la Sp\'eculation}, Bachelier (1870-1946), the founder of mathematical finance and Poincar\'e's student, arrived at the same ``displacement" law, not for the colloidal diffusion, but for the ``mean displacement" of stock prices over time \citep{bachelier}.} As we have seen in the previous chapter, it played an important role in the proof of the discontinuous feature of matter. Einstein expressed the diffusion coefficient of a Brownian particle in terms of the mean-square of its position\footnote{For further information on the history of Brownian motion, see \citet{perrin-1909}, \citet{brush-brownian}, and \citet{haw}. In addition, let us quote from \citet[p.~40-41]{pullman}:
\begin{quote}
It is appropriate to examine with greater attention these corpuscules, the disorderly motion of which can be observed in rays of sunshine: such chaotic movements attest to the underlying motion of matter, hidden and imperceptible. You will indeed observe numerous such corpuscules, shaken by invisible collisions, change path, be pushed back, retrave their steps, now here, now there, in all directions. It is clear that this to-and-fro movement is wholly due to atoms. First, the atoms move by themselves, then the smallest of the composite bodies, which are, so to speak, still within the reach of the forces of the atom, jostled by the invisible impulse from the latter, start their own movement; they themselves, in turn, shake slightly larger bodies. That is how, starting from atoms, movement spreads and reaches our senses, in such a way that it is imparted to these particles which we are able to discern in a ray of sunshine, without the collisions themselves which produce them being manifest to us. 
\end{quote}
At first sight, this quotation might be given by a 19th-century scientist. But it is quite amazing to know that it goes back to the famous Roman poet and philosopher Lucretius (99-55 BC).}
\begin{equation}
D=\lim_{t \to \infty} \frac{\left \langle \left[x(t) - x(0)\right]^{2}\right \rangle}
{2 \; t} \; .
\label{einstein.diffusion}
\end{equation}
More than putting simply in evidence the discrete character of the matter, he established a relationship between the diffusion coefficient of a Brownian particle and the spontaneous fluctuations intrinsic to the medium, which are due to the random collisions of particles of the surrounding fluid.\footnote{In fact, before 1905, Einstein already wrote three papers, sometimes called the {\em statistical trilogy} \citep{einstein_1902,einstein_1903,einstein_1904} which show how he reached the concepts developed in his 1905 paper. See \citet{kuhn_1978}, \citet{gearhart_1989}, and \citet{uffink_2006}. His papers did not draw much attention except some criticisms, especially due to \citet{hertz_1910a,hertz_1910b}. In a reply to him, Einstein confessed that
\begin{quote}
[...] the road taken by Gibbs in his book, which consists in one's starting directly from the canonical ensemble, is in my opinion preferable to the road I took. Had I been familiar with Gibbs's book at that time, I would not have published those papers at all, but would have limited myself to the discussion of just a few points. (Quoted from \citet{einstein_1911}, A. Beck's translation in \citet[p.~250]{einstein_collected_papers_vol3}.)
\end{quote}
For a comparison between Gibbs' and Einstein's approaches, see \citet{navarro_1998}.} Hence these natural fluctuations, generated by the microscopic dynamics at the thermodynamic  equilibrium, define a limit on the accuracy of the measuring instruments.\footnote{For further information, see \citet{bowling-barnes-silverman}.}

Different phenomena similar to the Brownian movement were discovered later. In particular, Shottky (1886-1976) observed that the thermionic current in a vacuum tube presented rapid and irregular changes in magnitude, due to the random emission from the cathode \citep{schottky}. It induces fluctuations of the voltage in any circuit in which the tube is connected, phenomenon now called \textit{Schottky effect} \citep{schottky}. But it is another phenomenon to which a special place is attributed in the development of statistical mechanics of irreversible processes. In 1927, Johnson (1887-1970) observed experimentally that spontaneous fluctuations of potential difference is produced in any electric conductor and concluded that the thermal agitation of the electric charges in the conductor is the cause of this phenomenon \citep{johnson1,johnson2}. The next year, Nyquist (1889-1976) obtained theoretically the same results \citep{nyquist} and is now considered at the origin of the fluctuation-dissipation theorem. Later, \citet{kirkwood-1946} studied the Brownian motion in liquids and derived a new formula for the friction constant which involves the autocorrelation of the intermolecular force acting on the Brownian particle at a certain time $t$ with its value a time $t + \tau$. But it is to Callen (1919-1993) and Welton that one must attribute the generalization of the Nyquist relation for any dissipative system. In this way, they established the well-known \textit{fluctuation-dissipation theory} \citep{callen-welton}.

Let us consider a charged Brownian particle in a liquid driven by an external electric field. The random collisions of the molecules of the liquid induce, on the one hand, a random driving force on the Brownian particle maintaining this in constant irregular motion (fluctuation). On the other hand, they imply a resistance to the driving motion, trying to slow down the charged particle (dissipation). Because of their common origin, i.e., the thermal agitation, these two effects are related. This relationship is precisely the aim of the so-called fluctuation-dissipation theory. These different works issued from Nyquist's discovery contributed to the establishment of the \textit{linear-response theory} developed in the fifties by \citet{green51,green60}, \citet{kubo57}, and \citet{mori58}, which relates the transport coefficients to the integral of time auto-correlation functions. In particular, by writing the microscopic expression of the stress tensor as\footnote{See \ref{micro_visc}.}
\begin{equation}
J_{ij}(t) = \sum_{a=1}^{N} \frac{1}{m} p_{ai} p_{aj} + \frac{1}{2} \sum_{a, b \ne a} F_i({\bf r}_a - {\bf r}_b) \left (r_{aj} - r_{bj}\right ) , 
\label{mic_stress_tensor}
\end{equation}
the shear viscosity is expressed in terms of the $xy$-component of Eq. (\ref{mic_stress_tensor}) as follows:
\begin{equation}
\eta = \lim_{V \to \infty} \frac{1}{Vk_B T} \int_{0}^{\infty} \left \langle J_{xy}(0) J_{xy}(t)\right \rangle \; dt \; ,
\label{viscosity-GK}
\end{equation}
while the bulk viscosity $\zeta$ is given by
\begin{equation}
\zeta + \frac{4}{3} \eta = \lim_{{n=N/V; \ N, V \to \infty}} \frac{1}{V k_B T} \; \times \int_{0}^{\infty} dt \left \langle \left ( J_{xx}(0) - \left \langle J_{xx} \right \rangle \right ) \left ( J_{xx}(t) - \left \langle J_{xx} \right \rangle \right ) \right \rangle \; .
\label{green-kubo-viscosity}
\end{equation}
The relations (\ref{viscosity-GK}) and (\ref{green-kubo-viscosity}) are nowadays called the \textit{Green-Kubo formulas}.

Notice that Mori has shown that, in the case of dilute gases, this expression reduces to the Chapman-Enskog results \citep{mori-58-2}.

The simplicity of Eq. (\ref{einstein.diffusion}) as well as as the fact that it expresses explicitly the diffusion coefficient as a non-negative quantity, present a particular interest. The extension of such a relation to the other transport coefficients could thus be useful. In this context, \citet{helf} proposed quantities associated with the different transport processes in order to establish Einstein-like relations such as Eq. (\ref{einstein.diffusion}) between the transport coefficients and associated quantities. In the case of self-diffusion, the associated Helfand moment is nothing but the position of one particle $x_{i}$. For the shear viscosity coefficient, we have
\begin{equation}
\eta=\lim_{t \to \infty} \frac{1}{2t V k_B T}\; \left \langle \left[G_{xy}
(t)-G_{xy} (0)
\right]^{2}\right \rangle \; ,
\label{Einstein.shear.viscosity}
\end{equation}
where $G_{xy}(t)$ is the so-called \textit{Helfand moment}
\begin{equation}
G_{xy}(t) = \sum_{a = 1}^{N} p_{ax} y_a
\label{helfand_moment_viscosity}
\end{equation}
associated with the shear viscosity.\footnote{Eqs. (\ref{viscosity-GK}-\ref{Einstein.shear.viscosity}), allowing to calculate numerically the viscosity in periodic fluid systems by molecular dynamics, have been and continue to be fundamental in the investigation of the properties of the bulk matter.  The first numerical calculation of viscosity was due to \citet{alder}. They proposed an algorithm based on generalized Einstein relations derived from the Green-Kubo formulas and adapted to the hard-disk fluid. The application of the pure Green-Kubo technique by equilibrium molecular dynamics to the Lennard-Jones fluid has been performed a short time afterward by \citet{LVK-73}, and later, among others, by \citet{SH-1985}, \citet{erpenbeck-88}, and \citet{heyes-88}. However, for using Eq. (\ref{Einstein.shear.viscosity}) in periodic systems, the Helfand moment (\ref{helfand_moment_viscosity}) should be modified in order to take into account the periodicity of the systems introduced for numerical reasons \citep{viscardy-gaspard1,VSG2007,viscardy_book}. As we shall see in Section {\ref{mnesm}}, the Helfand moment has taken an important place in recent advances in non-equilibrium statistical mechanics.} In \ref{appendixA}, it is shown that Eqs. (\ref{viscosity-GK}) and (\ref{Einstein.shear.viscosity}) are equivalent if drawing the attention on the relation: $J(t) = \frac{dG(t)}{dt}$.

Therefore fluctuations at equilibrium induced by the thermal agitation play a fundamental role in modern statistical mechanics of non-equilibrium processes. The advantages of such approaches to irreversible processes are that irreversible phenomena can be described  by the tools of equilibrium  statistical mechanics. Thereby, the construction of non-equilibrium distribution functions is not necessary. In addition, such relations are valid in general and can thus be applied to dilute gases as well as to dense gases and liquids.\footnote{For a general overview of these theories, see, for example, \citet{kubo-1966}. \citet{hoffman_1962} gives a non technical outline of the fluctuation-dissipation theorem.} 

The transport coefficients are usually calculated using some kinetic equation. However, each kinetic equation involves a stochastic assumption such as the \textit{Sto\ss zahlansatz} which is reached by some truncation of the evolution equation. Phenomena like the transport processes, playing such an important role in nature, must present a more fundamental justification in terms of the intrinsic properties of the underlying microscopic dynamics. The development of the dynamical systems theory (chaos theory), by introducing the hypothesis of the microscopic chaos, seems to provide the key  for a better understanding of the irreversibility of the macroscopic phenomena. The following section will thus be devoted to the development of the chaos theory.


\section{Theory of chaos}\label{micro-chaos}

\subsection{From Poincar\'e to Smale, Lorenz, and Ruelle}\label{history_chaos}

It is interesting to note that the idea of considering a gas as a disordered state is quite old. During the 16th century Paracelsus (1493-1541) studied gases and observed that the air and vapors did not have a fixed volume or shape. He then gave them the name of \textit{chaos} in analogy with \tg{q'aos} in the greek mythology, being the origin of the universe. Later, van Helmont (1577-1644), who was the first to make the distinction between the different gases and the air, invented the word \textit{gas} according to the flemish pronounciation of chaos \citep{van_helmont_1648}.

In the end of the 19th century, it was generally accepted that a large number of degrees of freedom is a necessary condition to observe unpredictable behavior. This great number was the necessary element to introduce a statistical description of phenomena having a deterministic underlying dynamics. However, in 1892, Poincar\'e showed that a low-dimensional deterministic system, namely the \textit{three-body problem}, exhibited an unpredictable behavior now commonly called \textit{chaos} \citep{poincare}. Following his work, a large number of nonlinear dynamical systems were discovered  so that the set of equations
\begin{equation}
\frac{d{\bf X}}{dt} = {\bf F}({\bf X}) 
\label{dyn_syst}
\end{equation}
rules the time evolution of their variables ${\bf X}$. Moreover, in these systems, chaos appears to be a generic phenomenon rather than an exotic one.\footnote{For further information, see \citet{nicolis}.} In the same year, Lyapunov (1857-1918) published his work on the stability of such systems \citep{lyapunov}.\footnote{The Lyapunov instability will be addressed in more detail in Section \ref{charact_chaos}.} 

As a matter of fact, in 1873, Maxwell himself had already emphasized the sense of unstable behaviors:
\begin{quote}
When the state of things is such that an infinitely small variation of the present state will alter only by an infinitely small quantity the state at some future time, the condition of the system, whether at rest or in motion, is said to be stable; but when an infinitely small variation in the present state may bring about a finite difference in the state of the system in a finite time, the condition of the system is said to be unstable.

It is manifest that the existence of unstable conditions renders impossible the prediction of future events, if our knowledge of the present state is only approximate, and not accurate.\footnote{Quoted from the text of Maxwell's conference reproduced in the 1997 Digital Preservation of {\em The Life of James Clerk Maxwell} \citep[p.~211]{garnett}. For further detail on Maxwell's avant-gardist ideas on unstable systems, see \citet{hunt-yorke}.} \end{quote}
\citet{maxwell_matter_motion} also remarked clearly that there is a maxim which is often quoted, that
\begin{quote}
``Like causes produce like effects".

This is only true when small variations in the initial circumstances produce only small variations in the final state of the system. In a great many physical phenomena this condition is satisfied; but there are other cases in which a small initial variation may produce a very great change in the final state of the system [...].\footnote{Quoted from the reprinted book \citep[p.~13-14]{maxwell_matter_motion_dover}.}
\end{quote}

Late in the 19th century, Hadamard (1865-1863) studied the geodesic flows on a negative-curvature surface and concluded that
\begin{quote}
each change of the initial direction, as small as possible, of a geodesic which remains at a finite distance is enough for implying an arbitrary variation of the behavior of the trajectory.\footnote{Quoted from \citet[p.~70-71]{hadamard}.}
\end{quote}
Considering Hadamard's work, \citet{duhem} also concluded in the chapter entitled \textit{Exemple de d\'eduction math\'e\-ma\-tique \`a tout jamais inutilisable} that, in Hadamard's billiard, the trajectory obtained mathematically becomes unusable for a physicist. Indeed, an experimental measure realized by any physical procedure as precise as possible is always endowed of an error growing with time. Furthermore, Poincar\'e clearly explained that
\begin{quote}
A very small cause which escapes our notice determines a considerable effect that we cannot fail to see, and then we say that the effect is due to chance.\footnote{Quoted from \citet[p.~68]{poincare-1908}.}
\end{quote}
This implies that determinism and chance may be combined thanks to the unpredictability. 

Birkhoff was one of the only western mathematicians to take into account what we now call the \textit{sensitivity upon the initial conditions}, property put in evidence by Poincar\'e and others. In constrast, Soviet scientists were particularly interested in the ideas emerging from Poincar\'e's works.\footnote{For a history of the extensive contribution of the russian schools, see, for example, \citet{diner_1992} and \citet{aubin_dahan_2002}.} For instance, Mandelshtam (1879-1944) aimed to elaborate a nonlinear physical thinking. In this context, his student Andronov (1901-1952) introduced the paradigm of {\em self-oscillations}. Concretely, \citet{andronov_1929} considered dissipative systems of radiophysics analogous to those of van der Pol (1889-1959) involving {\em relaxation oscillations} \citep{van_der_pol_1926}. He then recovered a behavior already introduced by \citet{poincare_1881,poincare_1882} in a mathematical context; namely, the {\em limit cycle}. 

Although questions on nonlinearity were less tackled outside of USSR at that time, World War II, nevertheless, favoured in the West a renewal of interest for this type of problems, especially due to the necessity of improving radio communication. Levinson (1912-1975) developed topological methods using the {\em Poincar\'e map}\footnote{For further information, see, for example, \citet{nicolis}.} and worked on nonlinear differential equations of the second order \citep{levinson_1944,levinson_1949}. These methods allowed the Americans Cartwright (1900-1998) and Littlewood (1885-1977) to study the equation of Li\'enard (1869-1958) 
\begin{equation}
\ddot{y} - k \left (1 - y^2 \right ) \dot{y} + y = b \lambda k \cos \left ( \lambda t + a \right )
\end{equation}
which is a generalization of the van der Pol equation \citep{lienard_1928a,lienard_1928b}. \citet{cartwright_littlewood_1945} observed the existence of a threshold at $b=2/3$. Whereas a periodic solution of limit-cycle type is found for $b>2/3$, below this value, irregular behaviors occur which involve a strange (but robust) set of nonperiodic trajectories. Later, in the sixties, these {\em fractal} sets influenced Smale to interpret	geometrically rather than analytically such a phenomenon, as he did by introducing his {\em horseshoe} (see below).

Nevertheless, in Soviet Union, a nonlinear culture was raising much more strongly. In this context, N. S. Krylov (1917-1947)\footnote{Do not confuse Nikolay Sergeevich Krylov with his comtemporary Nikolay Mitrofanovich Krylov (1879-1955) who also worked on nonlinear mechanics and published together with Bogoliubov an {\em Introduction to Non-linear Mechanics} \citep{krylov_bogoliubov_1943}.} played a tremendous role regarding the purpose of this paper, as we shall see in Section \ref{charact_chaos}. In his unfinished monograph,\footnote{N. S. Krylov deceased before he ended his ambitious book on the foundations of statistical mechanics. Even if his notes were published in 1950 in USSR, the rest of the world had to wait for the seventies for being aware of his remarkable work (see, for example, \citet[p.~6]{lebowitz_penrose_1973}). The english translation has actually been available only from 1979 \citep{krylov}. The reception of his masterpiece was thus particularly surprising.} he stated that the instability in a dynamical system should be the ingredient for allowing the application of statistical mechanics on mechanical systems \citep{krylov}.

The interest for such issues increased significantly in the West only around the early 1960s, among other things, by tackling forecast problems in meteorology.\footnote{Let us mention that such questions were already considered from the 1940s in the framework of the Meteorology Project promoted by von Neumann (1903-1957) and the meteorologist Charney (1917-1981). See \citet{aubin_dahan_2002}.} In this perspective, Lorenz (1917-2008) found that a system of three equations --- the so-called Lorenz model \citep{lorenz} ---  produced trajectories in phase space exhibiting the \textit{sensitivity upon the initial conditions} defining chaos, and drawing an \textit{attractor} --- a {\em strange attractor} according to the definition proposed later by \citet{ruelle_takens_1971}.

\begin{figure}
\centerline{\mbox{\scalebox{.43}{\rotatebox{270}{\includegraphics{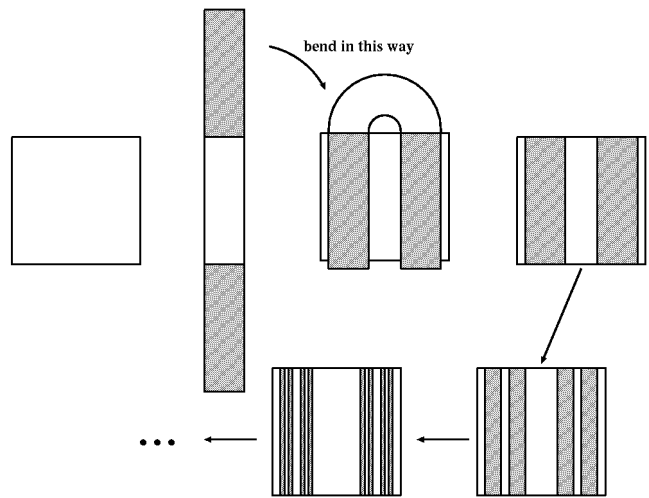}}}}}
\caption{\textit{Smale's horseshoe} introduced by \citet{smale_1965}: Succession of stretchings and foldings
illustrating the evolution of phase-space due to the chaotic dynamics.}
\label{smale}
\end{figure}

The real breakthrough was however due to Smale who, as he explained later,
\begin{quote}
[...] was lucky to find [himself] in Rio at the confluence of three different historical traditions in the subjects of dynamics. These three cultures, while dealing with the same subject, were isolated from each other, and this isolation obscured their development.\footnote{Quoted from \citet[p.~44]{smale_1998}.}
\end{quote}
The three traditions which he refered to were (i) the ideas developed by Poincar\'e and Birkhoff from the turn of the last century to the 1930s, (ii) the Gorki school growing especially with Andronov in the 1930s, and (iii) works on the van der Pol equation produced by Cartwright and Levinson. Smale considered global behavior of phase space rather than a particular trajectory,  and invented his \textit{horseshoe} (see Fig. \ref{smale}) to get a visual analogy to the sensitive dependence upon the initial conditions \citep{smale}. In the 1960s, Smale's advances were comprehensible only by his community, as the mathematical concepts he used were largely abstruse to members of the other disciplines. Contacts with Thom's (1923-2002) school around 1970 thereby encouraged Smale to apply his ideas to other areas of sciences such as economics, celestial mechanics, and biology.

Ruelle, one of Thom's colleagues, has to be associated with Lorenz and Smale as the dominant people in the emergence of chaos theory. In the sixties, he studied Landau (1908-1968)'s theory of modes describing the turbulence in hydrodynamics.\footnote{This theory was originally elaborated by \citet{landau_1944} (in Russian). A presentation in english is available in \citet{L-L}'s {\em Fluid Mechanics}, Chapter 27. Let us also mention that Hopf (1902-1983) independently proposed a similar theory of turbulence \citep{hopf_1948}. Hence, this is sometimes called the {Landau-Hopf theory}.} Not satisfied with this theory, \citet{ruelle_takens_1971} thus introduced a new description by establishing a rapprochement between Smale's topological approach of dynamical systems theory and hydrodynamics, which led to the introduction of {\em fractal} objects (See Section \ref{chaos_fractals}); namely, the famous {\em strange attractor}.\footnote{For further detail on the way followed by Ruelle toward this result, see his book {\em Chance and Chaos} \citep{ruelle_1991}.} Nevertheless, the reception of this new theory of turbulence was lukewarm. The turning point came around 1975 with verifications performed first numerically by \citet{mclaughtin_martin_1974}, and secondly experimentally by \citet{gollub_swinney_1975}. This was symbolically marked by the term {\em Chaos} explicitly used in \citet{li_yorke_1975}'s paper entitled {\em Period three implies chaos}.\footnote{Note that, in their paper, \citet{li_yorke_1975} chose the word \textquotedblleft Chaos\textquotedblright \ for qualifying a phenomenon different from the sensitivity upon initial conditions now commonly accepted as defining chaos.}

\begin{figure}
\centerline{\mbox{\scalebox{.55}{\rotatebox{0}{\includegraphics{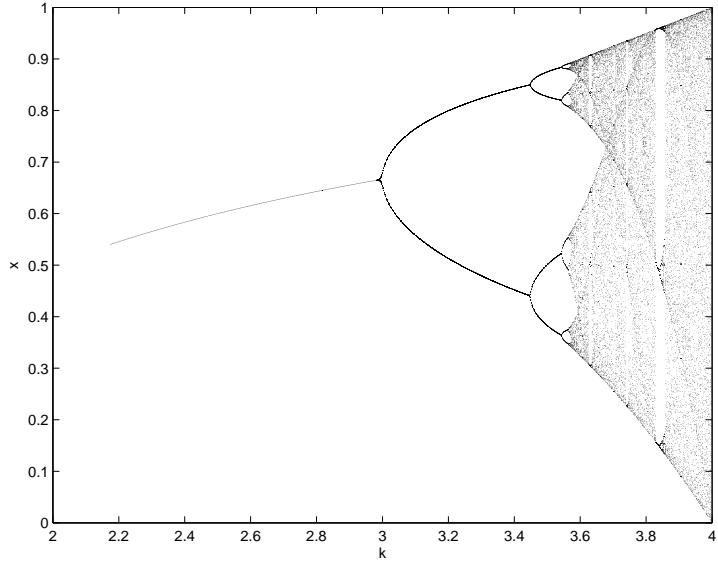}}}}}
\caption{Period doubling bifurcations cascade introduced by \citep{feigenbaum} for the logistic map (\ref{logistic_map}).}
\label{feigenbaum_fig}
\end{figure}

In the same year, Feigenbaum learned from Smale that the {\em logistic map}\footnote{The logistic map, in spite of its simplicity, was first introduced by Verhulst (1804-1849) as a demographic model describing the evolution of a population \citep{verhulst_1838}, and was popularized by \citet{may_1976}.}
\begin{equation}
x_{n+1} = k x_n (1-x_n)
\label{logistic_map}
\end{equation}
exhibited a succession of bifurcations with $k$ increasing from 0 to 4, and led to chaos beyond $k = 3.57 \dots$ He found in particular that
\begin{equation}
\lim_{m \to \infty} \frac{k_{m}-k_{m-1}}{k_{m+1}-k_m} \equiv \delta = 4.669201 \dots,
\end{equation}
where $k_m$ is the value of $k$ at the $m$th bifurcation \citep{feigenbaum}. This very general property is known as {\em self-similarity}. Furthermore, the most remarkable result is that this quantity  $\delta$ is identical for all dynamical systems of the same family. This is the so-called {\em universality of the period doubling bifurcations cascade} which is illustrated in Fig. \ref{feigenbaum_fig}.\footnote{For further detail, see, for example, \citet{ott}.}


\subsection{Characterization of chaos}\label{charact_chaos}

In the end of the 19th century, Lyapunov defended his doctoral thesis \textit{The general problem of the stability of motion} \citep{lyapunov}. He proposed a method which provided ways of determining the linear stability of sets of ordinary differential equations, and controled how an infinitesimal perturbation of a trajectory evolves in time. This perturbation may be evaluated by integration of the evolution equations (\ref{dyn_syst}) of two trajectories, the first considered as the reference and  passing by the point $\mathbf{X}$, the second being separated by an infinitesimal quantity $\delta \mathbf{X}$. We then have
\begin{equation}
\frac{d\delta \mathbf{X}}{dt} = \mathbf{F}(\mathbf{X} + \delta \mathbf{X}) - \mathbf{F}(\mathbf{X})
=\frac{\partial \mathbf{F}(\mathbf{X})}{\partial \mathbf{X}} \cdot \delta \mathbf{X} \label{ecart}
\end{equation}
at the first order. These vectors $\delta \mathbf{X}$ belong to a linear tangent space of the phase space $\cal M $ in each point $\mathbf{X}$. This space, noted $\mathcal{T}\: {\cal M} (\mathbf{X})$, is called \textit{tangent space}. Let us rewrite Eq. (\ref{dyn_syst}) in terms of the flow ${\bf \Phi}^t$ induced by these equations
\begin{equation}
\mathbf{X} = \mathbf{\Phi}^t \mathbf{X}_0 .
\label{flow}
\end{equation}
Since Eq. (\ref{ecart}) is linear, all its solutions are of the type
\begin{equation}
\delta \mathbf{X}_{t} = \frac{\partial \mathbf{ \Phi}  ^{t}(\mathbf{X}_{0})}{\partial \mathbf{X}_{0}}  \cdot \delta \mathbf{X}_{0}
= \mathbf{M}(t, \mathbf{X}_{0}) \cdot \delta \mathbf{X}_{0} ,
\end{equation}
with $\mathbf{X}_{0}$ and $\delta \mathbf{X}_{0}$, respectively the values of $\mathbf{X}$ and $\delta \mathbf{X}$ at $t=0$, and $\mathbf{M}(t, \mathbf{X}_{0})$ called \textit{fundamental matrix}.

The infinitesimal perturbation $\delta \mathbf{X}$ can growth exponentially, as we have already mentioned by the property of sensibility upon the initial conditions. This growth is characterized by the \textit{Lyapunov exponent} associated with an arbitrary tangent vector $\mathbf{\overline {e}}$ 
\begin{equation}
\lambda (\mathbf{X},\mathbf{\overline {e}}) = \lim_{t \to\ \infty } \frac{1}{t} \: 
\ln \: || \: \mathbf{M}(t, \mathbf{X}) \cdot \mathbf{\overline {e}} \: ||  .
\end{equation}
Moreover, \citet{oseledec} introduced what he called a \textit{multiplicative ergodic theorem} claiming that, for an ergodic system, the Lyapunov exponent in the direction $\mathbf{\overline{e}}$ is independent of the position $\mathbf{X}$ of the trajectory in the phase space
\begin{equation}
\lambda (\mathbf{X},\mathbf{\overline {e}}) = \lambda (\mathbf{\overline {e}}) .
\end{equation}
This allows to construct the so-called {\em Lyapunov spectrum}:
\begin{equation}
\lambda_1 > \cdots > \lambda_i > \lambda_r
\end{equation}
with multiplicities $m_1, \cdots, m_i, \cdots, m_r$, such that $\sum_{i=1}^{r} m_i = M$, i.e., the dimension of the tangent phase space (or phase space itself). The Lyapunov spectrum then characterizes the growth ($\lambda_i >0$) or decay ($\lambda_i <0$) rate
of the distance between two nearby trajectories in the phase space of dynamical systems.\footnote{For further information, see, for example, \citet{eckmann-ruelle}.}

Historically, N. S. Krylov was one of the most influential physicists in the 20th-century statistical mechanics. The title of his theses --- {\em Mixing Processes in Phase Space} \citep{krylov_1941} and {\em The Processes of Relaxation in Statistical Systems and the Criterion of Mechanical Instability} \citep{krylov_1942} --- showed early his leading ideas. Aiming at understanding how to justify the use of statistical tools on mechanical systems, he started from the existence of relaxation processes toward equilibrium \citep{krylov_1944}. He then came to the conclusion that
\begin{quote}
mechanical ergodicity is quite insufficient for statistical purposes ---and, in particular, for the definition of the fundamental notion of relaxation. \footnote{Quoted from \citet[p.~709]{krylov_1944}.}
\end{quote}
A stronger condition was needed, that is, the mixing property (see Section \ref{ergodicity_mixing}). Showing that local instability, i.e., the exponential divergence of the trajectories, is the source of mixing, he studied physical billiards in the continuity of the works carried out by \citet{hadamard}, \citet{hedlund_1939}, and \citet{hopf_1939} on the hyperbolic behavior of geodesic flow on compact manifolds of constant negative curvature. He then claimed that
\begin{quote}
The main condition of mixing, which ensures the fulfilment of this condition, is a sufficiently rapid divergence of the geodesic lines of this Riemann space (that is, of the paths of the system in the [$3N$]-dimensional configuration space), namely, an exponential divergence. \footnote{Quoted from \citet[p.~709]{krylov_1944}.}
\end{quote}

Krylov hence showed that hard ball systems were dynamically unstable. His studies have inspired generations of mathematical physicists. In particular, Sinai and his coworkers proved the ergodic hypothesis for billiard systems, and characterized the stochastic-like behavior of these deterministic systems \citep{sinai}. This type of behavior was then studied extensively for simple systems with few degrees of freedom. Among the most important contributions are those carried out by \citet{bunimovich1979}, and \citet{bunimovich-sinai80,bunimovich-sinai81}.

In the language of the Lyapunov exponents, the use of a statistical approach can be explained as follows. The positivity of such quantities expresses a dynamical instability and induces a sensitivity upon the initial conditions, property defining chaos. Two trajectories initially very close separate exponentially in time. This sensitivity upon the initial conditions limits the possible predictions on the trajectories because they are only known through a given precision $\epsilon_{0}$. By considering a rate of separation of very close trajectories given by the maximum Lyapunov exponent $\lambda_{\mathrm{max}}$, the error between  the predicted and the actual trajectories grows as $\epsilon_{t} \simeq \epsilon_{0} \exp(\lambda_{\mathrm{max}} t)$. After a finite time, the error becomes larger than the given allowed precision $\epsilon_{T}$, which then defines the Lyapunov time $t_{\mathrm{Lyap}} \simeq \left ( 1/\lambda_{\mathrm{max}} \right ) \ln \left ( \epsilon_{T}/ \epsilon_{0} \right )$. Given these initial and final precisions, predictions after the Lyapunov time are no longer relevant. This result is the requirement to adopt a statistical description.

The unstable character of dynamical systems is therefore such that even deterministic systems can generate random behaviors. On the other hand, this dynamical instability produces information in time allowing to reconstruct the system trajectory in phase space. The separation in time of nearby trajectories gives us the possibility to distinguish the trajectories. In this context, in the late fifties Kolmogorov (1903-1987) and Sinai applied to the dynamical systems the concept of entropy per unit time introduced a decade before with the information theory by \citet{shannon_1948_1,shannon_1948_2}. This entropy characterizes the {\em temporal disorder} by analogy to the entropy per unit volume (\ref{cg_gibbs_entropy}) which characterizes the spatial disorder.

\begin{figure}[t]
\centerline{\mbox{\scalebox{.5}{\rotatebox{0}{\includegraphics{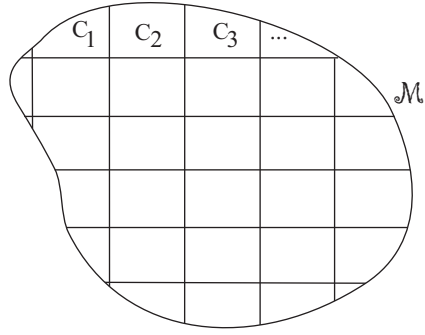}}}}}
\caption{Partition $\mathcal{P}$ of phase space $\mathcal{M}$ into cells $C_{\omega_{i}}$
with $\omega _{i} = 1, \dots , M$.}
\label{partition}
\end{figure}

As for the coarse-grained entropy (\ref{cg_gibbs_entropy}), a partition $\mathcal{P}$ is carried out of phase space $\cal M $ into $M$ cells (see Fig. \ref{partition})
\begin{equation}
\mathcal{P} = \lbrace C_{1},C_{2}, \dots , C_{M} \rbrace \; .
\end{equation}
Let $\mu_{\rm i}$ be an invariant probability measure (see \ref{measure}). The probability  $\mu_{\rm i} (\omega _{0} \omega _{1} \dots \omega _{n-1})$ of having a trajectory which visits successively the cells ($C_{\omega _{0}}, C_{\omega _{1}}, \dots, C_{\omega _{n-1}}$) at times $t= 0$, $\Delta t$, $\dots$, $(n-1)  \Delta t$ is given by a $n$-time correlation function\footnote{For further information, see, for example, \citet{eckmann-ruelle} and \citet{gasp-book}.}
\begin{equation}
\mu_{\rm i} (\omega _{0} \omega _{1} \dots \omega _{n-1}) = \int _{\cal M}^{} \mu_{\rm i} (d\mathbf{X})
\: \chi  _{\omega _{0}}(\mathbf{X})
\: \dots \: \chi  _{\omega _{n-1}}({\bf \Phi} ^{(n-1)\Delta t}\mathbf{X})
\end{equation}
where $\mathbf{\Phi}$ is the flow introduced in Eq. (\ref{flow}) and $ \chi _{\omega _{i}}$ is the indicator function of the cell $C_{\omega_{i}}$ defined as
\begin{equation}
\chi_{\omega_{i}}(\mathbf{X}) =
\left \lbrace 
\begin{array}{l}
1 \qquad if \; \mathbf{X} \in C_{\omega_{i}} \; ,\\
0 \qquad if \; \mathbf{X} \not\in  C_{\omega_{i}} \; .
\end{array}
\right.
\end{equation}
The entropy per unit time of this partition $\mathcal{P}$ is defined as
\begin{equation}
h(\mathcal{P}) = \lim_{n \to\ \infty } - \frac{1}{n\Delta t} \; \sum_{\omega _{0}, \dots,  \omega _{n-1} }^{} \mu_{\rm i} (\omega _{0} \dots \omega _{n-1}) \: \ln \: \mu_{\rm i} (\omega _{0} \dots \omega _{n-1}) \; . 
\label{entropy_unit_time}
\end{equation}
with the same structure as Gibbs' coarse-grained entropy. However, $h(\mathcal{P})$ depends on the selected partition. \citet{kolmogorov_1958,kolmogorov_1959} and \citet{sinai_1959} have then shown how to avoid such a restriction. By definition, one calls \textit{Kolmogorov-Sinai entropy} the \textit{supremum} over all the partitions $\mathcal{P}$
\begin{equation}
h_{{\rm KS}} = \mathrm{Sup} _{\mathcal{P}}\:  h(\mathcal{P}) \; ,
\end{equation}
which is independent of the partition and defines an intrinsic quantity to the dynamics of the system ${\bf \Phi} ^{t}$ and to the invariant measure $\mu_{\rm i} $. This new quantity measures the (exponential) rate at which information is obtained in time in random processes. As a result, the KS entropy per unit time can be viewed as the dynamical analogue of the Gibbs entropy per unit volume defined in statistical mechanics.

Among the various classes of dynamical systems, a specific one has particularly attracted the attention of physicists from the seventies onward; namely, the {\em Anosov systems} \citep{anosov_1967}, which play an important role for the use of dynamical systems in statistical mechanics (see \ref{anosov_systems} for some detail on the properties of the Anosov systems). The two aspects of chaos ---i.e., the \textit{dynamical instability} characterized by the Lyapunov exponents and the \textit{dynamical randomness} by the KS entropy--- are strongly related, the latter being the consequence of the former. \citet{pesin} proposed a theorem for {\em closed} Anosov systems\footnote{A system is closed when phase space is mapped onto itself according to Eq. (\ref{flow}), that is, without any escape of trajectories out of the initial phase space. In section \ref{escape_rate_formalism}, we shall see the case of open systems involving an escape process in the context of the escape-rate formalism.} which is now known under the name of \textit{Pesin's identity}. Concretely, this relates the KS entropy to the sum of all the positive Lyapunov exponents of the system\footnote{For further information on technical aspects of chaos, see, for example, \citet{ott}, \citet{gasp-book}, and \citet{dorf-book}. Further information on the history of chaos and	 dynamical systems theory are provided, for example, by \citet{aubin_dahan_2002}, \citet{dalmedico_1992}, and \citet{stewart_book}. \citet{gleick} also outlined the history of chaos theory. Nevertheless, his book does not at all cover the remarkable advances due to Soviet schools. This could be corrected and completed by refering, for example, to \citet{diner_1992}.}
\begin{equation}
h_{\rm KS} = \sum_{\lambda_i>0}^{}\lambda_i \; .
\label{pesin}
\end{equation}

Later, in the seventies, Sinai, Bowen, and Ruelle\footnote{See \citet{sinai_1972}, \citet{bowen_ruelle_1975}, and \citet{ruelle_1976,ruelle_1980}.} proposed the mathematical foundations of a new formalism for chaotic systems applying the techniques used in statistical thermodynamics, which is called the \textit{thermodynamic formalism}.\footnote{See, for example, \citet{ruelle} and \citet{beck-schlogl}.} In this formalism, a new quantity, the \textit{topological pressure}, was defined and played a role in dynamical systems very similar to that of the free energy for statistical-mechanical systems.


\subsection{Chaos and fractals}\label{chaos_fractals}

As seen above, the instability of the dynamics in phase space induces an exponential separation of trajectories characterized by the Lyapunov exponents. Because the phase-space volume accessible to the trajectories is finite (e.g. the volume defined by the energy of the system), they  have to fold onto themselves. We then observe successive stretchings and foldings of phase-space volumes, as Smale's horseshoe illustrates it in Fig. \ref{smale}, creating strange objects ---after an infinite number of such operations--- nowadays called \textit{fractals}.  In addition, the period-doubling bifurcations cascade investigated by Feigenbaum (see Section \ref{history_chaos}), exhibiting the property of self-similarity, also illustrates the deep connection between chaos and fractals.

\begin{figure}
   \begin{minipage}{.4\textwidth}
\hspace*{1cm}
       \mbox{\scalebox{.45}{\includegraphics{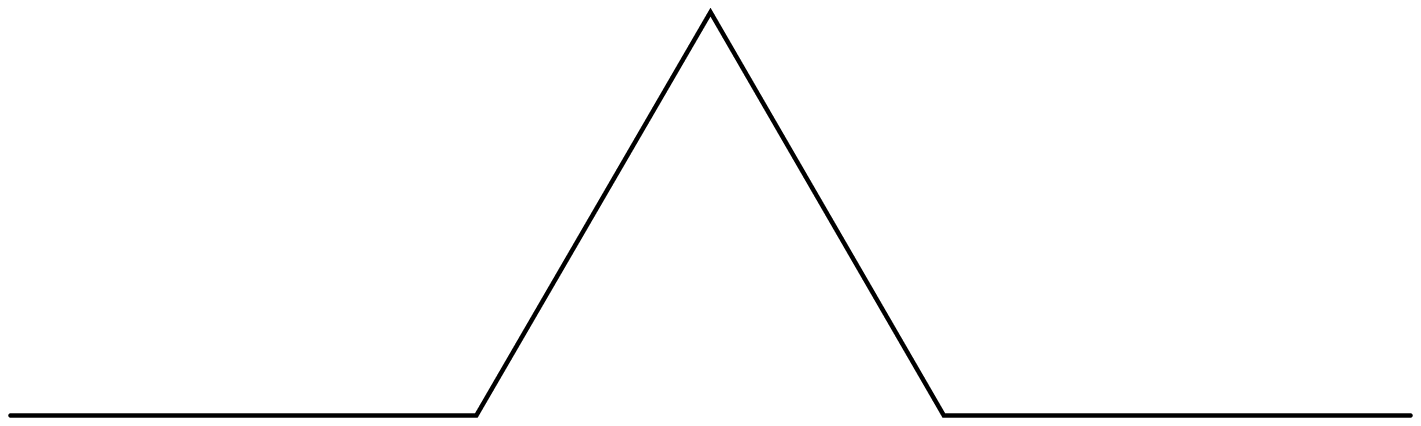}}}
   \end{minipage}
   \hfill
   \begin{minipage}{.4\textwidth}
\hspace*{-1cm}
       \mbox{\scalebox{.45}{\includegraphics{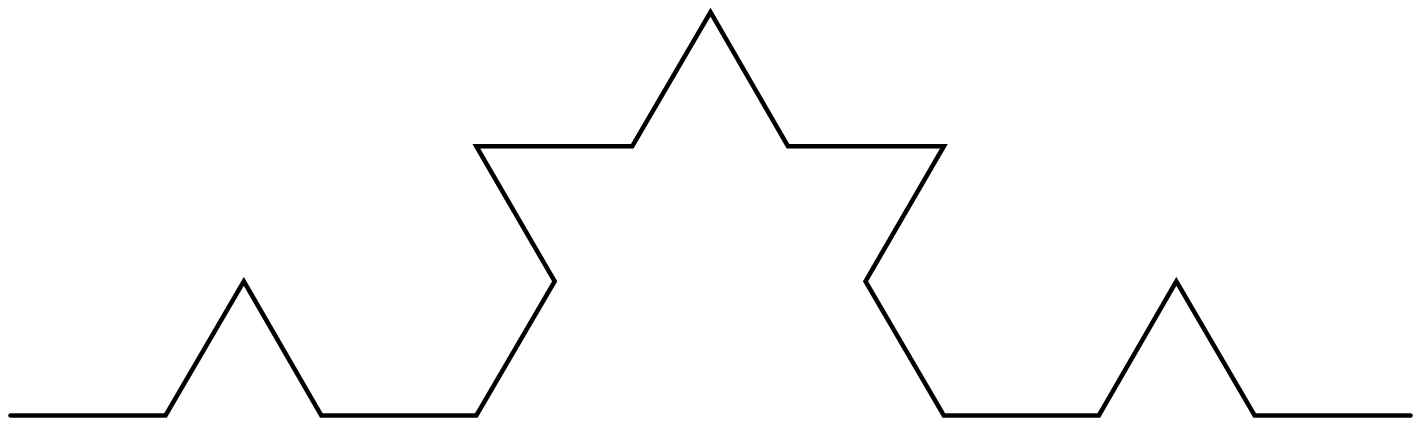}}}
   \end{minipage}
      \\
   \begin{minipage}{.4\textwidth}
\hspace*{1cm}
       \mbox{\scalebox{.45}{\includegraphics{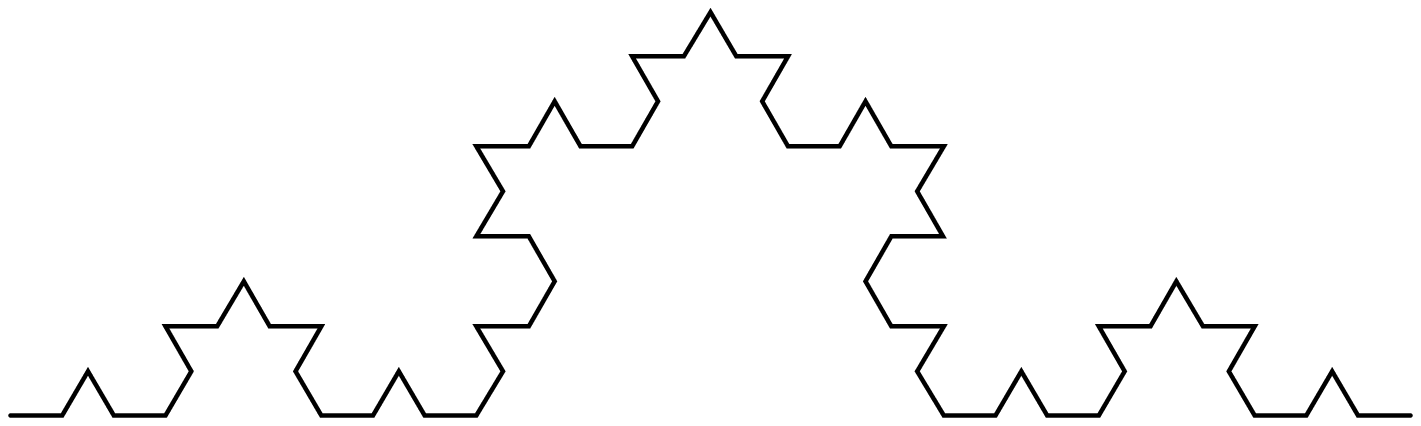}}}
   \end{minipage}
  \hfill
   \begin{minipage}{.4\textwidth}
\hspace*{-1cm}
       \mbox{\scalebox{.45}{\includegraphics{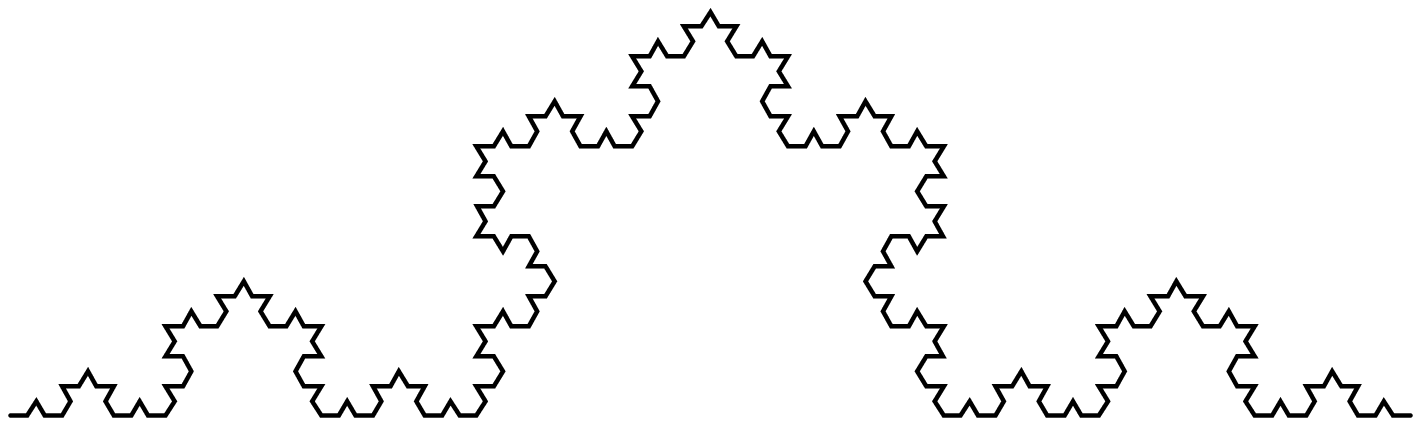}}}
   \end{minipage}
   \hfill
\caption{\textit{Koch's curve}: the first four steps of construction. The operation consists in dividing each segment into 3 parts, and on the one in the medium is constructed an equilateral triangle.}
\label{koch-curve}
 \end{figure}

Whereas the term \textit{fractal} is associated with Mandelbrot, the history of this intriguing discipline of mathematics began during the 1870s, when continuous functions without derivatives were discovered. For a long time, the idealization of nature implied a smooth and regular representation of real objects.\footnote{For further information, see, for example, \citet{fractal-early}.} In mathematics, one dealt with continuous functions such that a tangent could be drawn at (almost) each point. However, Riemann (1826-1866) already claimed a contradictory opinion.  In the 1870s, Weierstrass (1815-1897) gave an example having no derivative in any point.\footnote{The paper was read in 1872 in the Royal Prussian Academy of Sciences, but was only published on the original version in 1895 \citep{weierstrass}.} In geometry, mathematicians, such as Koch (1870-1924) in 1904, introduced continuous curves without a tangent at any point obtained by an elementary geometric construction \cite{koch}. \textit{Koch's curve} depicted in  Fig. \ref{koch-curve}
is a clear example of such objects. Moreover, the mathematician O. Takagi (1875-1960) working at G\"ottingen proposed in 1903 a simple example of continuous but nondifferentiable function known today as \textit{Takagi's function} \citep{takagi}. Furthermore, studying the Brownian motion, Perrin observed experimentally that the trajectory drawn by a Brownian particle was highly irregular. For that matter, he deduced that
\begin{quote}
Though derived functions are the simplest and the easiest to deal with, they are nevertheless exceptional; to use geometrical language, curves that have not tangents are the rule, and regular curves, such as the circle, are interesting though quite special cases.\footnote{Quoted from \citet[p.~viii-ix]{perrin-atomes_en}.}
\end{quote}
Therefore, according to the precision of the measure, the length of the trajectory is different. The larger the precision, the longer the trajectory. The relevance of the notion of length thus vanishes and has to be replaced by a new quantity characterizing such objects, that is the \textit{dimension}. In this context, Hausdorff (1868-1942) introduced in 1919 a new notion of dimension which was no longer a whole number, but can take noninteger values \citep{hausdorff}. For example, in the case of Koch's curve, its \textit{Hausdorff dimension} is neither equal to one --- the dimension of a line ---  or equal to two --- the dimension of the plane --- but between both of them, that is 1.26 . 

Introducing the term \textit{fractal},\footnote{Mandelbrot proposed this phrase from the Latin word \textit{fractus} which means \textit{irregular, broken}.} \citet{mandelbrot} brought back into fashion all the old works on these strange objects. He used them in a lot of different disciplines such as in physics (turbulence), geography (``How long is the Coast of Britain?" \citep{mandelbrot_1967}), astronomy (craters of the moon), biology, etc., in order to show that objects without a tangent at any point are nothing but the rule in nature. Hence, the \textit{self-similarity}\footnote{An enlarged part of a fractal object is similar to the whole object, which implies a scale invariance.} appears to be the property common to different objects, and thus the one implying a certain unity in nature. These mathematical concepts became helpful in areas as miscellaneous as macromolecular science \citep{witten1998}, neuroscience \citep{babloyantz_destexhe_1985,babloyantz_1986}, climatology \citep{nicolis_nicolis1984,nicolis_nicolis1986}, cosmology \citep{jones2005}, or even finance \citep{mandelbrot1997}.\footnote{Several fundamental papers on the development of theory of fractals are reprinted in \citet{classics-fractals}.}


\section{Irreversibility and microscopic chaos}\label{mnesm}

\subsection{Statistical mechanics of small systems}

In the light of the key concepts elaborated since Krylov (see Section \ref{charact_chaos}), microscopic chaos, which is induced by the defocusing character of successive collisions between atoms and molecules, is crucial to observe the property of mixing responsible of the relaxation processes at the macroscopic scale.

Let us consider the Brownian motion of a colloidal particle. The high-dimensional microscopic chaos of the surrounding fluid, characterized by the spectrum of Lyapunov exponents, induces a dynamical randomness given by the KS entropy, which is calculated by Pesin's identity (\ref{pesin}). This huge dynamical randomness appears to be at the origin of the erratic motion of the Brownian particle. Consequently it gives a new interpretation of the observed stochastic processes in terms of a high-dimensional chaos in the microscopic Hamiltonian dynamics.\footnote{An experimental work performed by \citet{gaspard-experiment} has shown the chaotic character of the microscopic dynamics.} Therefore the statistical mechanics of irreversible processes no longer needs stochastic models to describe the macroscopic irreversibility. The chaotic deterministic systems may exhibit a stochastic-like behavior without need of any stochastic assumption. Moreover, the sensitivity upon the initial conditions implying unpredictability for long times justifies the use of statistical mechanics, even for systems with a low-dimensional phase space, so that statistical mechanics no longer requires systems with a huge number of particles. As a result, the hypothesis of microscopic chaos allows to establish the foundations of non-equilibrium statistical mechanics at a more fundamental level \citep{GC1995,GC1995b}.

For several decades, computer tools has widely been exploited in statistical mechanics to calculate transport coefficients in different models. In this context, various few-degree-of-freedom systems appear very useful since they allow to considerably reduce the CPU time. Furthermore, numerical techniques applied to calculate quantities characterizing the microscopic chaos are much easier in systems with a small number of phase-space dimensions. 

The first class of models studied with the perspective of connecting transport and microscopic chaos brings together the so-called \textit{thermostatted systems} which have been developed in the eighties \citep{HLM_1982,evans_1983,nose84a,nose84b,hoover85,hoover-book,evans-cohen-morriss}.\footnote{For further information, see, for example, \citet{evans-morriss-book}.} These systems are composed of $N$ interacting particles. However, the particles are also submitted to a fictitious nonHamiltonian force to induce a specific transport process. In order to keep constant the energy, a thermostat must be implemented. As a consequence, the equations of motion are no longer Hamiltonian, and such systems no longer satisfy Liouville's theorem. This approach will be illustrated in Section \ref{Therm-syst-appr} in the case of the viscosity.

Beside systems made of a few particles such as hard disks or hard spheres, more abstract models have been used and developed for their simplicity. Among these is found the famous {\em Lorentz gas} initially introduced by Lorentz (1837-1923) to describe the diffusion of electrons in metallic bodies \citep{lorentz_1905_1,lorentz_1905_2,lorentz_1905_3}. Elaborating on the work of \citet{drude_1900_1,drude_1900_2}, he was able to derive the so-called Wiedeman-Franz law describing the temperature dependence of the ratio between the thermal and electrical conductivity of a metal \citep{wiedemann_franz_1853}. In the context of non-equilibrium statistical mechanics, the Lorentz gas rapidly appeared useful as a model of diffusion of light particles among heavy ones. Concretely, this model is made of hexagonal cells containing a fixed scatterer at their center onto which noninteracting pointlike particles bounce. Despite its simplicity, the dynamics of the pointlike particles is unstable because of the specular collisions on the scatterers, so that the Lorentz gas presents the mixing property needed to study relaxation processes such as transport. This feature combined to the simplicity of the model (few degrees of freedom) justify its popularity in non-equilibrium statistical mechanics.\footnote{Note that \citet{MHB_1987} proposed a thermostatted version of the Lorentz gas by introducing a constant electric field acting on the pointlike particles endowed with an electric charge. On the other hand, the Lorentz gas has also been widely used as one of the simplest models to test the ideas developed in the kinetic theory. See \citet{vanbeijeren_1982} for further information.}

\begin{figure}[t!]
\centerline{\mbox{\scalebox{.5}{\rotatebox{270}{\includegraphics{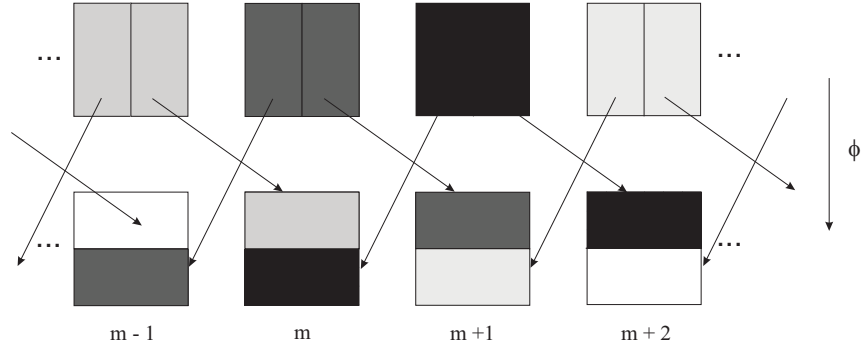}}}}}
\caption{Multibaker map developed by \citet{gaspard_1992}. Each cell is submitted to the usual baker map, except that the resulting parts are moved to their neighboring cells, which involves a diffusive process.}
\label{multibaker_map}
\end{figure}

In addition to the well-known Lorentz gas, very simple models are used to study the relationships between transport and microscopic chaos, such as the so-called {\em multibaker map} \citep{gaspard_1992} constructed as a chain of coupled {\em baker maps} introduced by \citet{seidel_1933}. The baker map applied on a unit square consists of two steps: (i) the square is contracted on the $y$-direction and stretched on the $x$-direction by a factor 2, so that this process is area-preserving; (ii) the rectangle obtained in this way is cut into two parts, and the right part is put on the left part.\footnote{See for example \citet{dorfman_vanbeijeren_1997}.} The multibaker map is made of a chain of such systems where the right (left) part of a specific square is moved to the top (bottom) of the unit square on its right (left) (see Fig. \ref{multibaker_map}), so that a diffusion process appears. This model has been widely used and elaborated to investigate remarkable properties in a time reversal, deterministic system, such as the relationships between entropy production and fractal structures in systems maintained in non-equilibrium steady states \citep{tasaki_gaspard_1995} or relaxing toward the equilibrium \citep{gilbert-dorfman-gaspard}. Furthermore, the multibaker map has also been modified to study conduction in an external field \citep{TVB_1996,VTB_1997,VTB_1998,BTV_1998,GFD_1998}, thermal conduction \citep{tasaki_gaspard_1999}, chemical reactions \citep{gaspard_1997}, and shear flow \citep{TVM2001,MTV2001}. 


\subsection{Existence of transport coefficients in few-degree-of-freedom systems}

Transport in the bulk of matter can be studied in such systems if the dynamics is spatially extended by introducing {\em periodic boundary conditions}. However, in most cases, the transport coefficients have not been rigorously proved to exist, i.e., to be finite, non-vanishing and positive as a result of the positivity of the entropy production (see \ref{phen_appr_visc}). The existence of strictly positive coefficients requires the establishment of a \textit{central limit theorem}.\footnote{For further information, see \citet{van_kampen_2007}.} Only a few small systems can claim to have such an advantage. The case of diffusion in \textit{periodic Lorentz gas} considered by \citet{bunimovich-sinai80,bunimovich-sinai81} appeared as the easiest since it requires a central limit theorem for the position of the point-like particle moving in the physical space. More recently, it has been suggested by numerical studies that the central limit theorem could also be satisfied for a polygonal billiard channel \citep{sanders-thesis,sanders05}. In the case of the viscosity, \citet{buni-spohn} proved that a periodic two-hard-disk model (assuming that the diameter of the particles is sufficiently large) satisfied such a theorem for the stress tensor, proving consequently the existence of viscosity coefficients already in this very simple model. In addition, \citet{ladd-hoover85} and, later, \citet{viscardy-gaspard1} numerically showed that viscosity already existed with only two particles.

The Lorentz gas as well as the multibaker map have the fundamental feature that the volume in phase space is conserved, as expected in Hamiltonian systems. However, the moving particles in the Lorentz gas, or their equivalents in the multibaker map, are {\em noninteracting}. This gave rise to some objections to the use of these models as diffusive systems \citep{cohen_rondoni_2002}. However, the mixing property of these models have been shown.  In the specific case of the Lorentz gas, this leads naturally to the establishment of a local equilibrium in velocity direction. Indeed, the dynamics of the pointlike particles bouncing onto the fixed scatterers involves the randomization of the velocity direction, so that the velocity distribution asymptotically becomes uniform in the velocity angle while the velocity remains constant in magnitude. Consequently, although the particles are noninteracting, the local equilibrium in velocity direction ensures the transport by diffusion, so that this does not contradict the thermodynamics of irreversible processes and are useful to develop tools required to treat more realistic fluids \citep{TVM2003,GND_2003}. 

During the last two decades, different theories have been elaborated which have established connections between large-deviations properties of the temporal dynamics and irreversible properties such as transport phenomena and entropy production.\footnote{Further information on this section dedicated to these recent advances can be found in \citet{dorfman_vanbeijeren_1997}, \citet{dorf-book}, and \citet{gasp-book,gaspard_review_2002,Gasp_05,Gasp_PhysA_06,Gasp_PTPS_06,Gasp_ACP_07,Gaspard_Rev2008}.} Some of these advances will be described after a quick review of the phenomenology of irreversible processes and an introduction to preliminary concepts on the analysis of the Liouville equation giving rise to the Policott-Ruelle resonances that control the exponential relaxation toward the equilibrium.


\subsection{Hydrodynamic modes and statistical mechanics}

\subsubsection{Macroscopic approach to relaxation processes}

At the macroscopic level, the time evolution of a one-component fluid is described by equations obtained from the laws of local conservation of mass, momentum and energy; namely, respectively the continuity equation (\ref{continuity-equation}), the Navier-Stokes equations (\ref{NS-equations-intro}) and the heat equation (\ref{hydro-eqs}):
\begin{eqnarray}
\frac{\partial \rho}{\partial t} & = & - \frac{\partial}{\partial r_j} \left ( \rho v_j \right ) \label{continuity_equation}\\
\frac{\partial v_i}{\partial t} & = & - v_j \frac{\partial v_i}{\partial r_j}
- \frac{1}{\rho} \left ( \frac{\partial P}{\partial \rho} \right )_T \frac{\partial \rho}{\partial r_i}
- \frac{1}{\rho} \left ( \frac{\partial P}{\partial T} \right )_\rho \frac{\partial T}{\partial r_i}
+ \frac{1}{\rho} \eta \frac{\partial^2 v_i}{\partial r_j \partial r_j} 
+ \frac{1}{\rho} \left ( \zeta + \frac{1}{3}\eta \right ) \frac{\partial^2 v_j}{\partial r_i \partial r_j}   \\
\frac{\partial T}{\partial t} & = & -\frac{T}{\rho c_V} \left ( \frac{\partial P}{\partial T} \right )_\rho \frac{\partial v_j}{\partial r_j}
- v_j \frac{\partial T}{\partial r_j} + \frac{\kappa}{\rho c_V} \frac{\partial^2 T}{\partial r_j \partial r_j}
+ \frac{\eta}{\rho c_V} \left ( \frac{\partial v_j}{\partial r_i} + \frac{\partial v_i}{\partial r_j}\right ) \frac{\partial v_j}{\partial r_i} 
+ \frac{1}{\rho c_V} \left ( \zeta - \frac{2}{3}\eta \right ) \left ( \frac{\partial v_j}{\partial r_j} \right )^2 \; .
\label{hydro-eqs}
\end{eqnarray}
where we have expanded the pressure $P$ in terms of the mass density $\rho$ and temperature $T$.

\begin{figure}[h!]
\centerline{\mbox{\scalebox{.4}{\rotatebox{270}{\includegraphics{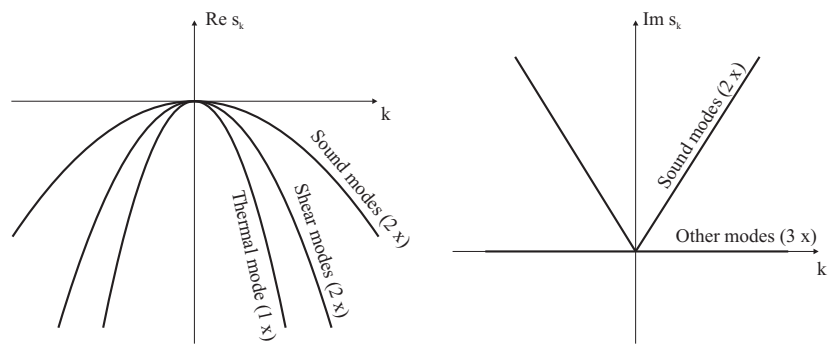}}}}}
\caption{Schematic representation of the dispersion relation of the five modes in fluids: 2 degenerate sound modes; two degenerate shear modes; one thermal modes. On the left: the real part. On the right: the imaginary part.}
\label{schema-disp-rel-fluids}
\end{figure}

The approach to the thermodynamic equilibrium can be described by the linearization of these equations around the equilibrium state.\footnote{For further detail on the linearization of this set of equations, see, for example, \citet{balescu-book} and \citet{resibois-book}.} The solutions of the linearized equations are obtained by using the principle of linear superposition. The general solution is expressed as a combination of spatially periodic solutions parameterized by the wavenumber $\bf k$:
\begin{eqnarray}
\rho({\bf r},t) & = & \rho_{\bf k} \exp(i {\bf k} \cdot {\bf r}) \exp(s_k t) + \rho_0 \label{linear_1}\\
{\bf v}({\bf r},t) & = & {\bf v}_{\bf k} \exp(i {\bf k} \cdot {\bf r}) \exp(s_k t) \label{linear_2} \\
T({\bf r},t) & = & T_{\bf k} \exp(i {\bf k} \cdot {\bf r}) \exp(s_k t) + T_0 \label{linear_3} \; .
\end{eqnarray}
These spatially periodic solutions are the so-called \textit{hydrodynamic modes}. The general solution is therefore given by a superposition of five independent hydrodynamic modes, each describing a motion among the five variables. The roots, namely the {\em dispersion relation}
\begin{equation}
s_{k} = {\rm Re} \; s_k + i \; {\rm Im} \; s_k \; ,
\end{equation}
are given in Table \ref{dispersion-relations-fluids} and schematically represented in Fig. \ref{schema-disp-rel-fluids}. We observe that the different modes are associated with different physical processes. They describe the exponential relaxation toward the equilibrium state in which the velocity vanishes while the temperature and the pressure become uniform, so that temperature and density reach respectively their equilibrium value $T_0$ and $\rho_0$. The first two modes are associated with the propagation of sound, in which the two viscosity coefficients $\eta$ and $\zeta$ appear. Two degenerate modes describe the dissipation due to the shear viscosity $\eta$. The fifth mode is associated with the heat diffusion through the presence of the heat conductivity coefficient $\kappa$. In the particular case of the two viscosity coefficients, we observe that the shear viscosity coefficient $\eta$ appears alone in the two degenerate shear modes and contributes together with the bulk viscosity $\zeta$ to the damping of sound $C$
\begin{equation}
C = \frac{1}{2 \rho_0} \left \lbrack \left ( \frac{1}{c_V} - \frac{1}{c_P} \right ) \kappa
+ \frac{4}{3} \eta + \zeta  \right \rbrack \; ,
\label{damping_coefficient}
\end{equation}
which has already been introduced in Section \ref{hydro-visc} on the experimental evidence and measuring of the bulk viscosity.

\begin{table} [!t]
\begin{center}
\begin{tabular}{|l|l|}
\hline
Hydrodynamic modes & Dispersion relations \\
\hline
\hline
Sound mode & $s_k = i v_s k - C  k^2 $ \\
Sound mode & $s_k = - i v_s k - C k^2$ \\
Shear mode & $s_k = - \frac{\eta}{\rho_0} k^2$ \\
Shear mode & $s_k = - \frac{\eta}{\rho_0} k^2$ \\
Thermal mode & $s_k = - \frac{\kappa}{\rho_0 c_P} k^2$\\
\hline
\end{tabular}
\end{center}
\caption{Dispersion relations of the decay rates $s_k$ associated with the five hydrodynamic modes in one-component fluids. $k$ is the wavenumber, $v_s$ the sound velocity (\ref{acoustic_velocity}), $\rho_0$ the mass density at equilibrium, $C$ the damping coefficient of sound (\ref{damping_coefficient}), $\eta$ the shear viscosity coefficient, $\kappa$ the heat conductivity, $c_P$ the heat capacity at constant pressure.} \label{dispersion-relations-fluids}
\end{table}

\subsubsection{Microscopic approach to relaxation processes}\label{micro_approach_relax_proc}

Time reversibility, which characterizes the Hamiltonian dynamics, implies that if the phase-space trajectory
\begin{equation}
{\cal C} =\lbrace {\bf \Gamma}_t =  {\bf \Phi}^t ({\bf \Gamma}_0) \colon t \in \mathbb{R} \rbrace
\end{equation}
is a solution of Hamilton's equations, then its time reversal
\begin{equation}
\Theta ({\cal C}) = \lbrace \tilde{\bf \Gamma}_{t'} =  {\bf \Phi}^{t'} \circ \Theta ({\bf \Gamma}_0) \colon t' \in \mathbb{R} \rbrace
\end{equation}
also satisfies them. However, both solutions are typically different trajectories:
\begin{equation}
{\cal C} \ne \Theta ({\cal C}) \; .
\end{equation}
Actually, an equation and its solutions do not have to share the same symmetry properties. While the ensemble of all the solutions of equations of motion are time-reversal, this is no longer the case of the unique trajectory followed by the system. This phenomenon is known under the name of {\em spontaneous symmetry breaking}, which is here induced by the selected initial conditions. In the light of this time-reversal symmetry breaking, if the irreversibility is associated with the trajectory instead of Hamilton's equations themselves, Hamiltonian microscopic dynamics appears to be compatible with the irreversible macroscopic processes.

At the statistical level, the probability distribution $f$ is the appropriate mathematical tool to describe the evolution of mechanical systems. It measures the statistical frequency of occurrence of each initial condition --- and thus each phase-space trajectory --- among a huge number of copies of the same system. Let us consider a non-equilibrium system composed of particles diffusing between two reservoirs at different concentrations (let ${\rm R_H}$ and ${\rm R_L}$ be respectively the reservoirs at high and low concentration). After a certain time, a non-equilibrium steady state is reached, which is described by a invariant probability distribution $f_{\rm neq}$ over which the average gives a mean current of diffusion from ${\rm R_H}$ to ${\rm R_L}$. This is a consequence of the fact that the probability associated with trajectories issued from ${\rm R_H}$ is greater than that of trajectories issued from ${\rm R_L}$. However, the set composed of the latter trajectories contains the time reversal of the former ones. As a result, the time-reversal symmetry breaking similar to that for Hamilton's equations is observed at the statistical level: $f_{\rm neq}(\Theta {\bf \Gamma}) \neq f_{\rm neq}({\bf \Gamma})$. Let us note that, if one closes the reservoirs, the system relaxes to the equilibrium state described by an invariant probability distribution $f_{\rm eq}$ which satisfies the time reversal symmetry: $f_{\rm eq}(\Theta {\bf \Gamma}) = f_{\rm eq}({\bf \Gamma})$.\footnote{For further information, see, for example, \citet{Gasp_05,Gaspard_Rev2008}.}

The issue of the emergence of irreversibility from the underlying microscopic dynamics can be investigated by means of the exponential decay of the modes (\ref{linear_1}-\ref{linear_3}) toward the equilibrium.
Until recently, the hydrodynamic modes were not described in terms of the Liouvillian dynamics, but at the intermediate level of the kinetic equations, such as the Boltzmann equation (\ref{boltzmann-equation}). However, the derivation of the latter is obtained by introducing assumptions such as the {\em Sto\ss zahlansatz} (see Section \ref{Boltzmann_stat_irrev}). On the other hand, the Liouville equation involves a $N$-particle probability distribution so that, in pratice, it is hard to solve. In addition, since the solution of the Liouville equation and the integration of Hamilton's equations are equivalent problems, the former equation has not been studied in great detail until, in particular, \citet{prigogine_1962}, and later, \citet{pollicott_1985,pollicott_1986} and \citet{ruelle_1986a,ruelle_1986b,ruelle_1987a,ruelle_1987b,ruelle_1989} who developed the concept of resonances for Axiom-A systems.\footnote{Some properties of the Axiom-A systems are given in \ref{anosov_systems}.} This gave rise to the so-called {\em Pollicott-Ruelle resonances} which have permitted to overcome these difficulties.

As seen above, the property of mixing is needed to have a relaxation to equilibrium. This property implies that the time-correlation functions 
\begin{equation}
C_{AB}(t) =  \langle A({\bf \Phi}^t {\bf \Gamma}) B({\bf \Gamma}) \rangle -
\langle A  \rangle \;  \langle B  \rangle
\label{TCF}
\end{equation}
between two observables $A$ and $B$ tend asymptotically to zero as a consequence of Eq. (\ref{mixing_def}). A spectral approach is used to characterize the relaxation.\footnote{For further information on the following concepts, see \citet{gasp-book}.} The spectral functions are defined as a Fourier transform of Eq. (\ref{TCF})
\begin{equation}
S_{AB}(\omega) = \int_{- \infty}^{+ \infty} \exp(i \omega t) C_{AB} (t) \; dt \; .
\label{spectral_functions}
\end{equation}
It is known that mixing systems have necessarily a continuous spectrum, which provides little information on the relaxation of these systems. To overcome this, the analytical continuation of the functions (\ref{spectral_functions}) is rather considered toward complex frequencies
$z = {\rm Re}\; z + i \; {\rm Im}\; z$, with $\omega = {\rm Re} \; z$.
The correlation function $C_{AB}(t)$ is recovered by the inverse Fourier transform
\begin{eqnarray}
C_{AB}(t) & = & \frac{1}{2 \pi} \int_{- \infty}^{+ \infty} \exp(-i \omega t) S_{AB} (\omega) \; d\omega \nonumber \\
& = & \frac{1}{2 \pi} \int_{\cal C} \exp(-izt) S_{AB} (z) \; dz \; .
\label{TCF_FT}
\end{eqnarray}
Different types of possible complex singularities have been studied, which describe how $C_{AB}(t)$ relaxes to equilibrium. They can be simple or multiple poles $z_\alpha$, or branch cuts $z_c$. 
\begin{figure}[t!]
\centerline{\mbox{\scalebox{.55}{\rotatebox{0}{\includegraphics{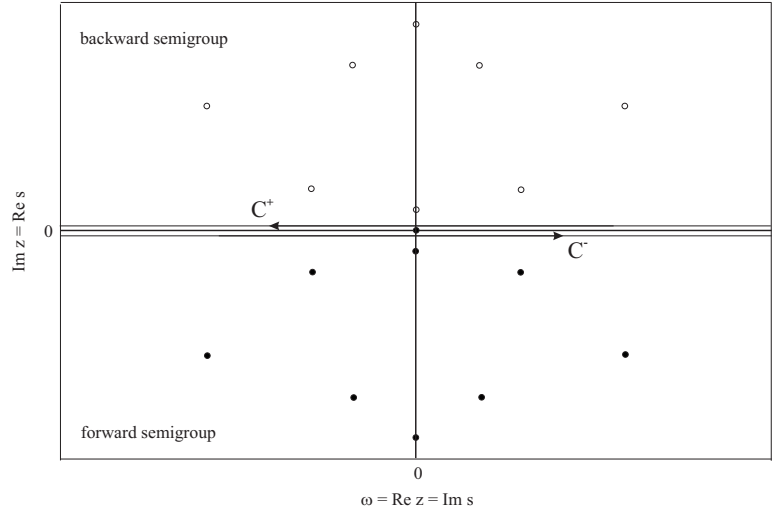}}}}}
\caption{Complex plane of the variable $s = -i z$. The vertical axis Re $s$ = Im $z$ is the axis of the relaxation rates. The contour $C^+$ ($C^-$)  is slightly above (below) the axis Re $s = 0$ and is deformed in the lower (upper) part of the complex plane in order to get the resonances ruling the relaxation for positive (negative) times. The branch cuts are not depicted in this figure.}
\label{pollicott-ruelle_resonances}
\end{figure}

Nevertheless, the complex singularities of the lower half plane and those of the upper half plane of the complex frequencies should be distinguished since they correspond to the time evolution toward positive and negative times, respectively. Therefore, for positive times, the integration contour C$^+$ has to be deformed to the lower half-plane of the complex variable $z$, as shown in Fig. \ref{pollicott-ruelle_resonances}. Eq. (\ref{TCF_FT}) can thus be rewritten as 
\begin{equation}
C_{AB}(t) = \sum_{\alpha:{\rm Im} \; z_\alpha < 0} C_{\alpha}(t) + \sum_{c:{\rm Im} \; z_c < 0} C_{c}(t) \; .
\end{equation}
The first term of the right-hand side consists in a sum of the contribution of the poles
\begin{equation}
C_{\alpha}(t) = \sum_{l=1}^{m_\alpha} \frac{a_l^{(\alpha)} \exp \left ( \frac{i \pi (l-m_\alpha -1)}{2} \right)}{(m_\alpha -l)!} t^{m_\alpha - l} \exp (-i z_\alpha t)
\end{equation}
where $m_\alpha$ is the multiplicity of the pole $z_\alpha$, while the second sum is the contribution of the branch cuts. Hence, the simple poles contributes to the correlation function as: $C(t) \sim \exp(-i z_\alpha t)$, i.e., as purely exponential decays. Here, we shall focus only on these simple poles, keeping in mind the possible existence of multiple poles and branch cuts.

Correlation functions or statistical averages of observables need to consider a statistical ensemble of phase-space trajectories described by probability distribution $f({\bf \Gamma})$ whose time evolution is ruled by the Liouville equation (\ref{liouville_equation1})
\begin{equation}
\frac{\partial f}{\partial t} = \hat{L} f
\label{liouville_equation2}
\end{equation}
where the {\em Liouville operator} ${\hat L}$ has been introduced. In Hamiltonian systems, ${\hat L}$ is given by the Poisson bracket of the Hamiltonian
$\hat{L}(\cdot) = \lbrace H, \; \cdot \; \rbrace$, so that the Liouville equation conserves the time-reversal symmetry. Since no assumptions are needed to derive it, Eq. (\ref{liouville_equation2}) has the status of the {\em fundamental equation} of statistical mechanics as \citet{gibbs_1884,gibbs} had already noticed (see Section \ref{gibbs_stat_mech}). As a result, this plays a central role in modern theories of non-equilibrium statistical mechanics.

Because the Liouville equation is linear, its solution for an invertible conservative system can be formally written as
\begin{equation}
f({\bf \Gamma},t) = \exp(\hat{L}t) f({\bf \Gamma},t) = \hat{P}^t f({\bf \Gamma},0)
\label{liouville_solutions}
\end{equation}
where $\hat{P}^t$ is the {\em Perron-Frobenius operator}. As already mentioned in Section \ref{ergodicity_mixing}, \citet{koopman_1931} brought a tremendous contribution to the study of classical systems by using the framework of the quantum-mechanical formalism. Let ${\hat G}$ a Hermitian operator:
\begin{equation}
{\hat G} = i {\hat L}
\end{equation}
defined on a Hilbert space of phase-space functions. Eq. (\ref{liouville_equation2}) can thus be expressed as a Schr\"{o}dinger-like equation
\begin{equation}
i\frac{\partial f}{\partial t} = {\hat G} f \; .
\end{equation}
 Accordingly, the solutions (\ref{liouville_solutions}) can be rewritten as 
\begin{equation}
f({\bf \Gamma},t) = \exp(-i\hat{G}t) f({\bf \Gamma},t) = \hat{U}^t f({\bf \Gamma},0) 
\end{equation}
where $\hat{U}^t$ is a unitary operator.

Instead of characterizing the decay of a time-correlation function depending on the observables considered, the complex singularities introduced above may turn out to be intrinsic to the dynamical system itself and characterize the properties of the Perron-Frobenius operator rather than of specific observables considered in the correlation functions. Consequently, the singularities are independent of the observables. Assuming that the right- and left-eigenstates $\Psi_\alpha$ and ${\tilde \Psi}_\alpha$ are such that
\begin{eqnarray}
\langle \tilde{\Psi}_\alpha \vert \Psi_{\alpha'} \rangle & = & \delta_{\alpha \alpha'} \nonumber \\
\int \vert \Psi_{\alpha'} \rangle \tilde{\Psi}_\alpha \vert & = & 1 \; ,
\end{eqnarray}
this leads us to consider the poles associated with the resonances as generalized eigenvalues of the Liouville operator associated with $\Psi_\alpha$ and ${\tilde \Psi}_\alpha$
\begin{eqnarray}
\hat{L} \; \vert \Psi_\alpha \rangle & = & s_\alpha \; \vert \Psi_\alpha \rangle	\nonumber \\
\langle {\tilde \Psi}_\alpha \vert \;  \hat{L} & = & s_\alpha \; \langle {\tilde \Psi}_\alpha \vert
\end{eqnarray}
where $s_\alpha = -i z_\alpha$.

The mathematical method used to determine the Pollicott-Ruelle resonances is based on the integration of the resolvent
\begin{equation}
\hat{R}(z) = \frac{1}{z - {\hat G}} = -i \int_{0}^{\infty} \exp(izt) \exp(-i{\hat G}t) \; dt \; .
\end{equation}
The separation of the upper and lower half parts of the complex plane leads to the construction of two distinct semigroups defined by the positive and negatives times, i.e., respectively a forward semigroup associated with the lower half plane of complex frequencies and a backward semigroup having its spectrum in the upper half plane.

The time evolution of the statistical average of an observable $A$ can be written as
\begin{equation}
\langle A \rangle_t = \int_\Gamma A({\bf \Gamma}) \; \exp (\hat{L} t) \; p_0 ({\bf \Gamma}) \; d {\bf \Gamma} \; .
\label{obs_stat_aver}
\end{equation}
The resonances obtained by the analytic continuation  toward the negative values of ${\rm Re}\ s_\alpha$ allow to expand Eq. (\ref{obs_stat_aver}) and define the forward semigroup:
\begin{equation}
\langle A \rangle_t   \simeq  \sum_\alpha \; \langle A \vert \Psi_\alpha \rangle \; \exp(s_\alpha t)
 \; \langle \tilde \Psi_\alpha \vert p_0 \rangle + \dots
\label{asymptotic_observable}
\end{equation}
wich is valid only for $t > 0$.\footnote{Let us mention that the dots in Eq. (\ref{asymptotic_observable}) indicate the other contributions besides simple resonances, such as multiple resonances with $m_\alpha > 1$ and branch cuts. For further detail, see, for example, \citet{gasp-book}.} The coefficients of the expansion are given by
\begin{eqnarray}
\langle A \vert \Psi_\alpha \rangle & = & \int A({\bf \Gamma})^* \Psi_\alpha ({\bf \Gamma}) \; d{\bf \Gamma} \\
\langle \tilde{\Psi}_\alpha \vert f_0 \rangle & = & \int \tilde{\Psi}_\alpha ({\bf \Gamma})^*
f_0 ({\bf \Gamma}) \; d{\bf \Gamma} \; .
\end{eqnarray}
In the same way, the backward semigroup can be written as
\begin{equation}
\langle A \rangle_t   \simeq  \sum_\alpha \; \langle A \vert \Psi_\alpha \circ \Theta \rangle \; \exp(s_\alpha t)
 \; \langle \tilde \Psi_\alpha \circ \Theta \vert p_0 \rangle + \dots
\label{backward_semigroup}
\end{equation}
for $t<0$.

Since the microscopic Hamiltonian dynamics is deterministic and unstable, the associated eigenstates $\Psi_\alpha$ of the Liouville operator are smooth in the unstable directions (associated with the positive Lyapunov exponents) and singular in the stable phase-space directions (associated with the negative Lyapunov exponents). As a result of the time-reversal symmetry, all the unstable directions can be mapped onto the stable directions. However, as the solutions of Hamilton's equations obtained by time reversal are not identical, the unstable directions are physically distinct from the stable ones.

In the following sections, three of recent approaches and methods establishing relationships between microscopic chaotic dynamics and irreversible processes will be presented. In Section \ref{Therm-syst-appr}, the so-called {\em thermostatted-system approach} is sketched, in which (i) the equations of motion are modified by introducing a force that maintains the system out of equilibrium, and (ii) a thermostat is introduced to keep constant the energy of the system. However, this breaks the Hamiltonian character and the volume-preserving property. By contrast, two other methods recently elaborated do not violate this fundamental property. Section \ref{escape_rate_formalism} is devoted to the first one called {\em escape-rate formalism} by Gaspard and Dorfman, which is based on the introduction of {\em absorbing boundary conditions} inducing an escape process. Section \ref{hydro_mode_method} outlines the second one known as the {\em hydrodynamic-mode method} or Liouville-equation approach, whose the purpose is to construct the hydrodynamic modes at the microscopic level.


\subsection{Thermostatted-system approach}\label{Therm-syst-appr}

\begin{figure}[h!]
\centerline{\mbox{\scalebox{.6}{\includegraphics{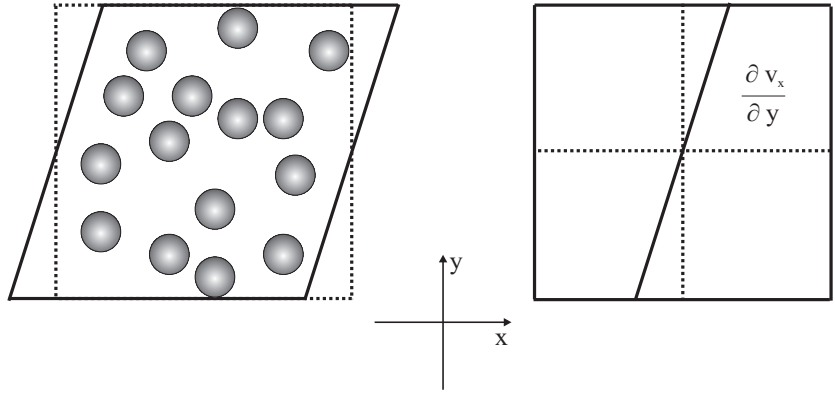}}}}
\caption{Illustration of the shearing of the system inducing a velocity gradient
$\frac{d v_{x}}{d y}$.}
\label{thermostatted-fluid}
\end{figure}

In the \textit{thermostatted-system approach}, non-equilibrium systems are composed of particles submitted to interparticle forces, to external forces, but also to a fictitious nonHamiltonian force modeling the coupling to some hypothetical thermostat.\footnote{For further information, see, for example, \citet{evans-morriss-book}.} For instance, in order to study the viscosity, the idea is to reproduce a Couette flow induced by a shearing force (see Fig. \ref{thermostatted-fluid}).
Hence a velocity gradient is established in the system \citep{evans-morriss-1984}. However, this gradient leads to considerable viscous heating of the fluid and the energy of the system does not remain constant. To deal with this problem, it is necessary to introduce an internal thermostat --- a fictitious frictional force --- in order to keep constant the energy. Formally, for a 3-D system, in addition to the shearing force, we have to add a term with a thermostating multiplier $\alpha$
\citep{evans-morriss-book}
\begin{equation}
\left \lbrace
\begin{array}{l}
\frac{d\mathbf{q}_{i}}{dt}  = \frac{\mathbf{p}_{i}}{m} + {\gamma} y_{i} \\
\\
\frac{d\mathbf{p}_{i}}{dt}  =  \mathbf{F}_{i}- {\gamma} p_{yi} - \alpha \mathbf{p}_{i} \; , 
\end{array}
\right.
\label{sllod}
\end{equation}
where ${\gamma} = (\frac{\partial v_{x}}{\partial y},0)$, and $\mathbf{v}$ the mean velocity. However, such dynamical systems violate Liouville theorem. This violation leads to fundamental problems for defining an entropy in non-equilibrium steady states. It thus appears as an artefact of a nonHamiltonian force that the phase-space volume visited by the system decreases in time and is expressed by the non-zero sum of Lyapunov exponents $\sum_{i=1}^{6N} \lambda_{i} < 0$. This phase-space contraction is introduced through the presence of the thermostat which takes away the energy given to the system by shearing. It can be shown that the relation between the viscosity and the maximum and minimum Lyapunov exponents becomes for large systems \citep{evans-cohen-morriss}
\begin{equation}
\eta({\gamma}) = \frac{-3 n k_{B} T}{\gamma^{2}} \left \lbrack \lambda_{\mathrm{max}}({\gamma})
+ \lambda_{\mathrm{min}}({\gamma}) \right \rbrack \; ,
\label{visc-thermostat}
\end{equation}
the $N$ dependence of $\eta$ and $\lambda$ disappearing. The shear viscosity occurring in the Navier-Stokes equation is given by $\eta = \lim_{\gamma \to 0}\eta({\gamma})$. This expression relates the Lyapunov exponents to the viscosity coefficient because of the violation of Liouville's theorem by the artifitial nonHamiltonian systems. Therefore, this method cannot be used for Hamiltonian systems. Several works have shown that other  ways exist to maintain a system out of equilibrium for instance by stochastic boundary conditions or by deterministic scattering \citep{klages-rateitschak-nicolis} in which cases the relation (\ref{visc-thermostat}) do not apply. Moreover, violating Liouville's theorem creates problems in defining the entropy for non-equilibrium steady states.


\subsection{Escape-rate formalism}\label{escape_rate_formalism}

As for the previous method, the escape-rate formalism introduces non-equilibrium conditions. Here we do not impose an external constraint like a shearing. Instead we open the system in order to generate an escape process. More precisely, we impose {\em absorbing boundary conditions} at the statistical level of description, keeping the Hamiltonian character of the equations of motion themselves. The so-called \textit{escape rate} is related to the studied transport coefficient on the one hand, and to the chaotic quantities of the microscopic dynamics on the other hand.  The method was first developed by \citet{gasp-nicolis} for the case of diffusion and was extended by Dorfman and Gaspard to the other transport processes \citep{dorf-gasp,gasp-dorf}. For pedagogical reasons we shall first expose the escape-rate formalism for diffusion.

\subsubsection{Escape-rate formalism and diffusion}

\begin{figure}[h!]
   \begin{minipage}{.6\textwidth}
\hspace*{2cm}
       \mbox{\scalebox{.4}{\includegraphics{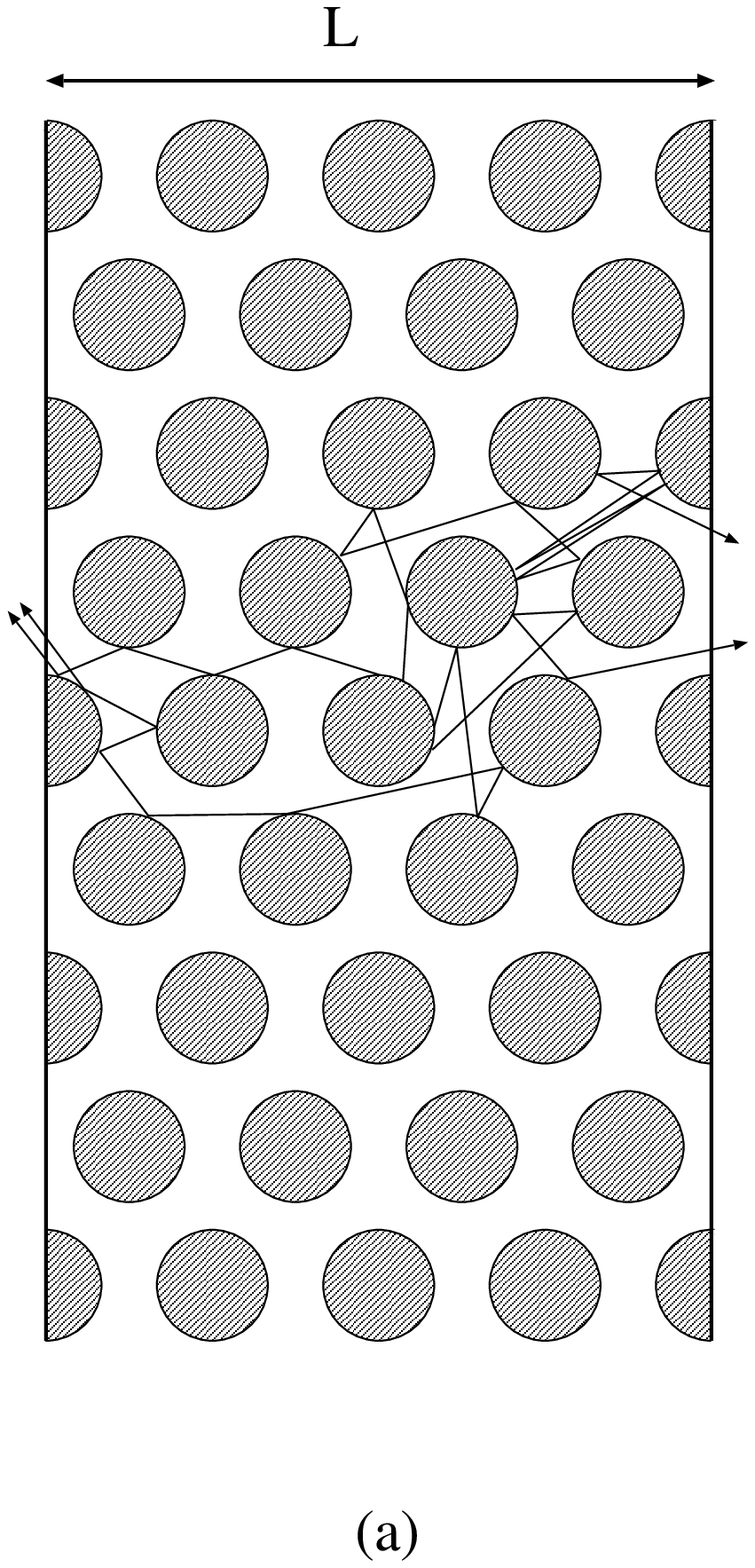}}}
   \end{minipage}
   \hfill
   \begin{minipage}{.6\textwidth}
\hspace*{-2cm}
       \mbox{\scalebox{.4}{\includegraphics{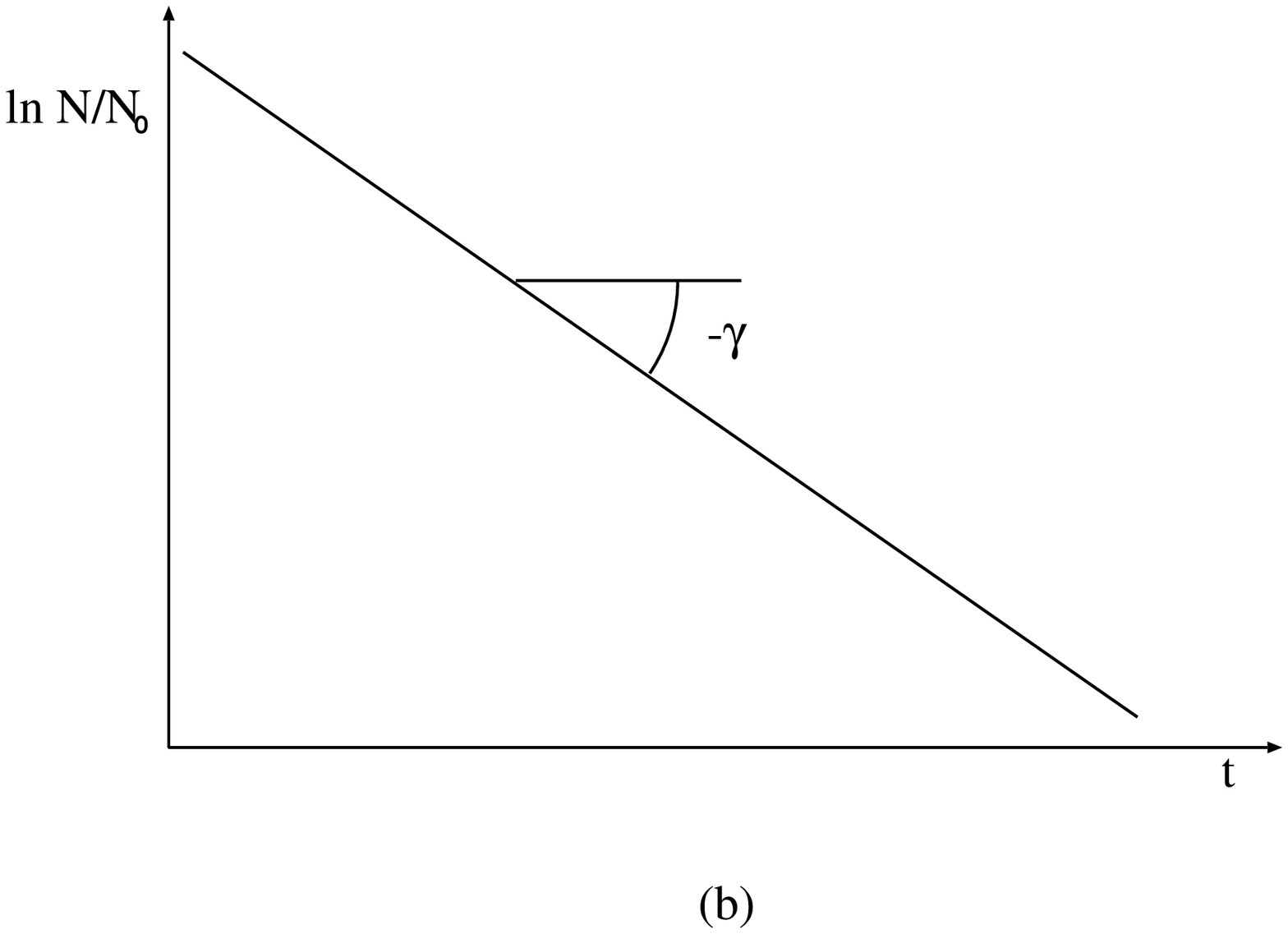}}}
   \end{minipage}
\caption{Lorentz gas with absorbing boundary conditions separated by the distance $L$. (a) Escape process of some particles after diffusion in the scatterer. (b) Exponential decrease in time of the number of particles lying into the limits defined by Eq. (\ref{diff-abs-limits}).} \label{Lorentz-absorbing}
\end{figure}

Let us take the Lorentz gas that consists of a particle of mass $m$ moving with energy $E$ among a fixed set of two-dimensional scatterers (see Fig. XX). Here, we consider a modification of the Lorentz gas, whose scatterers are confined within an infinite slab of width $L$ such that the scatterers are within the interval (see Fig. \ref{Lorentz-absorbing})
\begin{equation}
-\frac{L}{2} \le x \le \frac{L}{2} \; .
\label{diff-abs-limits}
\end{equation}
Absorbing walls are placed on the planes at $x=\pm \frac{L}{2}$.

The particles, initially introduced at the center of the system, evolve in the system by the successive collisions with the scatterers.

Let $\tilde{\rho}(\mathbf{r},t)$ denote the spatial distribution function. Its time evolution is ruled by the continuity equation (\ref{continuity_equation}). If $L$ is sufficiently large and for long times after some initial time, and if we introduce the phenomenological law of diffusion obtained by \citet{fick_1855_1,fick_1855_2}:
\begin{equation}
\tilde{\rho} v_j = - D \frac{\partial \tilde{\rho}}{\partial r_j} \; ,
\end{equation}
$\tilde{\rho}(\mathbf{r},t)$ is expected to be described by the diffusion equation
\begin{equation}
\frac{\partial \tilde{\rho}}{\partial t} = D \nabla^{2} \tilde{\rho} \; ,
\label{diffusion_equation}
\end{equation}
where the diffusion coefficient $D$ is assumed to be constant in space. The absorbing boundary conditions lead to the condition that $\tilde{\rho}(\mathbf{r},t)\vert_{{\rm boundary}}=0$. Then the probability for the distribution of particles in the $x$-direction can be expressed as a superposition of hydrodynamic modes
\begin{equation}
\tilde{\rho}(x,t) = \sum_{n=1}^{\infty} a_{n} \sin \left ( \frac{\pi n}{L} x \right ) \exp \left \lbrack - \left ( \frac{n \pi}{L}\right )^{2} Dt \right \rbrack ,
\end{equation}
where $a_{n}$ are numerical coefficients fixed by the initial profile of concentration. As time increases, each mode decreases exponentially and vanishes successively, the first ones having the greatest values of $n$. Consequently, for long times, the slowest decaying mode ($n=1$) describes the escape process and decays as $\exp (- \pi^{2} Dt / L^{2})$. As a consequence, for large systems we can define a \textit{macroscopic escape rate} as
\begin{equation}
\gamma_{\mathrm{mac}} = \left (\frac{\pi}{L}\right )^{2} D \; .
\label{escape-mac}
\end{equation}
In the following we shall call Eq. (\ref{escape-mac}) the {\em escape-transport formula}.

Let us consider the same process at the microscopic scale. Historically, the escape rate was introduced  in the context of the thermodynamic formalism (see \ref{section-escape-rate-formula}). We here propose a more intuitive presentation of this concept. In open systems such that depicted in Fig. \ref{Lorentz-absorbing}, particles go out of the boundaries and never go back into the system. But a set of trajectories remain forever (in the future and the past)  into the limits. A particle bouncing forever between two scatterers is the simplest example of such trajectories. This set is therefore the best candidate to be the appropriate support for a non-equilibrium invariant measure in order to evaluate the different quantities. This object in phase space presents a particular property: it has a zero Lebesgue measure and is of non-integer dimension. As seen in Section \ref{chaos_fractals}, such an object is called \textit{fractal}. Contrary to the situation in the thermostatted systems, this fractal is not an attractor but a \textit{repeller} because of the dynamical instabibility involving the escape of close trajectories from it. We denote this fractal repeller by the symbol ${\cal F}_{L}$~.

In Section \ref{micro-chaos}, we saw that the instability of the dynamics characterized by the positive Lyapunov exponents induces a gain of information on the initial conditions of the trajectories in phase space. This information grows exponentially in time and the exponential rate at which information is obtained is measured by the KS entropy $h_{\mathrm{KS}}$. Let us consider a certain region of phase-space points with a characteristic dimension of the order of $\delta$, which is the error on the observation. The different points in this region are not distinguishable at the initial conditions, but after a certain time $t$, the initial set will be stretched along the unstable directions (which correspond to the positive Lyapunov exponents) to a length of order of $\delta \exp \left (t \; \sum_{\lambda_{i} > 0} \lambda_{i} \right )$. Consequently, trajectories emerging from the initial set of points will be separated and we can easily resolve their images in the initial set. In  closed systems (without any escape condition) this gain of information is therefore evaluated as
\begin{equation}
\exp (h_{\mathrm{KS}}\; t) = \exp \left (t \; \sum_{\lambda_{i} > 0} \lambda_{i} \right ) \; ,
\end{equation}
which gives us Pesin's theorem (\ref{pesin}). On the other hand, in open systems, most of the trajectories escape the system because of the absorbing boundaries (\ref{diff-abs-limits}). The part of trajectories moving into the limits decays as $\exp (- \gamma_{\rm mic} t)$, $\gamma_{\rm mic}$ being the escape rate obtained by the microscopic approach. When a trajectory escapes the system it can no longer provide information on its origin by the aforementioned mechanism of dynamical instability. Accordingly, because of the absorbing conditions we loose in time a quantity of information (brought by the unstable character of the dynamics) versus the case of closed systems. This phenomenon induces a modification of Pesin's theorem by the introduction of the escape term
\begin{equation}
\exp (h_{\mathrm{KS}}\; t) = \exp (-\gamma_{\rm mic} t) \exp \left (t \; \sum_{\lambda_{i} > 0} \lambda_{i} \right )
\end{equation}
or simply
\begin{equation}
\gamma_{\mathrm{mic}}({{\cal F}_{L}}) = \sum_{\lambda_{i} > 0}^{} \lambda_{i}({{\cal F}_{L}}) - h_{\rm KS}({{\cal F}_{L}}) \; .
\label{escape-mic}
\end{equation}
This is the so-called \textit{escape-rate formula}.
However, the escape process being the same at the macroscopic and microscopic scales, the identity 
\begin{equation}
\gamma_{{\rm mac}} = \gamma_{{\rm mic}}
\end{equation}
is obtained in the limit $L \to \infty$. Consequently we can relate the two levels by combining Eqs. (\ref{escape-mac}) and (\ref{escape-mic}), and finally we have
\begin{equation}
D = \lim_{L\to\infty} \left ( \frac{L }{\pi  } \right )^{2} \left(\sum_{\lambda_i>0} \lambda_i - h_{\rm KS}\right)_{{\cal F}_{L}} .
\label{diff-haus}
\end{equation}

An equivalent formula can be obtained which involves the partial fractal dimensions of the repeller instead of the KS entropy.  Indeed, the fractal character of the repeller is a direct consequence of the escape of trajectories, so that the KS entropy is no longer equal to the sum of Lyapunov exponents but to \citep{young}
\begin{equation}
h_{\rm KS} = \sum_{\lambda_i>0} d_i \; \lambda_i \; ,
\end{equation}
where the coefficients are the partial information dimensions\footnote{It is known that the partial information dimension of the repeller can be approximated by the partial Hausdorff dimension if the escape rate is small enough and if Ruelle's topological pressure does not present a discontinuity \citep{gasp-book}. This last condition is fulfilled if the system does not undergo a dynamical phase transition. This is the case in the finite-horizon regimes of Sinai's billiard which controls the dynamics of the Lorentz gas \citep{gasp-baras}. Under these conditions, we can replace the partial information dimension $d_i$ by the partial Hausdorff dimension $d_{Hi}$.} of the repeller associated with each unstable direction of corresponding Lyapunov exponent $\lambda_i$ \citep{eckmann-ruelle}. These partial dimensions satisfy
\begin{equation}
0 \leq d_i \leq 1 \; ,
\end{equation}
so that the KS entropy is in general smaller than the sum of positive Lyapunov exponents. Accordingly, the escape rate (\ref{escape-mic}) can be expressed as
\begin{equation}
\gamma_{\mathrm{mic}}({\cal F}_{L}) = \left(\sum_{\lambda_i>0} c_i \; \lambda_i\right)_{{\cal F}_{L}}
\label{escape-rate-chaos}
\end{equation}
in terms of the partial codimensions defined as
\begin{equation}
c_i \equiv 1 - d_i \; .
\end{equation}
Finally, Eq. (\ref{diff-haus}) can be rewritten as
\begin{equation}
D = \lim_{L\to\infty} \left ( \frac{L }{\pi  } \right )^{2}
\left(\sum_{\lambda_i>0} c_i \lambda_i \right)_{{\cal F}_{L}} .
\end{equation}

This relation is fundamental: it clearly establishes  the link between the microscopic and macroscopic levels, between
diffusion as a transport process and the chaotic properties of the underlying microscopic dynamics. This formula allows us to understand how
chaos controls transport at the microscopic scale \citep{gasp-baras2,gasp-baras}.

\subsubsection{Escape-rate formalism and viscosity}

In 1995, Dorfman and Gaspard extented the escape-rate formalism to the other transport processes \citep{dorf-gasp,gasp-dorf}. Instead of imposing an external constraint like a shearing, as for the thermostatting-system method, the non-equilibrium conditions are introduced by imposing absorbing conditions
\begin{equation}
- \frac{\chi}{2} \le G_{xy}(t) \le + \frac{\chi}{2}
\label{abs-bound-cdts}
\end{equation}
characterized by an arbitrary parameter $\chi$, over the evolution of the \textit{Helfand moment} $G_{xy}(t)$ written in Eq. (\ref{helfand_moment_viscosity}). When the Helfand moment reaches these absorbing boundaries, the system is said to be escaped. Consequently, considering a statistical ensemble of systems, an escape process appears, which is characterized by the escape rate. Since the system is in  non-equilibrium, the support of the invariant measure is no longer the whole phase space, but a fractal repeller. This fractal is constructed  with the trajectories for which the Helfand moment remains forever into the absorbing boundaries (\ref{abs-bound-cdts}).

As a matter of fact, the escape-rate formalism displays the irreversibility in the leading Pollicott-Ruelle resonance. In Section \ref{micro_approach_relax_proc}, we saw that the leading resonance was $s_0 = z_0 = 0$ in closed systems. Here, since absorbing boundary conditions are introduced, this is no longer vanishing. The resonances are then expected to be given by
\begin{equation}
s_n \simeq -\eta \left ( \frac{\pi n}{\chi} \right )^2
\end{equation}
for $n = 1, 2, 3, \dots$ in hyperbolic systems, so that the smallest resonance $s_1$ gives the escape rate
\begin{equation}
\gamma = -\eta \left ( \frac{\pi}{\chi} \right )^2 
\end{equation}
for $t \to \infty $.

The escape rate plays the role of intermediate between the viscosity as a transport coefficient, and quantities of the microscopic chaos. Indeed, on the one hand, the viscosity can be written in terms of a \textit{phenomenological} escape rate by considering the macroscopic escape process. On the other hand, the theory of dynamical systems provides a relationship between the escape rate and quantities such as the Lyapunov exponents and KS entropy, typical quantities of chaos. By identifying both escape rates, a relationship between the viscosity and the microscopic chaos is established and is written as \citep{dorf-gasp,gasp-dorf}
\begin{equation}
\eta = \lim_{\chi\to\infty} \left ( \frac{\chi }{\pi  } \right )^{2}
\left ( \sum_{\lambda_i>0} \lambda_i -h_{KS} \right )_{{\cal F}_{\chi}}  ,
\label{visc-escape-rate-formula}
\end{equation}
with the Lyapunov exponents {$\lambda_i$} and the KS entropy computed with the fractal repeller ${\cal F}_{\chi}$ as the support of the invariant measure.

Similarly to the case of the diffusion, Eq. (\ref{visc-escape-rate-formula}) can be reformulated in terms of the partial codimensions of the fractal repeller as
\begin{equation}
\eta = \lim_{\chi\to\infty} \left ( \frac{\chi }{\pi  } \right )^{2}
\left ( \sum_{\lambda_i>0} c_i \lambda_i  \right )_{{\cal F}_{\chi}} \; .
\label{visc-codim}
\end{equation}

Hence we can show how the microscopic chaos, responsible of the fractal character of the support of the invariant measure, controls the viscosity  \citep{viscardy-gaspard2}, or also the diffusion \citep{gasp-baras} as well as the reaction-diffusion processes \citep{claus-gaspard2001,claus-gaspard-van-beijeren}.


\subsection{Hydrodynamic-mode method}\label{hydro_mode_method}

The third approach is the \textit{hydrodynamic-mode method} developed during the nineties by Gaspard and coworkers \citep{gasp-chaos,gaspard96}. As seen previously, hydrodynamics describes the macroscopic dynamics of fluids in terms of equations governing the evolution of mass density, fluid velocity, and temperature, such as the Navier-Stokes equations (\ref{Navier-Stokes-equations}) and the diffusion equation (\ref{diffusion_equation}). Thanks to the kinetic equation developed by Boltzmann, non-equilibrium statistical mechanics is able to derive these phenomenological equations, while the Boltzmann equation is itself derived from Liouvillian dynamics, using the \textit{Sto\ss zahlansatz}. In the case of diffusion, the solutions of Eq. (\ref{diffusion_equation}), namely the hydrodynamic modes (\ref{linear_1}), are of the form
\begin{equation}
\exp (s_{\mathbf{k}} t) \exp (i \mathbf{k}
\cdot \mathbf{r}) \; ,
\end{equation}
each mode being characterized by a wavenumber $\mathbf{k}$. The hydrodynamic modes are spatially periodic of wavelength $\lambda = 2 \pi / k$ with $k = \vert \vert {\bf k} \vert \vert$. They decay exponentially in time because the corresponding eigenvalues are real and negative. The smallest decay rate controls the long-time hydrodynamic relaxation of the ${\bf k}$-component of the density. It then gives the dispersion relation of the hydrodynamic modes of diffusion. This relation can be expressed in terms of the Van Hove function as \citep{van_hove_1954}
\begin{equation}
s_{\mathbf{k}} = \lim_{t \to \infty} \frac{1}{t}
\ln \langle \exp \lbrack i {\bf k} \cdot  ({\bf r} - {\bf r}_0) \rbrack  \rangle
= - D k^{2} \; .
\label{dispersion-relation}
\end{equation}
where $D$ is the diffusion coefficient.\footnote{For further detail on the Van Hove intermediate incoherent scattering function and its role in the kinetic theory, see, for example, \citet{resibois-book}.}
 
Until recently the hydrodynamic modes were not described in terms of the Liouvillian dynamics, but at the intermediate level of the kinetic equations. Here, we present a method based on the construction of hydrodynamic modes of diffusion in terms of the microscopic deterministic dynamics by using the concepts introduced in Section \ref{micro_approach_relax_proc}. 

Boundary conditions are required to solve the Liouville equation (\ref{liouville_equation2}). This method does not use the simple periodic boundary conditions. Although we consider here $N$-particle systems periodically extended in position space and forming a lattice, ${\cal L}^{N}$, such as the Lorentz gas, the distribution function is allowed to extend nonperiodically over the whole lattice so that the periodic boundary conditions do not apply and have to be replaced by the so-called \textit{quasiperiodic boundary conditions}.\footnote{If the time evolution of the coordinates $\mathbf{\mathbf{\Gamma}} = {\bf r}, {\bf p} = \mathbf{r}_{1},..., \mathbf{r}_{N}, \mathbf{p}_{1},...,\mathbf{p}_{N}$ of the $N$ particles is governed by a first-order equations $\dot{\mathbf{\mathbf{\Gamma}}} = \mathbf{F}(\mathbf{\mathbf{\Gamma}})$, the vector field $\mathbf{F}$ is therefore symmetric under discrete position translations: $\mathbf{F}(\mathbf{r},\mathbf{p}) = \mathbf {F}(\mathbf{r}+\mathbf{l},\mathbf{p})$ with $\mathbf{l} \in {\cal L}^{N}$.} A Fourier transform must be carried out in position space to reduce the dynamics to the cell at the origin of the lattice ($\mathbf{l} = 0$). A wavenumber $\mathbf{k}$ is introduced which varies continuously in a Brillouin zone reciprocal to the lattice,\footnote{For further detail, see, for example, \citet{ashcroft_1976}.} so that the $N$-particle probability distribution (\ref{liouville_solutions}) is rewritten as
\begin{equation}
f_{\mathbf{k}}(\mathbf{r}, \mathbf{p}, t) = \sum_{\mathbf{l} \in {\cal L}^{N}}^{} \exp (-i \mathbf{k} \cdot  \mathbf{l}) f
(\mathbf{r} + \mathbf{l} , \mathbf{p}, t) \; .
\label{fourier-transform}
\end{equation}
In particular the hydrodynamic mode of wavenumber $\mathbf k$ is an eigenstate of the operator $\hat{T}_{\mathbf{l}}$ of translation by the lattice vector $\mathbf{l}$
\begin{equation}
\hat{T}_{\mathbf{l}}\Psi_{\mathbf k}= \exp(i \mathbf{k} \cdot
\mathbf{l})\Psi_{\mathbf{k}} \; .
\end{equation}
The translation operator $\hat{T}_{\mathbf{l}}$ commutes with the Frobenius-Perron operator
\begin{equation}
\lbrack \hat{P}^t, \hat{T}_{\mathbf{l}} \rbrack = 0 \; ,
\end{equation}
so that an eigenstate $\Psi_{\mathbf k}$ common to these two operators may be found:
\begin{equation}
\hat{P}^t \Psi_{\mathbf{k}}  = \exp (s_{\bf k} t) \Psi_{\mathbf{k}} \; .
\end{equation}

The wavenumber $\mathbf k$ characterizes the spatial periodicity of the observables and of the distribution functions. Each Fourier component of the distribution functions evolves differently in time, which requires the introduction of a new Frobenius-Perron operator $\hat {R}_{\mathbf k}$ depending explicitly on the wavenumber $\mathbf k$. Since the operator of translation $\hat{T}_{\mathbf{l}}$ commutes with the Frobenius-Perron operator $R_{\mathbf k}$ we can find an eigenstate common to both the spatial translations and the time evolution
\begin{equation}
\hat{R}_{\mathbf k} \Psi_{\mathbf{k}} = \exp (s_{\mathbf k}t) \Psi_{\mathbf{k}} \; .
\end{equation}

At vanishing wavenumber, we recover the dynamics with periodic boundary conditions that admits an invariant probability measure describing the microcanonical equilibrium state. In contrast, an invariant probability measure no longer exists as soon as the wavenumber is non-vanishing. Instead, we find a complex measure which decays at a rate given by the Pollicott-Ruelle resonance $s_{\mathbf k}$. This measure defines the hydrodynamic mode of wavenumber $\mathbf k$ and the associated Pollicott-Ruelle resonance $s_{\mathbf k}$ gives the dispersion relation (\ref{dispersion-relation}) of the hydrodynamic mode.

As said in Section \ref{micro_approach_relax_proc}, such microscopic hydrodynamic modes present an important difference with the phenomenological hydrodynamics. This difference holds in the fact that they are mathematical distributions or singular measures. The impossibility of constructing eigenstates in terms of functions has its origin in the pointlike character of the deterministic dynamics and in the property of dynamical instability. 

The singular property of the eigenstates plays a fundamental role in the understanding of irreversible processes. Indeed it leads to the result that the cumulative function
\begin{equation}
F_{\bf k}(\xi) = \int_{0}^{\xi} \Psi_{\mathbf k} (\mathbf \Gamma_{\xi '}) \; d \xi' \; ,
\end{equation}
where $\mathbf \Gamma_{\xi '}$ is a curve of parameter $\xi$ in the phase space, is fractal. In the case of diffusion in the periodic Lorentz gas, thanks to the thermodynamic formalism, it has been shown that the diffusion coefficient $D$ is related to the Hausdorff dimension $d_H$ of the cumulative function $F_{\mathbf k} (\xi)$ \citep{gaspard-claus-gilbert-dorfman,GDG_2001} as
\begin{equation}
D = \lambda \lim_{\mathbf k \to 0} \frac{d_H (\mathbf k) - 1}{\mathbf{k}^2}
\end{equation}
where $\lambda$ is the positive Lyapunov exponent.\footnote{The dynamics of the pointlike particle in the two-dimensional Lorentz gas is described in a four-dimensional phase space ${\bf \Gamma} = (r_x, r_y, p_x, p_y)$. Because the system is Hamiltonian, the Liouville theorem must be satisfied. As a corollary, the sum of the four Lyapunov exponents must be vanishing. However, two of them are respectively associated with the flow direction and the perpendicular direction of the energy shell. As a result, both of them are zero, so that we have only one positive Lyapunov exponent $\lambda$.} A similar study has also been done in reactive-diffusion systems \citep{claus-gaspard-2002}.

The fractal structure of the diffusive modes also plays a crucial role in the positivity of the entropy production in Hamiltonian systems \citep{gaspard97,gilbert-dorfman-gaspard,DGG_2002}. Indeed when the distribution functions are smooth, there is no change in the Gibbs entropy and no positive irreversible entropy production. The presence of the singular character therefore appears to be the fundamental element for an understanding of the second law of thermodynamics in terms of fractals.

Beside the relaxation processes, non-equilibrium steady states are very useful to study the relationships of interest. A non-equilibrium steady state can be obtained in a system of diffusion between two reservoirs at different concentrations. For instance, let us consider the Lorentz slab depicted in Fig. \ref{Lorentz-absorbing}(a). Let us suppose that the slab separates two particle reservoirs differing in their respective phase-space densities $f_{+}$ and $f_{-}$, so that the absorbing boundary conditions introduced in the escape-rate formalism is here replaced by the so-called flux boundary conditions \citep{gaspard_1997}. The particles entering the slab moves according to the deterministic dynamics of elastic collisions and the particle distribution tends asymptotically to its stationary value imposed by the reservoirs. As a result of Liouville's theorem, the density at a specific point ${\bf \Gamma}$ inside the system is either $f_-$ or $f_+$. In order to know it, the phase-space trajectory can be computed backward until the boundary ($x = \pm \frac{L}{2}$) is reached, let us say at the time $- T(\bf \Gamma)$. The invariant measure can thus be obtained as \citep{gaspard_1997}:
\begin{equation}
f_{\bf g}({\bf \Gamma}) = \frac{f_+ + f_-}{2} + {\bf g} \cdot \left \lbrack {\bf r}({\bf \Gamma}) + \int_{0}^{- T(\bf \Gamma)} {\bf v}({\bf \Phi}^t {\bf \Gamma}) dt \right \rbrack 
\end{equation}
where ${\bf g} = {\bf e}_x \frac{f_+ - f_-}{L}$ is the density gradient.

If we suppose that $L \to \infty$ in keeping constant the density gradient, the time to reach the boundaries becomes infinite. We thus obtain \citep{gaspard96}
\begin{equation}
\Psi_{\mathbf g} = {\bf g} \cdot  \left \lbrack {\bf r} (\mathbf{\Gamma}) + \int_{0}^{- \infty} {\bf v} \left ( \mathbf{\Phi}^t \mathbf{\Gamma}\right ) dt \right \rbrack
\label{NESS_hydr_modes}
\end{equation}
Moreover, this invariant probability distribution can be equivalently obtained in terms of the hydrodynamic modes themselves as \citep{tasaki_gaspard_1995}
\begin{equation}
\Psi_{\mathbf g} = - i {\bf g} \cdot \frac{\partial \Psi_{\mathbf k}}{\partial {\bf k}} \Big \vert _{{\bf k} = 0} \;  . 
\end{equation}

As for the hydrodynamic modes, $\Psi_{\mathbf g}$ is singular as a result of the unstable character of the deterministic dynamics. This singular character can be displayed by considering the cumulative function of the non-equilibrium steady state. We then obtain the Takagi function
\begin{equation}
{\cal T}_{\mathbf g}(\xi) = \int_{0}^{\xi} \Psi_{\mathbf g} (\mathbf \Gamma_{\xi '}) \;
d \xi'
\end{equation}
where $\mathbf \Gamma_{\xi '}$ is a curve of parameter $\xi$ in the phase space.

The remarkable result expressed in Eq. (\ref{NESS_hydr_modes}) is that it does not refer to the Lorentz gas, so that this should be considered as a general expression still valid for any Hamiltonian system. As a corollary, the singular character of the non-equilibrium states also appears as a general result.

\begin{figure}[t!]
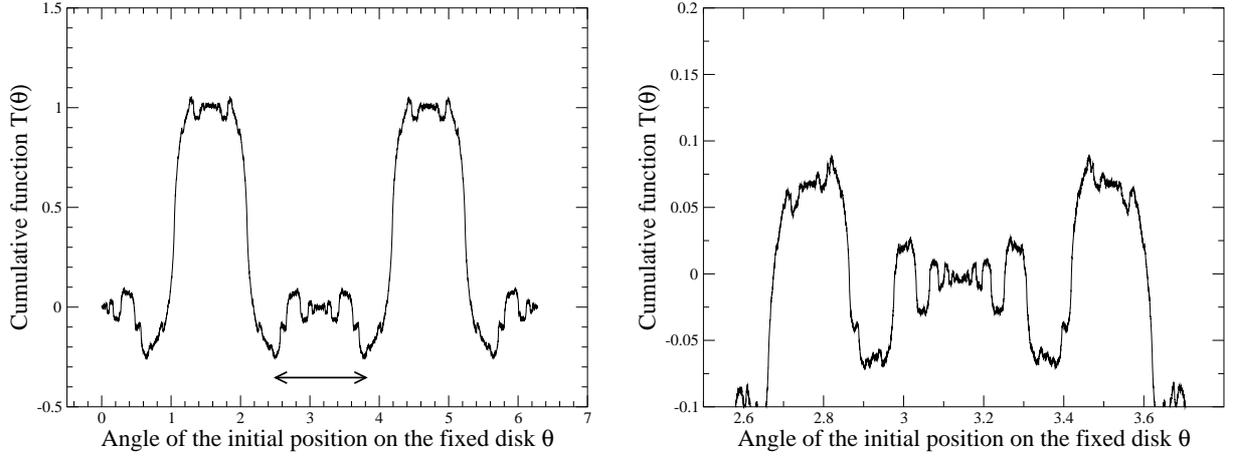

\begin{minipage}{.4\textwidth}
\hspace*{0.8cm}
\centerline{\scalebox{0.35}{\rotatebox{0}{{\includegraphics{fig11a.eps}}}}}
\end{minipage}
\hfill
\begin{minipage}{.4\textwidth}
\hspace*{-0.8cm}
\centerline{\scalebox{0.35}{\rotatebox{0}{{\includegraphics{fig11b.eps}}}}}
\end{minipage}
\caption{Left figure: fractal curve of the cumulative function of a non-equilibrium steady state corresponding to a velocity gradient in the two-hard-disk model. Right figure: enlargement of the domain underlined in the left figure. The self-similar character clearly appears and is the signature of the fractality of the cumulative function.}
\label{frac_cumul_visc}
\end{figure}

In the case of viscosity, the non-equilibrium steady state associated with a velocity gradient $g$ can be written similarly as \citep{gaspard96}
\begin{equation}
\Psi_{g}^{(\eta)} (\mathbf{\Gamma}) = g \left \lbrack  G^{(\eta)}
(\mathbf{\Gamma}) + \int_{0}^{- \infty} J^{(\eta)}
\left ( \mathbf{\Phi}^t \mathbf{\Gamma}\right ) dt \right \rbrack
\label{ness}
\end{equation}
where $G^{(\eta)} (\mathbf{\Gamma})$ is the Helfand moment, and $J^{(\eta)} = dG^{(\eta)}/dt$ the associated microscopic current. By resolving the integral, Eq. (\ref{ness}) formally becomes
\begin{equation}
\Psi_{g}^{(\eta)} (\mathbf{\Gamma}) = \lim_{t \to \infty} g \; G(\mathbf{\Phi}^{-t}
\mathbf{\Gamma}) \; .
\end{equation}
Here we want to obtain a representation of such a non-equilibrium steady state corresponding to a velocity gradient. Because of its singular character, the steady state can be represented by its cumulative function. Here we are concerned with a two-hard-disk system with periodic boundary conditions. According to \citet{buni-spohn}, it consists in the system having the lowest number of degrees of freedom in which viscosity already exists. The dynamics of the disks may be reduced to that of a pointlike particle moving in the periodic Sinai billiard, which has the coordinates of the relative position (${\bf r} = {\bf r_1} - {\bf r_2}$) and velocity (${\bf v} = {\bf v_1} - {\bf v_2}$). Let us define $\theta$ as the angle of the relative position ${\bf r}$ between both disks. Hence the Takagi function ${\cal T}_g(\theta)$ is written as
\begin{eqnarray}
{\cal T}_g(\theta) & = & \int_{0}^{\theta} \Psi_g (\theta')d\theta' \nonumber \\
& = & \lim_{t \to \infty}g \; \int_{0}^{\theta}G(\mathbf{\Phi}^{-t} \theta') d\theta' = g {\cal T}_g(\theta) 
\end{eqnarray}
and allows to reveal the singular character of the non-equilibrium steady state with a continuous curve instead of a distribution. In Fig. \ref{frac_cumul_visc} we plot ${\cal T}_g(\theta)$ in function of $\theta$. The self-similar character of the fractal is shown by zooming a domain of the complete cumulative function \citep{viscardy_book}. This clearly suggests that the extension of results obtained in the case of diffusion is justified and that the construction of hydrodynamic modes associated with viscosity at the microscopic level would lead to the expression of the positivity of the entropy production in terms of the dimension of a fractal structure.


\section{Conclusions}\label{conclusions}

In this paper we have outlined the history of statistical mechanics focusing especially on the emergence of the irreversible character of macroscopic processes from the reversible Newtonian dynamics. Viscosity has been used as a leading concept insofar as this is a quite convenient approach to draw the main steps from Maxwell's and Boltzmann's kinetic theories of gases to modern non-equilibrium statistical mechanics (see Table \ref{table_viscosity}).

\begin{table*}[t!]
\begin{center}
\begin{tabular}{|l|r|l|l|}
\hline
Viscosity expressions & Eq. & Year	& Approach\\
\hline
\hline
$\eta_M = \frac{1}{3} \; \left \langle v \right \rangle \; \rho \; \left \langle l \right \rangle $ & 
(\ref{maxwell-viscosity}) & 1860 & Mean-free path \\
$ \eta_B = 1.0162 \frac{5}{16 \sigma^2} \sqrt{\frac{m k_B T}{\pi}} $ & (\ref{chapman-enskog-viscosity})
& 1917 & Boltzmann equation \\
$ \eta_E =\eta_{B} \left ( \frac{1}{\chi} + \frac{4}{5} \; b_0 n + 0.7614 \; b_0^2
n^2 \chi \right ) $  &  (\ref{enskog-viscosity}) & 1922 & Boltzmann equation \\
$\eta_{DC} = \eta_B + \eta_1 n + \eta_2 n^2 \ln n + \eta_3 n^3 + \dots $ & (\ref{viscosity-expansion-log-density})
& 1967 & Boltzmann equation \\
$\eta_{GK} = \lim_{V \to \infty} \frac{1}{Vk_B T} \int_{0}^{\infty} \left \langle
J_{xy}(0) J_{xy}(t)\right \rangle \; dt $ & (\ref{viscosity-GK}) & 1950s & Fluctuation-dissipation theorem \\
$ \eta_{H}=\lim_{t \to \infty}\frac{1}{2t V k_B T}\; \left \langle \left[G_{xy} (t)-G_{xy}(0) \right]^{2}\right \rangle \; $
& (\ref{Einstein.shear.viscosity}) & 1960 & Fluctuation-dissipation theorem \\
$\eta_{ECM}(N,\gamma) = \frac{-3 n k_{B} T}{\gamma^{2}} \left \lbrack \lambda_{\mathrm{max}}(\gamma)
+ \lambda_{\mathrm{min}}(\gamma) \right \rbrack $  &  (\ref{visc-thermostat})  & 1990 & Dynamical systems theory \\
$\eta_{DG} = \lim_{\chi\to\infty} \left ( \frac{\chi }{\pi  } \right )^{2}
\sum_{\lambda_i>0} \left ( 1 - d_i \right ) \; \lambda_i $ & (\ref{visc-codim}) & 1995 & Dynamical systems theory \\
\hline
\end{tabular}
\end{center}
\caption{Viscosity relationships described in this paper. Four main approaches can be distinguished: (i) the mean-free path approach; (ii) the Boltzmann equation; (iii) the fluctuation-dissipation theorem; and (iv) the dynamical systems theory. The first formula (\ref{maxwell-viscosity}) derived by \citet{maxwell-1860} for dilute gases was based on \citet{clausius-1858}'s mean-free path $\left \langle l \right \rangle$. Since the latter is inversely proportional to the density, Maxwell's gas viscosity does not depend on $\rho$. Boltzmann's equation, which is central in the kinetic theory of gases, was solved for the viscosity the same year by \citet{chapman} and \citet{enskog-1917} who found another constant factor. A few years later, \citet{enskog-dense} modified the Boltzmann equation by replacing collisions between point-like particles by collisions between particles separated by a non-zero distance. He hence derived the so-called Enskog viscosity (\ref{enskog-viscosity}) which agreed with experimental measures in moderately dense gases. Later, \citet{dorfman-cohen-67} showed that, when the viscosity (\ref{viscosity-expansion-log-density}) is expressed in a power series in terms of the density $n$, the quadratic term has to be multiplied by the logarithm of the density as a consequence of correlations between molecules over large distances. In the framework of the fluctuation-dissipation theorem elaborated during the first part of the 20th century, \citet{green51,green60}, \citet{kubo57}, and \citet{mori58} established the linear-response theory which led to the Green-Kubo viscosity (\ref{viscosity-GK}) written in terms of the auto-correlation function of the microscopic flux $J_{xy}$ associated with viscosity. Furthermore, \citet{helf} derived an Einstein-like formula (\ref{Einstein.shear.viscosity}) related to the variance of the so-called Helfand moment $G_{xy}$. This relation and the Green-Kubo formula are found to be equivalent. The theory of chaos and dynamical systems theory elaborated from Poincar\'e in the 1890s provided a new approach of the greatest importance in modern non-equilibrium statistical mechanics. Assuming that the microscopic dynamics is chaotic, recent theories provided relationships between the viscosity and quantities characterizing the microscopic chaos. In particular, by means of the thermostatted-system approach, viscosity (\ref{visc-thermostat}) was expressed in terms of the maximum and minimum Lyapunov exponents ($\lambda_{\rm max}$, $\lambda_{\rm min}$). On the other hand, the espace-rate formalism proposed a formula (\ref{visc-codim}) depending on the positive Lyapunov exponents $\lambda_{i}$and the dimensions $d_i$ of the fractal resulting from the chaotic dynamics. The latter approach presents the fundamental advantage of conserving the Hamiltonian character of the microscopic dynamics.
\label{table_viscosity}}
\end{table*}

Classical mechanics was born in the 17th century, mainly with  \citet{newton-principia-1ed}'s works. During the following two centuries, other formalisms were constructed and proved to be equivalent to that of Newton, such as the Hamiltonian formalism based on the mechanical energy of the system. In this framework, the energy conservation is thus associated with the time reversible character of the Hamiltonian equations (\ref{hamilton_equations1}-\ref{hamilton_equations2}). Furthermore, Euler's hydrodynamic equations (\ref{euler_equations}), based on Newton's equations and derived in the 18th century, also exhibit the property of reversibility.

A tremendous change of paradigm started to rise in the first part of the 19th century thermodynamics pioneered by \citet{carnot}. According to the thermodynamic laws enunciated later, the conservation of the total energy (first law) was rather associated with the irreversibility of physico-chemical processes (second law). As a matter of fact, the same transition was observed in hydrodynamics. The Euler equations, which did not exhibit any dissipative character, were modified to take into account the viscous effects. This gave rise to the well-known Navier-Stokes equations (\ref{NS-equations-intro}).

At the same time, the revival of the atomic theory in chemistry initiated by \citet{Dalton1808} was more and more able to explain chemical properties of matter. On the other hand, many experimental studies performed in the 1840s refuted the previous theory of heat largely accepted since Lavoisier, namely the caloric theory. \citet{joule} and others discovered that heat (i.e. caloric) was actually not conserved, but could be converted into mechanical works and vice versa. This led to the idea that heat was nothing but a mechanical energy. This was the fundamental purpose of the kinetic theories of gases. In this context, considerable advances were due to \citet{maxwell-1860}. He showed for the first time that, contrary to the common opinion, the velocities of the microscopic particles, instead of being equal, satisfied a {\em Maxwellian distribution}. Furthermore, he predicted theoretically the startling result that {\em gas} viscosity (\ref{maxwell-viscosity}) would not be dependent on pressure or density -- prediction he later confirmed himself experimentally. This allowed \citet{loschmidt} to give the first convincing estimate of molecular properties such as their size and the Avogadro number.

\citet{carnot} founded thermodynamics with the aim of optimizing steam engines which were able to produce work by means of the heat transfered from a hot reservoir to a colder one. In this context, \citet{clausius-1854,clausius-1865} defined the entropy in terms of heat transferred and thus showed that the entropy always increased whenever heat flowed from a hot body to a cold body. Later, \citet{boltzmann-1872} derived his famous kinetic equation and introduced a $H$ function in terms of a one-particle distribution and which could only decrease in systems out of equilibrium. In response to many objections,  \citet{boltzmann_1877} proposed a more general definition of entropy (\ref{boltzmann_entropy}) measuring the {\em probability} of finding a macrostate in terms of the number of corresponding microstates. In other words, Eq. (\ref{boltzmann_entropy}) is a measure of the disorder of the system. It implies that the thermodynamic equilibrium is reached when this probability is maximum, i.e., when the disorder is maximum.

Although Boltzmann obtained his kinetic equation in 1872, some mathematical difficulties prevented scientists until \citet{chapman} and \citet{enskog-1917} from deriving transport coefficients at the kinetic level. Like \citet{maxwell-1860}'s viscosity (\ref{maxwell-viscosity}), the so-called {\em Boltzmann viscosity} (\ref{chapman-enskog-viscosity}) also exhibits a dependence on the square root of the temperature. On the other hand, it does not depend on density. In order to extend the density range of validity, \citet{enskog-dense} modified the binary collision term by rejecting the pointlike character of the particles, and then obtained the Enskog viscosity in dense gases (\ref{enskog-viscosity}). Moreover, \citet{dorfman-cohen-67} highlighted the existence of correlations at long distances -- larger than the range of the intermolecular interactions -- which contributed to a logarithmic-density dependence of the viscosity (\ref{viscosity-expansion-log-density}).

In 1884, Gibbs thought that some advances realized in the kinetic theory of gases, especially due to Boltzmann between 1868 and 1871, were such that a new phrase would have to be used \citep{gibbs_1884}; namely, {\em statistical mechanics}. He also emphasized the importance of Liouville's theorem that he rewrote as the well-known Liouville equation (\ref{liouville_equation1}). This is on the basis of this fundamental equation that he built his abstract mathematical theory describing a series of ensembles of similar systems at equilibrium. Instead of using the ergodic hypothesis underlying Boltzmann's approach, he introduced the concept of mixing to garantee the relaxation of systems toward the equilibrium. In addition, mixing condition is stronger than ergodicity and implies it.

The erratic motion of small grains immerged in a fluid was already discovered at the end of the 17th century \citep{gray}. A vital force was thought to be at the origin of the this behavior until \citet{brown} who rejected this biological cause in favor of a physical cause. However, it is to \citet{einstein} that we owe the modern explanation of the phenomenon, initiating thereby the fluctuation theory. In this framework, the so-called {\em fluctuation-dissipation theorem} was developed and led to expressions of viscosity (\ref{viscosity-GK}) in terms of the autocorrelation of the associated microscopic flux \citep{green51,green60,kubo57}. In addition, \citet{helf} obtained an Einstein-like relation (\ref{Einstein.shear.viscosity}) which involved the variance of the so-called {Helfand moment} (\ref{helfand_moment_viscosity}). Both expressions (\ref{viscosity-GK}) and (\ref{Einstein.shear.viscosity}) are exactly equivalent (see \ref{appendixA}).

At the end of the 19th century, a new paradigm emerged with Maxwell, Poincar\'e, Hadamard, and others who already concluded that the principle ``like causes produce like effects" was not satisfied in systems that are said dynamically {unstable}. However the actual breakthrough did not appear before the 1960s with the seminal works of Lorenz, Smale, Ruelle, and many others, which gave rise to the {\em Chaos theory}. Although classical physics opposed determinism and probability, chaos theory showed that both were compatible due to the unstable character of --- even simple --- deterministic systems.

In the middle of the 20th century, a milestone was reached with \citet{krylov_1944} who connected the need of mixing property as a relaxation condition on the one hand, and the fact that the dynamical systems had to be unstable to observe relaxation. This idea was developed by Sinai and others, which thus defined the basis of modern non-equilibrium statistical mechanics.

In this context, we have outlined recent theories of non-equilibrium statistical mechanics. The widespread thermos\-tatted-system approach is based on systems with non-equilibrium constraints. In the case of the viscosity, the particle systems is sheared, so that the phase space is reduced to a fractal attactor which is the signature of chaotic, dissipative systems. Hence the shear viscosity (\ref{visc-thermostat}) is expressed in terms of the sum of the Lyapunov exponents which is no longer vanishing. On the other hand, it means that this phase space contraction violates the Liouville theorem (\ref{volume_preserve}) which is a fundamental property of Hamiltonian systems.

By contrast, the escape-rate formalism, initiated by \citet{gasp-nicolis} for diffusion and later extended for the other transport coefficients \citep{dorf-gasp,gasp-dorf}, introduces non-equilibrium (absorbing boundary) conditions which induces an escape process of the phase-space trajectories. Despite this process, an infinite number of trajectories remains forever within the escape limits and formed a fractal repeller. This is therefore used as a support of an invariant measure. Viscosity (\ref{visc-codim}) is then expressed as the difference between the sum of the positive Lyapunov exponents and the Kolmogorov-Sinai entropy. In contrast to the thermostatted-system approach, the dynamics is still Hamiltonian, so that the Liouville theorem remains satisfied. 

These new approaches of the emergence of irreversibility provide remarkable advances. While Boltzmann proposed a pure {\em probabilistic} interpretation of irreversibility of relaxation processes, the dynamical systems theory has brought crucial elements in order to interpretate this emergence in a {\em dynamical} way. Furthermore, with the escape-rate formalism and the hydrodynamic mode method, irreversibility emerges from dynamical systems without violating the fundamental Liouville theorem, that is to say, without loosing the expected Hamiltonian character of the microscopic dynamics.\\

\noindent
{\bf Acknowledgments:} The author thanks Massimiliano Esposito and Pierre Gaspard for support and helpful discussions.


\appendix


\section{Phenomenological approach to viscosity}\label{phen_appr_visc}

The Navier-Stokes equations (\ref{NS-equations-intro}) introduced in Section \ref{hydro-visc} are the central equations in hydrodynamics. First, let us write them in the modern way. Developing the total differential $\rho \frac{d v_i}{dt}$  and using the continuity equation (\ref{continuity-equation}), we get the general form
\begin{equation}
\frac{\partial \rho v_{i}}{\partial t}
=-\frac{\partial \Pi_{ij}}{\partial r_{j}} \; ,
\label{navier-stokes}
\end{equation}
where the \textit{momentum flux density tensor} $\Pi_{{ij}}$ is written as
\begin{equation}
\Pi_{ij}= \rho \, v_{i}v_{j} + P \; \delta_{ij} - \sigma'_{ij}\; .
\end{equation}

Let us consider a $d$-dimensional fluid which may be anisotropic, so that the friction may depend on the selected direction. The viscous stress tensor  depends linearly on the velocity gradient tensor in the framework of Newtonian fluids. The most general quantity relating two second-order tensors is a fourth-order tensor
\begin{equation}
\sigma'_{ij}=\eta_{ij,kl} \; \frac{\partial v_{k}}{\partial r_{l}} \; .
\label{visc-stress-tensor}
\end{equation}
$\eta_{ij,kl}$ is the \textit{viscosity tensor}. This is the most general expression for the viscous stress tensor including anisotropic as well as isotropic systems for Newtonian fluids. The theory of Cartesian tensors shows that the basic isotropic tensor is the Kronecker tensor $\delta_{ij}$ and that all the isotropic tensors of even orders can be written like a sum of products of tensors $\delta_{ij}$
\begin{equation}
\eta_{ij,kl}=a \; \delta_{ij} \;\delta_{kl}
+ b \; \delta_{ik} \;\delta_{jl} + c \; \delta_{jk} \;\delta_{il}\; ,
\end{equation}
where $a,b$ and $c$ are scalars.\footnote{See  \citet{aris}.} Since the viscous stress tensor is symmetric $\sigma'_{ij}=\sigma'_{ji}$, we find that $b=c$ so that only two of these coefficients are independent. After a rearrangement we obtain the expression (\ref{isotropic-stress-tensor}). The coefficients $\eta=b$ and $\zeta=a+(2/d)b$ are respectively the \textit{shear and bulk viscosities} and they can be expressed in terms of the elements of the fourth-order viscosity tensor as
\begin{eqnarray}
\eta &=& \eta_{xy,xy}\; ,\nonumber \\
\zeta &=& \frac{1}{d}\; \eta_{xx,xx} + \frac{d-1}{d}\;
\eta_{xx,yy}\; .
\end{eqnarray}
As a result, the viscous stress tensor $\sigma'_{ij}$ in an isotropic $d$-dimensional fluid is written as
\begin{equation}
\sigma'_{ij}=\eta \left ( \frac{\partial v_{i}}{\partial r_{j}}
+ \frac{\partial v_{j}}{\partial r_{i}} - \frac{2}{d}\;
\delta_{ij}\, \frac{\partial v_{l}}{\partial r_{l}} \right )
+ \zeta \; \delta_{ij} \,  \frac{\partial v_{l}}{\partial r_{l}}\; ,
\label{isotropic-stress-tensor}
\end{equation}

The two viscosity coefficients must respect thermodynamic laws, especially the second law. According to \citet{prigogine_1947}, the variation of entropy $dS$ of a system may be written as the sum of two terms:
\begin{equation}
dS = d_{\rm e} S + d_{\rm i} S
\label{prigogine_entropy}
\end{equation}
which represent respectively the entropy exchange with the surroundings and the internal production of entropy satisfying the second law of thermodynamics:
\begin{equation}
\frac{d_{\rm i} S}{dt} \geq 0 \; .
\label{pos_entr_pdt}
\end{equation}

Let us rewrite the entropy in terms of the entropy per unit mass $s$:
\begin{equation}
S = \int_V \rho s dV
\end{equation}
and the time variation of entropy produced inside the system in terms of the entropy production $\sigma$:
\begin{equation}
\frac{d_{\rm i}S}{dt} = \int_V \sigma dV \; ,
\end{equation}
The time variation of the entropy of the system due to the exchange with its environment may take the form:
\begin{equation}
\frac{d_{\rm e}S}{dt} = - \int_\Sigma J_{s,i}^{\rm tot} \ n_i \ d\Sigma = - \int_V \frac{\partial J_{s, i}^{\rm tot}}{\partial r_i}  \ dV \; ,
\end{equation}
where ${\bf J}_{s}^{\rm tot}$ is the total entropy flow per unit area and unit time. The Gauss-Ostrogradsky theorem has been used to obtain the second equality.  However, Eq. (\ref{prigogine_entropy}) as well as Eq. (\ref{pos_entr_pdt}) must remain valid for any volume $V$, so that Eq. (\ref{prigogine_entropy}) may be rewritten as the balance equation for entropy density $\rho s$:
\begin{equation}
\frac{\partial \rho s}{dt} = - \frac{\partial}{\partial r_i} \left ( J_{s, i} + \rho s v_i \right) + \sigma \; 
\end{equation}
where
\begin{equation}
\sigma \geq 0
\end{equation}
and the total entropy flow has been replaced by the sum of the diffusive and the convective terms, respectively ${\bf J}_{s}$ and $\rho s {\bf v}$.

Let us consider the entropy production for an isotropic non-reactive system composed of only one component\footnote{See \citet{vidal_1994}}
\begin{equation}
\frac{\partial (\rho s)}{\partial t} + \frac{\partial}{\partial r_j} (\rho s v_j)= \frac{1}{T} \sigma_{ij}^{'} \frac{\partial v_i}{\partial r_j}
+ \frac{1}{T^2} \kappa \left ( \frac{\partial T}{\partial r_i} \right )^2 \geq 0
\label{balance-eq-entropy1}
\end{equation}
where $\kappa$ is the heat conductivity. Let us consider the second term introducing the viscous stress tensor for isotropic systems. By replacing $\sigma_{ij}^{'}$ by Eq. (\ref{visc-stress-tensor}), we have
\begin{eqnarray}
\sigma_{ij}^{'} \frac{\partial v_i}{\partial r_j} & = & \eta \left ( \frac{\partial v_i}{\partial r_j} \frac{\partial v_i}{\partial r_j} + \frac{\partial v_j}{\partial r_i} \frac{\partial v_i}{\partial r_j} -\frac{2}{d}\delta_{ij}
\frac{\partial v_i}{\partial r_j} \frac{\partial v_l}{\partial r_l} \right )
+ \zeta \delta_{ij} \frac{\partial v_i}{\partial r_j} \frac{\partial v_l}{\partial r_l}\nonumber \\
& = & \eta \left \lbrack \left ( \frac{\partial v_i}{\partial r_j} \right )^2 + \frac{\partial v_i}{\partial r_j} \frac{\partial v_j}{\partial r_i} \right \rbrack - \frac{2}{d} \eta \left ( \frac{\partial v_l}{\partial r_l} \right )^2 + \zeta \left ( \frac{\partial v_l}{\partial r_l} \right )^2
\end{eqnarray}
where the product $\delta_{ij} \frac{\partial v_i}{\partial r_j}$ gives the divergence of the velocity vector
\begin{equation}
\delta_{ij} \frac{\partial v_i}{\partial r_j} = \frac{\partial v_l}{\partial r_l} \; .
\end{equation}
By decomposing the first term as follows:
\begin{equation}
\eta \left \lbrack \left ( \frac{\partial v_i}{\partial r_j} \right )^2 + \frac{\partial v_i}{\partial r_j} \frac{\partial v_j}{\partial r_i} \right \rbrack
= \frac{\eta}{2}  \left ( \frac{\partial v_i}{\partial r_j} + \frac{\partial v_j}{\partial r_i} \right )^2 \; ,
\end{equation}
we obtain
\begin{equation}	
\sigma_{ij}^{'} \frac{\partial v_i}{\partial r_j} = \frac{\eta}{2}   \left ( \frac{\partial v_i}{\partial r_j} + \frac{\partial v_j}{\partial r_i} \right )^2 - \frac{2}{d} \eta \left ( \frac{\partial v_l}{\partial r_l} \right )^2 + \zeta \left ( \frac{\partial v_l}{\partial r_l} \right )^2,
\end{equation}
and this allows us to rewrite the balance equation for the entropy density (\ref{balance-eq-entropy1}) as
\begin{eqnarray}
\frac{\partial (\rho s)}{\partial t} + \frac{\partial}{\partial r_j} (\rho s v_j) = \frac{\eta}{2T} \left ( \frac{\partial v_i}{\partial r_j} + \frac{\partial v_j}{\partial r_i} \right )^2 - \frac{2 \eta}{T d} \left ( \frac{\partial v_l}{\partial r_l} \right )^2 
+ \frac{\zeta}{T} \left ( \frac{\partial v_l}{\partial r_l} \right )^2 +  \frac{1}{T^2} \kappa \left ( \nabla T \right )^2 \geq  0 \; .
\label{entropy-balance1}
\end{eqnarray}

Let us consider the terms where the shear viscosity $\eta$ appears and let us put $\frac{\eta}{2 T}$ in evidence:
\begin{equation}
\frac{\eta}{2 T} \left \lbrack \left ( \frac{\partial v_i}{\partial r_j} + \frac{\partial v_j}{\partial r_i} \right )^2 
- \frac{4}{d} \left ( \frac{\partial v_l}{\partial r_l} \right )^2 \right \rbrack \equiv \frac{\eta}{2 T} \; \mathcal{G} \; .
\end{equation}
We then have
\begin{eqnarray}
\mathcal{G} & = & \left ( \frac{\partial v_i}{\partial r_j} + \frac{\partial v_j}{\partial r_i} \right )^2 
- \frac{2}{d} \delta_{ij} \left ( \frac{\partial v_i}{\partial r_j} + \frac{\partial v_j}{\partial r_i} \right )
\left ( \frac{\partial v_l}{\partial r_l} \right ) \nonumber \\
& = & \left ( \frac{\partial v_i}{\partial r_j} + \frac{\partial v_j}{\partial r_i} \right )^2 
- \frac{4}{d} \delta_{ij} \left ( \frac{\partial v_i}{\partial r_j} + \frac{\partial v_j}{\partial r_i} \right )
\left ( \frac{\partial v_l}{\partial r_l} \right ) +  \left (\frac{2}{d} \; \delta_{ij} \frac{\partial v_l}{\partial r_l} \right )^2
\end{eqnarray}
where we have used the fact that
\begin{equation}
\delta_{ij} \left ( \frac{\partial v_i}{\partial r_j} + \frac{\partial v_j}{\partial r_i} \right ) = 2 \; \frac{\partial v_l}{\partial r_l}
\end{equation}
and
\begin{equation}
\delta_{ij} \; \delta_{ij} = d \; .
\end{equation}
We may then gather the terms as
\begin{equation}
\mathcal{G} = \left ( \frac{\partial v_i}{\partial r_j} + \frac{\partial v_j}{\partial r_i}
- \frac{2}{d} \; \delta_{ij} \frac{\partial v_l}{\partial r_l}\right )^2 \; .
\end{equation}
Finally Eq. (\ref{entropy-balance1}) can be rewritten as
\begin{equation}
\frac{\partial (\rho s)}{\partial t} + \frac{\partial}{\partial r_j} (\rho s v_j)= \frac{\eta}{2T} \left (  \frac{\partial v_i}{\partial r_j} + \frac{\partial v_j}{\partial r_i}
- \frac{2}{d} \; \delta_{ij} \frac{\partial v_l}{\partial r_l} \right )^2 + \frac{\zeta}{T} \left ( \frac{\partial v_l}{\partial r_l} \right )^2
+ \frac{1}{T^2} \kappa \left ( \nabla T \right )^2 \geq 0 \; .
\label{entropy-balance2}
\end{equation}
Hence, we obtain the condition of positivity for $\eta$, $\zeta$, and $\kappa$:
\begin{equation}
\eta \geq 0, \quad \zeta \geq 0, \quad \kappa \geq 0 \; .
\label{transport-coefficients-positivity}
\end{equation}


\section{Probability measure} \label{measure}

A probability measure $\mu$ can be associated with the probability density $f$. This probability measure over any volume $\cal A$ of the phase space $\cal M$ assigns nonnegative numbers to any set in ${\cal M}$, is countably additive, and assigns the number 1 to ${\cal M}$. 
\begin{eqnarray}
\forall {\cal A} \subset {\cal M} \ & : & \ \mu({\cal A}) > 0 \nonumber \\
\forall {\cal A}, {\cal B} \subset {\cal M} \vert {\cal A} \cap {\cal B} = \emptyset \ & : & \ \mu ({\cal A} + {\cal B}) = \mu ({\cal A}) + \mu ({\cal B}) \\
&& \mu ({\cal M}) = 1 \nonumber
\end{eqnarray}
The relation between a measure and its corresponding probability density is given by
\begin{equation}
f({\bf X}) = \frac{d\mu}{d{\bf X}}
\end{equation}

A measure is said to be {\em invariant} if it is stationary under the time evolution of the system
\begin{equation}
\mu_{\rm i} (d{\bf X}_t) = \mu_{\rm i} (d{\bf X}_0)
\end{equation}
Accordingly, such a measure corresponds to a probability density $f_{\rm i}$ which a stationary solution of the Liouville equation (\ref{liouville_equation1})
\begin{equation}
\frac{\partial f_{\rm i}}{\partial t} = \hat{L} f_{\rm i} = 0
\end{equation}
where $\hat{L}$ is the so-called {\em Liouville operator}
\begin{equation}
\hat{L}(f) = \lbrace H,f \rbrace = \frac{\partial H}{\partial {\bf q}}\frac{\partial f}{\partial {\bf p}} - \frac{\partial H}{\partial {\bf p}}\frac{\partial f}{\partial {\bf q}}
\end{equation}

An example of such a measure is the Liouville measure $d{\bf X} = d{\bf q} d{\bf p}$ since Hamiltonian systems have the foundamental property of phase-space volume conservation (i.e, the Liouville theorem (\ref{volume_preserve})). In addition, for isolated Hamiltonian systems (i.e., systems with a fixed energy $H = E$), an invariant measure corresponds to the microcanonical probability density $f_{\rm mc} ({\bf \Gamma})$ given in Eq. (\ref{mc_dens_prob}):
\begin{equation}
d \mu_{\rm mc} = \frac{1}{\omega (E)} \; \delta ( H({\bf \Gamma}) - E) \; d{\bf \Gamma}
\end{equation}
and is thus usually called {\em microcanonical invariant measure}.


\section{Microscopic expression of the viscosity}\label{micro_visc}

At the microscopic level, atoms and molecules evolve in time according to Newton's equation of motion
\begin{eqnarray}
\frac{d \mathbf r_a}{dt} & = & \frac{\mathbf p_a}{m} \nonumber \\
\frac{d \mathbf p_a}{dt} & = & \sum_{b \neq a}^{} \mathbf F(\mathbf r_a - {\bf r}_b)
\label{motion-equation}
\end{eqnarray}
where $a, b = 1, \ldots N$ ($N$ being the number of particles in the system).
We may express the momentum density in terms of microscopic variables as
\begin{equation}
\hat{\mathbf g}(\mathbf r) = \sum_{a = 1}^{N} \mathbf p_a \delta (\mathbf r - \mathbf r_a) \; .
\label{momentum-density}
\end{equation}

If we introduce a smooth test function $f(\mathbf r)$ which is time independent, Eq. (\ref{momentum-density}) becomes
\begin{eqnarray}
\int_{}^{}d \mathbf{r} f(\mathbf r) \hat{\mathbf g}(\mathbf r)
& = & \int_{}^{}d \mathbf r f(\mathbf r)
\sum_{a}^{} \mathbf p_a \delta (\mathbf r - \mathbf r_a) \nonumber \\
& = & \sum_{a}^{} \mathbf p_a f(\mathbf r_a) \; .
\label{f-intro}
\end{eqnarray}

Let us take the following definition for the \textit{microscopic momentum current density} $\hat{\tau}_{ij}$
\begin{equation}
\frac{\partial \hat{g}_i}{\partial t} + \frac{\partial \hat{\tau}_{ij}}{\partial r_j} = 0
\label{stress-tensor-definition}
\end{equation}
appearing in the equation of momentum conservation. By multiplying this equation by $f(\mathbf r)$ and by integrating
over $\mathbf r$, one has
\begin{eqnarray}
\int_{}^{} d \mathbf r f(\mathbf r) \frac{\partial \hat{g}_i}{\partial t} & = & - \int_{}^{} d \mathbf r f(\mathbf r) \frac{\partial \hat{\tau}_{ij}}{\partial r_j}
\nonumber \\
& = & - \int_{}^{} d \mathbf r \left \lbrack \frac{\partial}{\partial r_j} \left ( f \hat{\tau}_{ij} \right )
- \frac{\partial f}{\partial r_j} \hat{\tau}_{ij} \right \rbrack \nonumber \\
& = & - \int_{}^{} f \; \hat{\tau}_{ij} \; d A_j  + \int_{}^{} d \mathbf r \frac{\partial f}{\partial r_j} \hat{\tau}_{ij}
\end{eqnarray}
where $d A_j$ is an element of area perpendicular to the axis $r_j$. This boundary term vanishes because $f(\mathbf r) \to 0$ for $\mathbf r \to \infty$.
By using Eq. (\ref{f-intro}), we then get
\begin{eqnarray}
\int_{}^{} d \mathbf r f(\mathbf r) \frac{\partial \hat{g}_i}{\partial t} & = & \int_{}^{} d \mathbf r \frac{\partial f}{\partial r_j} \hat{\tau}_{ij}\nonumber \\
& = & \frac{d}{dt} \sum_{a}^{} f(\mathbf r_a)p_{ai}\nonumber \\
& = &\sum_{a}^{} \frac{df(\mathbf r_a)}{dt} p_{ai} + \sum_{a}^{} f(\mathbf r_a)\frac{dp_{ai}}{dt} \; .
\label{impulsion-dvlpmt}
\end{eqnarray}
First, consider the first term
\begin{equation}
\sum_{a}^{} \frac{df(\mathbf r_a)}{dt} p_{ai} = \frac{1}{m} \sum_{a}^{} \nabla f(\mathbf r_a) \cdot \mathbf{p}_a \; p_{ai} 
\end{equation}
knowing that $f$ is a time-independent function and that $\frac{d \mathbf r_a}{dt} = \frac{1}{m}\mathbf{p}_a$. The second term of Eq. (\ref{impulsion-dvlpmt}) is developed as follows
\begin{eqnarray}
\sum_{a}^{} f(\mathbf r_a)\frac{dp_{ai}}{dt} & = & \sum_{a, b \neq a} f(\mathbf r_a) F_i (\mathbf r_a - \mathbf r_b)  \\
& = & \frac{1}{2} \sum_{a, b \neq a} \left \lbrack f(\mathbf r_a) - f(\mathbf r_b) \right \rbrack
F_i(\mathbf r_a - \mathbf r_b) \nonumber
\end{eqnarray}
where we use the equation of motion Eq. (\ref{motion-equation}), and $F_i(\mathbf r_a - \mathbf r_b) = - F_i(\mathbf r_b - \mathbf r_a)$ since the forces of interaction between particles are central. Hence Eq. (\ref{impulsion-dvlpmt}) becomes
\begin{equation}
\frac{d}{dt} \sum_{a}^{} f(\mathbf r_a)p_{ai} = \frac{1}{m} \sum_{a}^{} \nabla f(\mathbf r_a) \cdot \mathbf{p}_a \; p_{ai} + \frac{1}{2} \sum_{a, b \neq a} \left \lbrack f(\mathbf r_a) - f(\mathbf r_b) \right \rbrack F_i(\mathbf r_a - \mathbf r_b)
\label{impulsion-dvlpmt-2}
\end{equation}

Let us consider a pair of particle $a \neq b$. We introduce an arbitrary smooth curve $\lambda \rightarrow \mathbf r_{ab}(\lambda)$ such that
$\mathbf r_{b} = \mathbf r_{ab}(0)$ and $\mathbf r_{a} = \mathbf r_{ab}(1)$. Thanks to this curve, we can express
$f(\mathbf r_a) - f(\mathbf r_b)$ as
\begin{eqnarray}
f(\mathbf r_a) - f(\mathbf r_b) & = & f( \mathbf r_{ab}(1)) - f(\mathbf r_{ab}(0) ) \nonumber \\
& = & \int_{0}^{1} \frac{d f(\mathbf r_{ab}(\lambda))}{d \lambda} d \lambda \nonumber \\
& = & \int_{0}^{1} \nabla f(\mathbf r_{ab}(\lambda)) \cdot \frac{d \mathbf r_{ab}}{d \lambda}d \lambda \; .
\end{eqnarray}
Consequently, Eq. (\ref{impulsion-dvlpmt-2}) is rewritten as
\begin{equation}
\frac{d}{dt} \sum_{a}^{}f(\mathbf r_a)p_{ai}  =  \frac{1}{m} \sum_{a}^{} \nabla f(\mathbf r_a) \cdot \mathbf{p}_a \; p_{ai} + \frac{1}{2} \sum_{a, b \neq a} \int_{0}^{1}d \lambda \frac{d \mathbf r_{ab}}{d \lambda} \cdot \nabla f(\mathbf r_{ab}(\lambda)) F_i(\mathbf r_a - \mathbf r_b) \; .
\end{equation}

Let us introduce the delta function $\delta (\mathbf r - \mathbf r_a)$ as follows
\begin{equation}
\frac{d}{dt} \sum_{a}^{}f(\mathbf r_a)p_{ai} = \frac{1}{m} \int_{}^{}d \mathbf r \nabla f(\mathbf r) \cdot \sum_{a}^{} \mathbf p_a p_{ai} \; \delta(\mathbf r - \mathbf r_a) + \frac{1}{2} \int_{}^{} d \mathbf r \sum_{a, b \neq a} \int_{0}^{1} d \lambda \frac{d \mathbf r_a}{d \lambda} \cdot
\nabla f(\mathbf r) \times F_{i}(\mathbf r_a - \mathbf r_b) \; \delta (\mathbf r - \mathbf r_{ab} (\lambda)) 
\end{equation}
so that
\begin{equation}
\frac{d}{dt} \sum_{a}^{}f(\mathbf r_a)p_{ai} = \int_{}^{}d \mathbf r \frac{\partial f(\mathbf r)}{\partial r_j}
\left \lbrace \frac{1}{m} \sum_{a}^{} p_{ai} p_{aj} \delta (\mathbf r - \mathbf r_a)
+ \frac{1}{2} \sum_{a, b \neq a} \int_{0}^{1} d \lambda \frac{d r_{abj}}{d \lambda} F_{i}(\mathbf r_a - \mathbf r_b)
\delta (\mathbf r - \mathbf r_{ab}(\lambda)) \right \rbrace \; .
\end{equation}
With Eq. (\ref{stress-tensor-definition}), we then find the following expression for the microscopic momentum current density
\begin{equation}
\hat{\tau}_{ij} = \frac{1}{m} \sum_{a}^{} p_{ai} p_{aj} \delta (\mathbf r - \mathbf r_{a})
+ \frac{1}{2} \int_{0}^{1} d \lambda \;  F_{i}(\mathbf r_a - \mathbf r_b) \frac{d r_{abj}}{d \lambda} \delta (\mathbf r - \mathbf r_{ab}(\lambda)) \; . 
\end{equation}

Let us take the integral over a volume $V$ of $\tau_{ij}$
\begin{eqnarray}
J_{ij}(t) & = & \int_{V}^{} d\mathbf r \tau_{ij}(\mathbf r,t) \nonumber \\
& = & \sum_{a}^{} \frac{1}{m} p_{ai} p_{aj} \int_{V}^{} d \mathbf r \delta(\mathbf r - \mathbf r_{a})
+ \frac{1}{2} \sum_{a, b \neq a} F_i(\mathbf r_a - \mathbf r_b)
\int_{0}^{1} dr_{abj}  \int_{V}^{} \frac{d \mathbf r}{d \lambda} \delta(\mathbf r - \mathbf r_{ab}(\lambda)) \; .
\end{eqnarray}
Finally we get the microscopic current
\begin{equation}
J_{ij}(t) = \sum_{a} \frac{1}{m} p_{ai} p_{aj} + \frac{1}{2} \sum_{a, b \neq a} F_i(\mathbf r_a - \mathbf r_b) (r_{aj} - r_{bj})
\label{flux}
\end{equation}
which enters in the Green-Kubo formula for the shear viscosity (\ref{viscosity-GK}).


\section{Proof of the equivalence between Green-Kubo and Einstein-Helfand formulas}
\label{appendixA}

Our aim is here to deduce the Green-Kubo formula (\ref{viscosity-GK}) from the Einstein-Helfand formula (\ref{Einstein.shear.viscosity}), proving the equivalence between both formulas under the conditions that (i) the Helfand moment is defined by the time integral of the microscopic current (\ref{mic_stress_tensor})
\begin{equation}
G_{ij}(t) = G_{ij}(0) + \int_0^t J_{ij}(\tau) \; d\tau
\label{helfand-integral}
\end{equation}
and (ii) the further condition that the time auto-correlation functions decrease fast enough.

The viscosity tensor is expressed in terms of the generalized Helfand moment $G_{ij}$ through the Einstein-Helfand formula
\begin{equation}
\eta_{ij,kl}=\lim_{t \to \infty} \frac{\beta}{2t V}\; \left[ \langle
G_{ij}(t)G_{kl}(t)
\rangle - \langle G_{ij}(t) \rangle \langle G_{kl}(t) \rangle \right]
\label{Einstein}
\end{equation}
Let us start from Eq. (\ref{Einstein}) with
\begin{equation}
\delta G_{ij}(t) = \int_0^t \delta J_{ij}(\tau) \; d\tau \; ,
\end{equation}
$\delta J_{ij}$ being defined by
\begin{equation}
\delta J_{ij}(t) = \int_V d{\bf r} \; \delta\tau_{ij}({\bf r},t) =
J_{ij}(t) - \langle
J_{ij}\rangle_{\rm eq} \; ,
\label{dJ}
\end{equation}
and supposing for simplicity that $\delta G_{ij}(0)=0$.  Accordingly, we have successively from Eq. (\ref{Einstein}) that
\begin{eqnarray}
\eta_{ij,kl} &=& \lim_{T\to\infty} \frac{\beta}{2TV} \; \langle \delta G_{ij}(T)
\delta G_{kl}(T) \rangle \nonumber\\
&=& \lim_{T\to\infty} \frac{\beta}{2TV} \; \int_0^T dt_1 \int_0^T dt_2 \;
\langle \delta J_{ij}(t_1) \delta J_{kl}(t_2) \rangle \nonumber\\
&=& \lim_{T\to\infty} \frac{\beta}{2TV} \; \int_{-T}^{+T} dt 
\int_{\vert t\vert/2}^{T-\vert t\vert/2} d\tau
\; \langle \delta J_{ij}(0) \delta J_{kl}(t) \rangle \nonumber\\
&=& \lim_{T\to\infty} \frac{\beta}{2V} \; \int_{-T}^{+T} dt 
\left( 1 - \frac{\vert t\vert}{T}\right) 
\; \langle \delta J_{ij}(0) \delta J_{kl}(t) \rangle \nonumber\\
&=&  \frac{\beta}{V} \; \int_{0}^{+\infty} dt 
\; \langle \delta J_{ij}(0) \delta J_{kl}(t) \rangle \; ,
\end{eqnarray}
where we have performed the change of integration variables
\begin{eqnarray}
t & = & t_2-t_1 \; , \nonumber \\
\tau & = & \frac{t_1+t_2}{2} \; ,
\end{eqnarray}
and supposed that
\begin{equation}
\lim_{T\to\infty} \frac{1}{T} \; \int_{-T}^{+T} dt \; \vert t\vert \; \langle \delta J_{ij}(0) \delta J_{kl}(t) \rangle = 0 \; ,
\end{equation}
which requires that the time autocorrelation functions decrease faster than
$\vert t \vert^{-1-\epsilon}$ with $\epsilon>0$. Q.E.D.


\section{Anosov systems and Axiom-A systems}\label{anosov_systems}

We here present some properties of dynamical systems which are useful to investigate non-equilibrium processes in statistical mechanics.\footnote{For further information on these properties, see, for example, \citet{ruelle_1989b} and \citet{gasp-book}.} Among those properties is found the concept of wandering motions in phase space which has been introduced by \citet{birkhoff_1927b,birkhoff_1927a}. A point $\bf X$ in phase space is {\em wandering} if an open set $\cal U \ni {\bf X}$ can be found such that $\cup_{|t| > t_0} {\bf \Phi}^t {\cal U} \cap {\cal U} = \emptyset $ for some $t_0 > 0$. This condition means that, after a certain time $t_0$, all points belonging to the neighborhood $\cal U$ escape it forever. Accordingly, the subset made up of all wandering points is called {\em wandering set}. At the opposite, the complementary set is the {\em nonwandering set} $\cal W_{\bf \Phi}$ of points $\bf X$ which satisfy the following condition: if, for any set $\cal U$ around $\bf X$ and for any $t_0 > 0$, there exists a time $t> t_0$ such that ${\bf \Phi}^t {\cal U} \cap {\cal U} \not= \emptyset$. In other words, after a finite time, the phase-space trajectory starting from such a point $\bf X$ comes back arbitrarily close to $\bf X$ as a consequence of Poincar\'e's recurrence theorem already mentioned in Section \ref{Boltzmann_stat_irrev}. In terms of the measure, this condition can be reformulated  as follows: $\mu({{\bf \Phi}^t \cal U}) \mu({\cal U}) > 0$, which thus implies that the condition of nonwandering set is required to have mixing (\ref{mixing_def}).

A subset of phase space is said to be {\em hyperbolic} when all the points of this subset are of saddle type, i.e., the tangent space of a trajectory in each point $\bf X$ of the subset can be decomposed into the stable and unstable subspaces together with a centre subspace containing only the direction of the flow: $\mathcal{T} \bf \mathcal{M}(\bf X) = \mathcal{E}_s(\bf X) \oplus \lbrace \bf F (\bf X)\rbrace \oplus \mathcal{E}_u(\bf X)$. The property of hyperbolicity implies that trajectories of such systems have nonvanishing Lyapunov exponents. A system is thus hyperbolic when it has a single hyperbolic invariant subset.

Moreover, a system is {\em continuous} if the stable and unstable manifolds of its trajectories extend without rupture in phase space.

\citet{anosov_1967} proposed a class of systems which are today known as {\em Anosov systems} and exhibit the following properties:
\begin{enumerate}
	\item [(i)]   the nonwandering set ${\cal W}_{\bf \Phi}$ is continuously hyperbolic;
	\item [(ii)]  the fixed points and periodic orbits is dense in ${\cal W}_{\bf \Phi}$;
	\item [(iii)] the whole phase space is made up of nonwandering points: ${\cal W}_{\bf \Phi} = {\cal M} \; .$
\end{enumerate}

The interest induced by the Anosov systems has given risen to more general hyperbolic systems, such as the so-called {\em Axiom-A} systems which do not require the condition (iii).


\section{Escape rate and escape-rate formula}\label{section-escape-rate-formula}

In section \ref{micro-chaos}, we mentionned the necessity of introducing techniques used in statistical thermodynamics for the study of chaotic dynamical systems. The theory obtained by Bowen and Ruelle is the so-called \textit{thermodynamic formalism}\footnote{See \citet{ruelle,beck-schlogl}}. The idea is to introduce a functional of physical observables which is the generating functional of the average and of the time correlation functions of the given observable $A (\mathbf X)$. This observable has to be averaged over given orbits of the invariant set $\cal A$ considered. With this aim, we introduce the notion of separated subsets.\footnote{For further information on this Appendix, see \citet{gasp-book}.}

A \textit{separated subset} ${\cal S} = {\mathbf Y_1, \ldots, \mathbf Y_s} \subset \cal A$ is composed of points which are separated by a distance
$d_T$ larger than $\epsilon$ over a time interval $[-T, +T]$, that is
\begin{equation}
d_T (\mathbf Y_i, \mathbf Y_j) = \max{-T \le t \le +T} \left | \left | \mathbf \Phi^t \mathbf Y_i -
\mathbf \Phi^t \mathbf Y_j \right| \right| > \epsilon, \; 
\end{equation}
$\forall i \not= j \in {1, \ldots, S}$.
If the invariant set $\cal A$ is bounded, one can always find a subset $\cal S$ with a finite number of points. This set is called an $(\epsilon, T)$-separated subset of the invariant set $\cal A$.\\

A central function for a given observable $A (\mathbf X)$ in this formalism is the \textit{topological pressure} which is defined as
\begin{equation}
\mathcal P(A) = \lim_{\epsilon \to 0} \; \lim_{T \to \infty} \frac{1}{2T} \ln \mathcal Z (\epsilon, T, A) \; ,
\end{equation}
with the partition functional
\begin{equation}
\mathcal Z (\epsilon, T, A) = \mathrm{Sup}_{\mathcal S} \sum_{\mathbf Y \in \mathcal S}^{} \exp \left ( \int_{-T}^{+T} A(\mathbf \Phi^t \mathbf Y)\; dt \right ) ,
\end{equation}
where $\mathcal S$ is a $(\epsilon, T)$-separated subset of the invariant subset $\mathcal A$.

If $B(\mathbf X)$ is another observable of the system, its average is defined as
\begin{eqnarray}
\left \langle B \right \rangle_{\mu_{A}} \equiv \mu_{A} (B) & = & \frac{d}{d \nu}\mathcal P (A + \nu B) | \nu=0 \nonumber \\
& = & \int_{}^{} B(\mathbf X) \mu_{A} (d\mathbf X) \; .
\label{B-average}
\end{eqnarray}
Using the definition of the pressure, we obtain an expression of this \textit{dynamical measure} $\mu_{\mathcal A}$
\begin{equation}
\mu_{A}(d \mathbf X)  = \lim_{\epsilon \to 0} \lim_{T \to \infty} \mathrm{Sup}_{\mathcal S}
\sum_{\mathbf Y \in \mathcal S}^{} \frac{\exp \left ( \int_{-T}^{+T} A (\mathbf{\Phi}^t \mathbf Y) \; dt \right )}{\mathcal Z(\epsilon, T, A)}
\frac{1}{2T} \int_{-T}^{+T} \delta (\mathbf X - \mathbf \Phi^t \mathbf Y) \; dt \; d\mathbf Y \; .
\end{equation}
Each trajectory of the subset $\mathcal S$ is weighted by a Boltzmann-type probability given by
\begin{equation}
\pi_{A}(\epsilon, T, \mathbf Y) = \frac{\exp \left ( \int_{-T}^{+T} A(\mathbf \Phi^t \mathbf Y) \; dt \right )}{\mathcal Z(\epsilon, T, A)} \; .
\label{pi-probability}
\end{equation}

The Kolmogorov-Sinai entropy per unit time with respect to this invariant measure $\mu_A$ is defined by
\begin{equation}
h_{\mathrm{KS}} (\mu_A) = - \lim_{\epsilon \to 0} \lim_{T \to \infty} \frac{1}{2T} \; \mathrm{Sup}_{\mathcal S}
\sum_{\mathbf Y \in \mathcal S}^{} \pi_A (\epsilon, T, \mathbf Y) \ln \pi_A (\epsilon, T, \mathbf Y) \; . 
\end{equation}
From Eqs. (\ref{B-average}) and (\ref{pi-probability}), we can deduce the important identity 
\begin{equation}
h_{\mathrm{KS}} (\mu_A) = -\mu_A (A) + \mathcal{P}(A) \; .
\label{entr-KS-pressure}
\end{equation}

An important particular choice for the observable $A (\mathbf X)$ is the following. $\beta$ being a real parameter, we take
\begin{equation}
A (\mathbf X) = - \beta \sum_{\lambda_i > 0}^{}\chi_i (\mathbf X)
\end{equation}
where $\chi_i$ are the local stretching rates related to the Lyapunov exponents by \citep{gasp-book}
\begin{equation}
\lambda_i (\mathbf X) = \lim_{t \to \infty} \frac{1}{t} \int_{0}^{t} \chi_i (\mathbf \Phi^\tau \mathbf X) \; d \tau \; .
\label{lyap-stretching}
\end{equation}
Using this observable, we observe that, for $\beta > 0$, the probability (\ref{pi-probability}) associated with a trajectory is larger for the more stable trajectories. The pressure functional becomes the \textit{pressure function} $P(\beta)$
\begin{equation}
P(\beta) = \mathcal P  \left \lbrack - \beta \sum_{\lambda_i > 0}^{} \chi_i (\mathbf X) \right \rbrack
\end{equation}
which defines an invariant probability measure $\mu_\beta$ depending on the parameter $\beta$. Since the local stretching rates and the Lyapunov exponents are related by Eq. (\ref{lyap-stretching}), and using the time invariance of the measure $\mu_\beta$ we have
\begin{equation}
\mu_\beta(\chi_i) = \mu_\beta(\lambda_i) \; .
\end{equation}
Therefore, Eq. (\ref{entr-KS-pressure}) becomes in this case
\begin{equation}
h_{\mathrm{KS}} (\mu_\beta) = \beta\sum_{\lambda_i > 0}^{} \mu_\beta (\lambda_i) + P(\beta) \; .
\label{futur-gamma}
\end{equation}

\subsection{Closed systems}

A closed system is a system in which any trajectory can escape. A time-independent Hamiltonian system presents the \textit{microcanonical measure} $\mu_e(d \mathbf X)$ as an appropriate invariant measure, which is given by
\begin{equation}
d\mu_e = \mathcal N \delta (H - E) \; d\mathbf q \; d\mathbf p \; .
\end{equation}
It can be shown that this measure corresponds to the measure associated with the observable $A = - \sum_{\lambda_i >0}^{} \chi_i (\mathbf X)$,
that is \citep{gasp-book}
\begin{equation}
\mu_e = \mu_{\beta = 1} \; .
\end{equation}
Furthermore, it can also be shown that, for closed systems, we have $P (\beta = 1) =0$, so that Eq. (\ref{entr-KS-pressure}) for $\beta = 1$ becomes
\begin{equation}
h_{\mathrm{KS}} (\mu_e) = \sum_{\lambda_i > 0}^{} \mu_e (\lambda_i) \; .
\end{equation}
Hence the Pesin's identity (\ref{pesin}) is recovered.

\subsection{Open systems}

Contrary to the closed systems, the open systems allow the escape of the trajectories out of the bounded phase-space domain $\mathcal M$. Furthermore, in such systems, an escaped trajectory is not allowed to reenter $\mathcal M$. The boundaries are therefore considered as absorbing and maintain the system in nonequilibium. As we will see below the support of the invariant measure that we may here choose is a \textit{fractal repeller}.

The construction of this measure is the following: let us consider a probability measure $\nu_0 (d \mathbf X)$ corresponding to an initial statistical
ensemble ${\mathbf X_0^{(i)}}$ on the phase-space domain $\mathcal M$. The measure $\nu_0$ is written as
\begin{equation}
\nu_0 (d \mathbf X) = \lim_{N_0 \to \infty} \frac{1}{N_0} \sum_{i=1}^{N_0} \delta (\mathbf X - \mathbf X_0^{(i)}) \; d \mathbf X \; .
\end{equation}
Because of the escape, after a time $T$ , only $N_T$ points from the initial ensemble are still in $\mathcal M$. The ratio of such points is given by
\begin{equation}
\lim_{N_0 \to \infty} \frac{N_T}{N_0} = \int_{\Upsilon_{\mathcal M}^{(+)}(T)}^{} \nu_0 (d \mathbf X)
\end{equation}
where $\Upsilon_{\mathcal M}^{(+)}(T)$ is the set of all the initial conditions $\mathbf X$ which escape out of $\mathcal M$ after a time $T_{\mathcal M}^{(+)} (\mathbf X)$ larger than $T$ (that is, the initial conditions of the trajectories still inside the absorbing boundaries at time $T$)
\begin{eqnarray}
T_{\mathcal M}^{(+)} (\mathbf X) & = & \max \left \lbrace T > 0 : \mathbf \Phi^t \mathbf X \in \mathcal M \right \rbrace \; ,\\
&& \forall t \in [0, T[ \;  \nonumber \\
\Upsilon_{T}^{(+)} (\mathbf X) & \equiv & \left \lbrace \mathbf X \in \mathcal M : T  < T_{\mathcal M}^{(+)} (\mathbf X) \right \rbrace \; .
\end{eqnarray}
The equivalent set of initial conditions $\Upsilon_{T}^{(-)} (\mathbf X)$ for backward evolution, is similarly obtained replacing $T$ by $-T$, $\mathbf \Phi^t$ and by $\mathbf \Phi^{-t}$.
 
The decay $\frac{N_T}{N_0}$ of the number of trajectories still in $\mathcal M$ is exponential since all the trajectories of the repeller are exponentially unstable. The exponential decay is characterized by an \textit{escape rate} given by
\begin{equation}
\gamma = - \lim_{T \to \infty} \frac{1}{T} \ln \nu_0 \left \lbrack \Upsilon_{\mathcal M}^{(+)} (T) \right \rbrack \; .
\end{equation}
If the system is ergodic, the time average of a dynamical quantity equals the its ensemble average. This is expressed as
\begin{eqnarray}
\mu_{\rm ne}(A) & = & \lim_{T \to \infty} \lim_{N_T \to \infty} \frac{1}{N_T} \sum_{i=1}^{N_T} \frac{1}{2T} \int_{-T}^{+T} A(\mathbf \Phi^t \mathbf X^{(i)}) \; dt
\nonumber \\
& = & \int_{}^{} A(\mathbf X) \mu_{\rm ne} (d \mathbf X)
\end{eqnarray}
where $N_T$ is the number of phase-space points remaining in the system during the time interval \break{$] -T, +T[$}. This allows us to write $\mu_{\rm ne}$ as
\begin{equation}
\mu_{\rm ne}(d \mathbf X) = \lim_{T \to \infty} \frac{1}{\nu_0 [\Upsilon_{\mathcal M} (T)]}
\int_{}^{} \nu_0 (d \mathbf y) I_{\Upsilon_{\mathcal M}(T)} (\mathbf y)
\frac{1}{2T} \int_{-T}^{+T} \delta (\mathbf X - \mathbf \Phi^T \mathbf y) \; dt \; d\mathbf X
\end{equation}
backward as well as forward evolution in order to get the invariant measure.

It can be shown that $\mu_{\rm ne}$ corresponds to the invariant measure associated with the observable $A = - \sum_{\lambda_i > 0}^{} \chi_i (\mathbf X)$ for $\beta = 1$ \citep{gasp-book}
\begin{equation}
\mu_{\rm ne} = \mu_{\beta = 1} \; .
\end{equation}
The escape rate is related to the topological pressure by
\begin{equation}
P(\beta = 1) = -\gamma \; .
\end{equation}
Thanks to this result, Eq. (\ref{futur-gamma}) becomes the generalized Pesin's identity available even for open systems. Hence we obtain the
so-called \textit{escape-rate formula} \citep{gasp-book,eckmann-ruelle}
\begin{equation}
\gamma = \sum_{\lambda_i > 0}^{} \mu_{\rm ne} (\lambda_i) - h_{\mathrm {KS}}(\mu_{\rm ne})
\label{escape-rate-formula-theorique}
\end{equation}
which plays an important role in the escape-rate formalism.


\bibliographystyle{model5-names}
\bibliography{bibliobibtex}

\end{document}